# Renormalization of Electromagnetic Quantities in Small Josephson Junctions

Doctorate Thesis:

# Renormalization of Electromagnetic Quantities in Small Josephson Junctions

by

**Godwill Mbiti Kanyolo**

Defense Committee Members (Thesis Advisors):
*Prof. Hiroshi Shimada, Prof. Yoshinao Mizugaki*

Defense Committee Members (others):
*Prof. Nobuhito Kokubo, Prof. Yuki Fuseya, Prof. Takeo Kato*

Date of Defense: August 11, 2020
Date of Thesis Submission: September 25, 2020

The University of Electro-Communications
1-5-1 Chofugaoka, Chofu, Tokyo 182-8585, Japan

**Dedication**

This doctorate thesis is dedicated to my immediate family
(P. K. Ngumbi, J. M. Mbiti, T. N. Kanyolo and B. M. Kanyolo) and friends for all their well wishes and support over the years.

**Acknowledgements**

I wish to thank my thesis advisors (Prof. Shimada and Prof. Y. Mizugaki) as well as the thesis committee (Prof. N. Kokubo, Prof. Y. Fuseya and Prof. T. Kato) for their invaluable guidance during the course of this work, Dr. T. Masese and K. Takeda for their insightful discussions, B. M. Kanyolo and N. Ito for help in simulations, and the members of the Shimada, Mizugaki and Kokubo laboratories at The University of Electro-Communications for their valuable suggestions. I appreciate the technical assistance by J. Kamekawa, H. Nishigaki, T. Suzuki and thank W. Kuo, Y. Nakamura and Y. Iwazawa for discussion and support. This work was supported by JSPS KAKENHI Grants Number 24340067 and 18H05258. Part of this work was conducted at the Coordinated Center for UEC Research Facilities, The University of Electro-Communications, Tokyo, Japan. The stable supply of liquid Helium from it is also acknowledged.

**Related Publications**

1) Godwill Mbiti Kanyolo, Kouchi Takeda, Yoshinao Mizugaki, Takeo Kato and Hiroshi Shimada. Cooper-Pair Tunneling in Small Josephson Junction Arrays Under Radio-Frequency Irradiation. In: *Journal of Low Temperature Physics* (2020) **210**, pp. 269. [arxiv:1911.02519];
2) Godwill Mbiti Kanyolo, Hiroshi Shimada. Rescaling of Applied Oscillating Voltages in Small Josephson Junctions. In: *Journal of Physics Communications* (2020) **4**, 10 pp. 105007. [arXiv:1911.10899].

**Availability of Thesis**

1) The University of Electro-Communications library (version, Sept. 2020)
2) By requesting G. M. Kanyolo via email: gmkanyolo@gmail.com (latest version)

**Copyright**

**Disclaimer**

This doctorate thesis includes whole/partial sections of the works submitted for peer review and publication to journals as partial fulfilment of PhD graduation requirements and in accordance with standard journal policies e.g. Elsevier. Journal information *etc.* is provided in the information section of their preprints (arXiv:1911.10899 and arXiv:1911.02519) in the ar$\chi$iv repository.

# Contents

# List of Figures

# List of Tables

# Abstract

A Josephson junction is a Superconductor/Insulator/Superconductor tunnel junction that admits a supercurrent across it even in the absence of applied voltages. Due to its non-linear response to applied electromagnetic fields, Josephson junctions have found varied metrology and detector applications. Moreover, the junction size determines the current ($I$)–voltage ($V$) characteristics observed in experiments. The small Josephson junction can be thought of as a dual system to the large junction due to the interchange of quantities such as charge and flux or current and voltage in the characteristics of the junctions. In particular, small (mesoscopic) junctions are known to be greatly susceptible to quantum fluctuations and changes in the electromagnetic environment compared to large junctions. This means that standard (quantum) phase dynamics inadequately describes dual characteristics such as Coulomb blockade. Consequently, the standard theory of Coulomb blockade in superconducting and normal junctions is formulated instead on the basis of phase-phase correlations. This leads to complex theoretical considerations albeit richer physics such as circuit-quantum electrodynamics (QED) in the small junction whose features are often captured by the so-called $P(E)$ theory. This complexity scales with the number of junctions connected in series forming an array.

Due to the aforementioned non-linearity, predictions with $P(E)$ theory has been marred by several predictive limitations which require the rescaling of measurable quantities appearing in the $P(E)$ function. Consequently, a handful of experimental and theoretical studies with oscillating electromagnetic fields has been conducted with small junctions to date due to their strong coupling to the electromagnetic environment, since the coupling significantly modifies the $I$–$V$ characteristics compared to the large junction.

In this thesis, we focus on the effect of the electromagnetic environment on applied electromagnetic fields in single small junctions as well as arrays. We apply microwaves (RF) in the sub-gigahertz frequency range on a one-dimensional array of small Josephson junctions exhibiting distinct Coulomb blockade characteristics. We observed a gradual lifting of Coulomb blockade with increase in the microwave power which we interpret is due to photon-assisted tunneling of Cooper pairs in the classical (multi-photon absorption) regime. We observe that, due to its high sensitivity to microwave power, the array is well-suited for in situ microwave detection applications in low temperature environments. A detailed analysis of the characteristics in the classical (multi-photon absorption) limit reveals that the microwave amplitude is rescaled (renormalized), which we attribute to the difference in dc and ac voltage response of the array.

We proceed to rigorously consider the origin of the aforementioned renormalization effect by considering the effect of the electromagnetic environment of the Josephson junction on applied oscillating voltages. We theoretically demonstrate that its effect is simply to renormalize the amplitude of oscillation in a predictable manner traced to the physics of wavefunction renormalization (Lehmann weights) consistent with circuit-QED. Such Renormalization implies that the sensitivity of the single junction and the array to oscillating electromagnetic fields (e.g. microwaves) is modulated and depends on the environmental impedance as well as the junction capacitance.

# Preface

Since the pioneering theoretical work by Likharev and co-workers [1, 2], small Josephson junctions have been thought of as a dual system to large Josephson junctions – the roles of current and voltage are interchanged. In the case of large Josephson junctions, their effective interaction with oscillating electromagnetic fields has intensively been studied, demonstrating their unique suitability for microwave-based applications such as the metrological standard for the *Volt* (in terms of voltage ($V$) Shapiro steps) and other microwave-based devices.[3–5] Thus, the dual system holds enormous promise for complementary applications such as a metrological standard for the *Ampere* in terms of the current ($I$) Shapiro steps.[1, 2] However, due to their mesoscopic size, microwave based studies with oscillating electromagnetic fields face daunting experimental and theoretical challenges due, in part, to the lack of a theory that consistently covers both regimes.[6]

In particular, small junctions are prone to quantum and thermal fluctuations, thus a well-known (fluctuation-dissipation) theorem applies.[7–9] Due to their dissipative nature, the characteristics of small junctions generally cannot be analyzed separate from their fluctuative environment.[10, 11] Heuristically, as a consequence of Heisenberg uncertainty principle, their $I$–$V$ characteristics is highly sensitive to energy changes in the environment.[12] For instance, tunneling of a single charge $e$ across a tunnel junction of capacitance $C$ and conductance $1/R$ is restricted unless the maximum energy $\hbar/RC$ it can absorb from the electromagnetic vacuum through zero-point oscillations is sufficient to offset its own charging energy $e^2/2C = E_\text{c}$ in the absence of other energy sources. Here, $e$ and $\hbar$ respectively denote the elementary charge and the reduced Planck constant. Thus, the necessary condition for Coulomb blockade to occur is $2\pi\hbar/RC < E_\text{c}$. On the other hand, lifting of Coulomb blockade occurs when other sources of energy are present. For instance, energy is easily supplied by thermal fluctuations $k_\text{B}T > 0$, large Josephson coupling energy $E_\text{J}$ across the junction $E_\text{J} \gg E_\text{c}$ or external voltages $V_\text{x} > V_\text{cb} \sim N_0 E_\text{c}$ above the Coulomb blockade threshold voltage $V_\text{cb}$, where $N_0$ is the number of junctions in an array. This results in $I$–$V$ characteristics highly dependent on these environmental parameters. Formally, within the context of circuit-quantum electrodynamics (QED), this implies that the classical action for small Josephson junctions is effective – it emerges from tracing out irrelevant degrees of freedom.[*]

This leads to complex theoretical considerations albeit richer physics such as circuit-QED in the small junction whose features are often captured by the so-called $P(E)$ theory.[13–16] This complexity scales with the number of junctions connected in series forming an array. In particular, the $P(E)$ theory of dynamical Coulomb blockade in single small Josephson junctions is formulated on the basis of phase correlation functions[15, 16] where tunneling across the barrier is influenced by a high impedance environment treated within the Caldeira-Leggett model.[10] $P(E)$ theory has successfully been tested to a great degree of accuracy in a myriad of experiments.[17–21] This has lead to its widespread application in describing progressively complex tunneling processes such as dynamical Coulomb blockade in small Josephson junctions and quantum dots.[17, 22] Moreover, owing to significant improvement in microwave precision measurement technology such as near-quantum-limited amplification[23, 24] and progress in theory, recently published works suggest novel features in the $P(E)$ framework ranging from time reversal symmetry violation[25] and Tomonaga-Luttinger Liquid (TLL) physics,[26] to renormalization of electromagnetic quantities appearing in the $P(E)$ function.[27–31] Despite this progress, aspects of the theory remain elusive especially in the case of one dimensional arrays.

---

[*] In the case of the large junction, it is necessary to trace out the environmental degrees of freedom that act as energy sources for tunneling charges.

$P(E)$ theory was recently extended to account for the effect of excited environmental modes by an alternating voltage in single normal junctions.[27] Novel features in the theory not yet observed experimentally include the renormalization of the radio frequency (RF) power absorbed by the junction as well as higher harmonic modifications of the time-averaged current.[32, 33] Moreover, the Caldeira-Leggett form of the environmental impedance neglects the back-action of the Josephson junctions on the environment (with the bath and the junction becoming entangled) which has been reported to dramatically change the predictions of the $P(E)$ theory.[31, 34, 35] This back-action manifests through the non-linear inductive response of the junction where the Josephson coupling energy is renormalized and the insulator-superconductor phase transition conditions for the single Josephson junction are altered.[35]

Due to this high sensitivity of small junctions to such environmental parameters, special considerations and techniques are required to observe such dual characteristics as Coulomb blockade and Bloch oscillations in single small Josephson junctions.[24, 34–37] However, for a one-dimensional array, Coulomb blockade is easily observed when the junction parameters such as the tunnel resistance $R_\mathrm{T}$ and the capacitance $C$ of each junction are adequately chosen, such that $E_\mathrm{J} < E_\mathrm{c} = e^2/2C$ and $R_\mathrm{T} > R_\mathrm{Q} = \hbar/e^2$, without any other special considerations of the environment.[38] Thus, one-dimensional arrays are more suitable than single junctions to studying the interaction of small Josephson junctions with RF electromagnetic fields (microwaves). Moreover, the environment of the superconducting array (as well as the single junction) is susceptible to an externally-applied magnetic field $H$ through the quotient $E_\mathrm{J}(H)/E_\mathrm{c}$ that governs the dynamics of the quasi-charge of each Josephson junction in the array within their respective Brillouin zone of the Bloch energy band.[2] In particular, the energy band gap, which is comparable to $E_\mathrm{J}(H)$, is diminished by applying a magnetic field $H \leq H_\mathrm{max}$ where $H_\mathrm{max}$ is the value of the magnetic field that leads to the largest Coulomb blockade of Cooper-pairs.

Pioneering experimental work by Delsing intensively studied the weak Josephson coupling limit $E_\mathrm{J} \ll E_\mathrm{c}$ and high tunnel resistances $R_\mathrm{T} > R_\mathrm{Q}$, where Cooper-pair tunneling is virtually non-existent and the quasi-particle current is dominant.[32] This work observed the lifting of Coulomb blockade of the quasi-particle current by a thermal bath and external microwaves, in addition to small peak structures in the differential conductance at currents $I_n = nef$, with $f$ being the frequency of the microwave and $n$ integer, corresponding to microwave phase-locked narrow-band single electron tunneling oscillations previously predicted by Averin and Likharev.[39] Recently, Billangeon et al. studied the ac Josephson effect and Landau-Zener transitions by detecting, through photon-assisted quasi-particle tunneling in a superconductor-insulator-superconductor junction, the microwave emission by a single Cooper-pair transistor (SCPT) in the Cooper-pair transport dominant regime.[40] In turn, by using photo-resistance measurement, Liou et al. studied the modulation of the $I$–$V$ characteristics of a 1D array of small dc superconducting quantum interference devices (SQUIDs) with the application of RF electromagnetic fields (microwaves) in the phase-charge crossover regime.[33, 41]

## Motivation

However successful, $P(E)$ theory has not been comprehensively tested especially with Josephson junction arrays. This is because of the aforementioned complexity due to the strong coupling of junctions with the electromagnetic environment, which scales with the number of junctions. This makes the determination of the environmental impedance of the junction taxing and sometimes impossible. Moreover, $P(E)$ theory cannot account for key photon-assisted tunneling features of Cooper pairs and quasi-particles, even for the simplest case of two small Josephson junctions in series forming a superconducting single electron transistor (SSET).[42] Here, Cooper-pair tunneling is incoherent while tunneling events are expected to be uncorrelated[43] leading to a dissipative processes such as the Josephson quasi-particle cycle (JQP).[44] Strong coupling to the electromagnetic environment has broad implications for the interaction of arrays with electromagnetic fields. For instance, a voltage biased array with no special coupling, fabricated adjacent to another unbiased array with a similar structure, has been shown to induce a strongly correlated current through the latter.[45, 46] The current has the characteristic that reversing the polarity of the bias voltage does not reverse the polarity of the induced current. Thus, such effects show the need for further theoretical and experimental studies pertaining the interaction of single Josephson junctions as well as arrays with the electromagnetic environment. This serves as the motivation for the research presented herein.

## Experimental work

We conducted an experiment to examine the effect of microwaves on an array of small Josephson junctions satisfying $0.1 < E_J/E_c < 1$ and $R_T > R_Q$ by measuring its current voltage ($I$–$V$) characteristics. Under these conditions, the tunneling of charges at small voltages is dominated by Cooper pairs, and the characteristics exhibited are in the charge regime, dual to the phase regime. However, Cooper-pair tunneling can easily be precluded by the electromagnetic environment of the array, leading to Coulomb blockade. In our experiment, the Coulomb blockade of tunneling Cooper-pairs was steadily diminished when radio-frequency electromagnetic radiation was applied, independent of frequency $f = \Omega/2\pi$ in the sub-gigahertz band $1\,\text{MHz} \leq f \leq 1000\,\text{MHz}$ with $\hbar\Omega \leq k_B T$. The observed diminishing of Coulomb blockade with microwave radiation is dual to the phase diffusion effect reported in Liou et al in ref. [41] for a one dimensional array of Josephson junctions in the regime, $E_J/E_c > 1$.

In the experiment, a substantial non-varying magnetic field, $H_{max}$ = 500 Oe is applied perpendicular to the unirradiated array in order to raise the value of the Coulomb blockade (threshold) voltage $V_{cb}$ to its maximum. This corresponds to a factor of approximately 1.4 its original value for $H$ = 0 Oe.[38, 47] Nonetheless, the $V_{cb}$ versus $V_{ac}$ characteristics of the irradiated array when $H$ = 500 Oe coincide with those for $H$ = 0 Oe when both axes of the $V_{cb}$–$V_{ac}$ plots are rescaled by the aforementioned factor, 1.4.

The experimental results are analysed by simulating the $I$–$V$ characteristics of the irradiated array using well-known equations[15, 48, 49] describing Photon-assisted tunneling in the multi-photon absorption regime $\hbar\Omega \ll 2eV_{ac}$. Comparing the simulated curves with the experimental results by plotting $V_{cb}$ versus $V_{ac}$ curves, we discover that, a mismatch of a factor, 0.87 persists between the values of the absorbed microwave power by the array in the experiment and the values corresponding the simulated curves with the same Coulomb blockade threshold voltage even after calibration of the microwave line. This factor is neither dependent on frequency nor the applied magnetic field after rescaling the $V_{cb}$–$V_{ac}$ axes by 1.4.

We discuss other possible origin of this mismatch by considering the uncertainties relating to the microwave generator, transmission line calibration procedure and the influence of electron heating at the islands of the array by ruling all out. Consequently, we conclude that a possible voltage division effect in the array leads to the renormalization of the microwave amplitude by a factor, $\Xi_A \sim \exp(-\Lambda^{-1}) \simeq 0.89$ comparable to the observed mismatch, where $\Lambda$ is the length over which the applied microwave is damped from the edge into the array (soliton length).[41, 50–52]

## Theoretical work

We focus on the effect of the electromagnetic environment on applied electromagnetic fields in one dimensional arrays of Josephson junctions. In particular, we apply path integral formalism to re-derive the Cooper-pair current and the BCS quasi-particle current in single small Josephson junctions and apply it to long Josephson junction arrays. We consider rescaling (renormalization) effects of applied oscillating voltages due to the impedance environment of a single junction as well as its implication to the array. For the single junctions, the amplitude of applied oscillating electromagnetic fields is renormalised by the same complex-valued weight $\Xi(\omega) = |\Xi(\omega)| \exp i\eta(\omega)$ which rescales the environmental impedance in the $P(E)$ function. Since the quasi-particle current naturally reduces to the normal current and the Cooper pair current vanishes when the superconducting gap vanishes $\Delta = 0$, the final expression of the tunneling current with renormalization effects in the single junction is essentially the recently proposed time averaged current result.[27] Our approach which, is restricted to the damping of the microwave amplitude for multi-photon-assisted tunneling in single junctions as well as arrays, qualitatively differs from other approaches e.g. reported in ref. [28, 29] and ref. [53] where amplification effects are also discussed.

In the approach herein, the weight acts as a linear response function for applied oscillating electromagnetic fields driving the quantum circuit, leading to a mass gap in the thermal spectrum of the electromagnetic field. The mass gap can be modeled as a pair of exotic 'particle' excitation with quantum statistics determined by the argument $\eta(\omega)$. We also consider arrays of small Josephson junctions, where the dynamics of Cooper-pair/charge solitons[50] become important. In the case of the array, this pair corresponds to a bosonic charge soliton/anti-soliton pair injected into the array by the electromagnetic field. When an infinitely long array[32, 51] is modeled as half the infinite array interacting with two junctions, one at the array edge and the other at the center, we find an additional Lehmann weight $\Xi_A = \exp(-\Lambda^{-1})$ compared to the case of the single junction. This requires that applied oscillating voltages are damped by this factor across a range $\Lambda$ along the array given by the soliton length of the array.

## Significance

The above experimental results demonstrate pristine Josephson junction arrays are poised for microwave detection applications in a wide range of environments such as on-chip detection schemes[54] due to their high sensitivity to low-power, of order $10^6$ V/W, whereas the renormalization effect can be exploited to configure 'opaque' $\Xi = 1$, 'translucent' $0 < \Xi < 1$ or 'transparent' $\Xi = 0$ quantum circuits to microwave radiation.

# Convention, Notation and Units

Throughout the sections of the theoretical work, we set reduced Planck's constant and Swihart velocity[55] $\hbar = \bar{c} = 1$ respectively to unity. Whenever Greek indices appear, Einstein summation convention is used with diag $\{\eta_{\mu\nu}\} = (1, -1, -1, -1)$ the Minkowski space-time metric and $\eta_{\sigma\mu}\eta^{\sigma\nu} = \delta_\mu^\nu$ the Kronecker delta symbol. Define integrals in the interval $[-\infty < t < +\infty]$ and $[-\infty < \omega < +\infty]$ are displayed as $\int dt$ and $\int d\omega$ without including the infinite signs. Other relevant notations and units are summarized in the Table below:

# Quantities, Symbols and SI units

| quantity [SI units] | quantity [SI units] |
|---|---|
| voltage, $V$ [V] | current, $I$ [A] |
| microwave amplitude, $V_{ac}$ [V] | Coulomb blockade voltage, $V_{cb}$ [V] |
| Boltzmann constant, $k_B$ [JK$^{-1}$] | reduced Planck's constant, $\hbar$ [Js] |
| charging energy, $E_c$ [J] | Josephson coupling energy, $E_J$ [J] |
| quantum phase, $\phi$ | charge, $Q$ [C] |
| vacuum speed of light, $c$ [ms$^{-1}$] | Swihart velocity, $\bar{c}$ [ms$^{-1}$] |
| electromagnetic tensor, $F_{\mu\nu}$ | Minkowski metric tensor, $\eta_{\mu\nu}$ |
| Levi-Civita symbol, $\varepsilon_{ijk}$ | |
| electric field, $F_{0i} \equiv \vec{E}$ [Vm$^{-1}$] | magnetic field, $\frac{1}{2}\varepsilon_{ijk}F^{ij} \equiv \vec{B}$ [Am$^{-1}$] |
| quantum resistance (electron), $R_Q$ [Ω] | electron charge, $e$ [C] |
| Fourier transform, $F(\omega) = \frac{1}{2\pi}\int dt f(t)\exp(i\omega t)$ | |
| Inverse Fourier transform, $f(t) = \int d\omega F(\omega)\exp(-i\omega t)$ | |
| Fourier transform angular frequency, $\omega$ [Hz] | microwave frequency, $f = \Omega/2\pi$ [Hz] |
| permittivity (vacuum), $\varepsilon_0$ | relative permittivity $\varepsilon_r$ |
| inductance, $L$ [H] | capacitance, $C$ [F] |
| impedance (environment), $Z(\omega)$ [Ω] | impedance ($LC$ circuit), $1/i\omega C + i\omega L$ [Ω] |
| inverse temperature, $\beta$ [K$^{-1}$] | temperature, $T$ [K] |

# Introduction 1.

## 1. A brief history: Josephson junctions

The 20th century saw impressive advancements in physics and technology. In the heart of this advancement is quantum mechanics and quantum field theory which explain with stunning accuracy microscopic phenomena and their macroscopic manifestations. The macroscopic manifestation related to this treatise is superconductivity. In the advent of discoveries related to superconductivity, such as perfect conductivity and perfect diamagnetism (Meissner effect), little was known as to what actually produced these effects in specific metals, more so how superconducting electrons differed from the normal electrons. Theories by Fritz and Heinz London, Anderson et al. [56] did a great deal to better our understanding on phenomenological superconductivity. Feynman et al. tried applying perturbation techniques to the behaviour of electrons in superconductivity using Feynman diagrams with limited success[57]. The theory was finally realized by the now well-known trio, Bardeen, Cooper and Schrieffer (BCS)[58]. In the language of quantum fields (second quantization), the BCS theory incorporates paired electrons (with opposite momenta and spin) that are weakly coupled to each other by phonons. These phonons can be exchanged between the coupled electrons in momentum space near the Fermi-surface via the crystal lattice, thus conserving their momenta and avoiding any effective scattering. This creates a bound state of paired electrons known as Cooper pairs. The consequence of this state is the coherence of the superconducting state across large distances (long range order). The Cooper pairs take the character of Bosons and, like photons (unlike electrons) can condense to form a degenerate ground state known as a Bose Einstein condensate, which is phenomenologically described by the Ginzburg-Landau theory of second order phase transitions.

In this picture, the Cooper pair condensate is a charged superfluid described by a macroscopic wavefunction $\psi = \sqrt{n}\exp(i2\varphi)$, where $n$ is the number of Cooper pairs and $2\varphi$ is the quantum phase of the coherent superconducting state. Microscopically, the BCS ground state has an energy gap $\Delta$ that represents the minimum energy required to break a Cooper pair into quasi-particles without breaking the superconducting state. Then the diamagnetic nature of superconductivity (Meissner effect) can be explained by the decrease of the superconducting gap by external magnetic fields. BCS theory, however, does little to explain a new class of high-temperature superconductors, where gapless superconductivity[56] and other (strong) coupling mechanisms, whose origin may be topological, have been observed.

# 2. Introduction to single Josephson junctions

## Large Josephson junctions

A subsequent discovery was made by then a PhD student, Josephson[59], who showed that the superconducting state remains coherent even across thin dielectric insulators (Fig. 1.1) by introducing the now well-known equations,

$$I_S = -2eE_J \sin 2\phi, \tag{1.1}$$

$$\frac{\partial \phi}{\partial t} = eV, \tag{1.2}$$

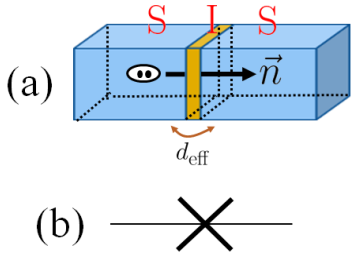

**Figure 1.1.:** Josephson junction: (a) A schematic of the Superconducting–Insulating–Superconducting, SIS tunnel junction (Josephson junction) depicting a Cooper pair tunneling across a barrier of effective thickness $d_{\text{eff}}$ along a direction $\vec{n}$ perpendicular to the barrier. (b) The electrical circuit symbol for a single Josephson junction.

where $\phi = \varphi_2 - \varphi_1$ is the quantum phase difference of the two superconductors, $I_S$ is the superconducting current, $E_J$ is known as the Josephson coupling energy and $V$ the applied biasing voltage through the junction.

Thus, the Josephson junction is a Superconductor/Insulator/Superconductor (SIS) tunnel junction that admits a supercurrent across it. The Josephson coupling energy depends on temperature via the BCS energy gap, $\Delta(T) = 1.74\Delta(0)(1-T/T_c)$,

$$E_J(T) = G_N \frac{\pi \Delta(T)}{4e^2} \tanh \frac{\Delta(T)}{2k_B T}, \tag{1.3}$$

where $\Delta(0)$ is the BCS energy gap at zero absolute temperature, $G_N$ is the normal conductance of junction measured at high bias voltages, $K_B$ is Boltzmann's constant and $T$ is the temperature.[60]

Due to its non-linear response to applied electromagnetic fields, Josephson junctions have found varied applications ranging from detection of electric and magnetic fields, low signal amplification, single electron transistors to metrology.[4, 54, 61] This versatility arises due to the quantum nature of the non-linear Josephson current, $I_S = 2eE_J \sin 2\phi$ arising from Cooper-pair tunneling across the junction, where $\phi$ is the quantum phase across the junction. The Josephson current couples to the electric field $\vec{E}$ via the quantum phase $\partial \phi/\partial t = -ed_{\text{eff}} \vec{n} \cdot \vec{E}$ where $V = -d_{\text{eff}} \vec{n} \cdot \vec{E}$ is the bias voltage across the junction, $d_{\text{eff}}$ is the effective thickness of the barrier and $\vec{n}$ is the unit normal vector in the tunneling direction across the barrier. This implies that the supercurrent, $I_S$ persists as a direct current even when when the electric field vanishes, $\vec{E} = 0$, the hallmark of superconductivity. In turn, the presence of electromagnetic fields causes the supercurrent $I_S(t) = 2eE_J \sin \omega_J t$ to oscillate with the Josephson frequency $\omega_J = 2eV$. Respectively, these are the direct current (dc) Josephson and the alternating current (ac) Josephson effects.

The exploitation of the dc and ac effects has led to the high precision standardization of the Volt [V] [62] through the observation of Shapiro steps[63, 64] by irradiating the large junction with microwaves whose oscillation frequency $\Omega$ phase-locks with the Josephson frequency at integer multiple values values, $\omega_J = n\Omega$ corresponding to the voltage values of the dc Josephson effect. This can be seen by plugging in $\partial \phi/\partial t = eV - eV_{\text{ac}} \cos \Omega t$ into $I_S = 2eE_J \sin 2\phi$ and using the identities, $\sin(x \sin y) = \sum_{n=-\infty}^{\infty} J_n(x) \sin ny$ and $\cos(x \sin y) = \sum_{n=-\infty}^{n=\infty} J_n(x) \cos ny$ where $J_n(x) = (-1)^{-n} J_{-n}(x) = \frac{1}{2\pi} \int ds \times \exp i(x \sin s - ns)$ is the Bessel function of the first kind and $n$ are integers, to yield $I_S = 2eE_J \sum_{-\infty}^{\infty} J_n(2eV_{\text{ac}}/\Omega) \sin(\omega_J t - \Omega nt + \phi_0)$ where the condition for the dc supercurrent is at the resonant frequency modes $\Omega_J = n\Omega = 2eV_n$ as shown in Fig. 1.2.

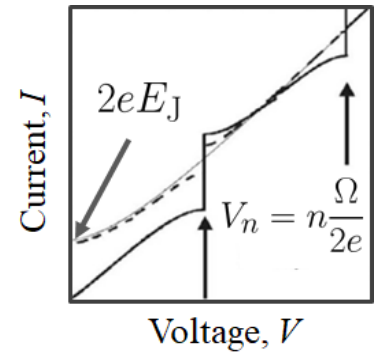

**Figure 1.2.:** Shapiro steps in a large Josephson junction with critical current $I_c = 2eE_J$. The dc Josephson currents occur at voltages $V_n = n\Omega/2e$, where $\Omega$ is the frequency of the ac voltage applied across the junction.

### Superconducting quantum interference device (SQUID)

The superconducting quantum interference device (SQUID) is a device composed of two parallel Josephson junctions as shown in Fig. 1.3. Because of the loop formed by such a design, and eq. (1.2), the current across each junction, $I_1 = E_J \sin 2\phi_1$ and $I_2 = E_J \sin 2\phi_2$, is not independent but are related by a gauge transformation due to the gauge invariance of the Cooper pair quantum phases, $\phi_1 = \phi + e \int_1^2 \vec{dl} \cdot \vec{A}$ and $\phi_2 = \phi + e \int_2^1 \vec{dl} \cdot \vec{A}$, where $\vec{l}$ is the length along the loop. Thus, the total current across it is given by,

$$I = E_J \sin 2\phi_1 + E_J \sin 2\phi_1$$
$$= 2E_J \cos(\phi_2 - \phi_1) \sin(\phi_1 + \phi_2) = E_J^* \sin 2\phi,$$

where $E_J^* = 2E_J \cos(eBA)$ and we have used, $\int_2^1 \vec{dl} \cdot \vec{A} = -\int_1^2 \vec{dl} \cdot \vec{A}$ and $\int_1^2 \vec{dl} \cdot \vec{A} - \int_2^1 \vec{dl} \cdot \vec{A} = \oint \vec{dl} \cdot \vec{A} = \int B dA$ where $A$ is the area enclosed by the SQUID loop and $B$ an applied magnetic field through the loop. Using the modulus, $|E_J^*|$, the condition that the current is maximum is $BA = n2\pi/2e$, where $n$ is the number of quantized flux $2\pi/2e$ stored in the loop. This quantization makes the SQUID an extremely sensitive detector of magnetic fields.[65] Finally, it is worth noting that the SQUID can merely be treated as a Josephson junction with a renormalized coupling energy $E_J \rightarrow E_J^* = E_J \cos(eBA)$.[66] Thus, in the discussions that follow, we shall not distinguish between the SQUID and the Josephson junction.

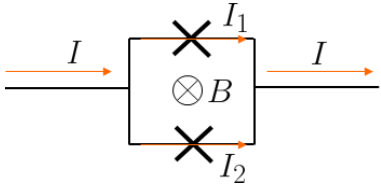

**Figure 1.3.:** A schematic of a superconducting quantum interference device (SQUID): two Josephson junctions (labeled by a cross) connected in parallel with a magnetic field $B$ applied through the loop in the direction of the circled cross $\otimes$. Each Josephson junction admits a current $I_1$ and $I_2$ respectively. The total current of the circuit is given by $I = I_1 + I_2$.

### Josephson Hamiltonian

The Josephson equations can be expressed using a conserved Hamiltonian,

$$H = \frac{Q^2}{2C} - E_J \cos 2\phi, \qquad (1.4)$$

where $Q = \varepsilon_0^{-1} \varepsilon_r^{-1} \int d\mathcal{A}\, \vec{n} \cdot \vec{E} = CV$ is the charge stored by the junction of capacitance $C = \varepsilon_0 \varepsilon_r \mathcal{A}/d_{\text{eff}}$ and cross-sectional area $\mathcal{A}$ and $\partial Q/\partial t = I_S$.[1] The Josephson equations are then given by Hamilton's classical equations of motion,

$$\frac{\partial H}{\partial \phi} = e^{-1}\frac{\partial Q}{\partial t},\ \frac{\partial H}{\partial Q} = e^{-1}\frac{\partial \phi}{\partial t},$$

respectively. Much of the aforementioned success is traced to the successful exploitation of the duality between the quantum phase $\phi$ and the stored charge, $Q$, which act as quantum

---

1: These conditions are simply Maxwell's equations, $\vec{\nabla} \cdot \vec{E} = \rho_s/\varepsilon_0 \varepsilon_r$ and charge conservation law (equation of continuity), $\partial \rho_s / \partial t = -\vec{\nabla} \cdot \vec{J}_s$ where $\rho_s, J_s$ are the charge and current densities respectively.

mechanical conjugate variables satisfying the commutation relations, $[Q, \phi] = -i2e$ where $e$ is the unit charge of a single electron. This is the origin of much of the interesting physics of the large and small Josephson junction.

## Small Josephson junctions: phase and charge duality

Duality refers to two related concepts: 1) invariant symmetry i.e. two variables (conjugate or otherwise) that when interchanged leave the Hamilton's equations of motion invariant; 2) The existence of two quantum mechanical conjugate pairs related by Heisenberg's uncertainty principle.[67] It is clear that as Josephson equations stands, there is an apparent lack of duality type 1 with respect to the current, $I_S$ and voltage, $V$ across the junction. Nonetheless, duality type 1 and type 2 are related through second quantization using the Josephson Hamiltonian given eq. (1.4).

In particular, taking the charge $Q = -i2e\partial/\partial\phi$ and phase $\phi$ as quantum mechanical conjugate variables, we can express eq. (1.4) as,

$$H = -4E_c \frac{\partial^2}{\partial \phi^2} - E_J \cos 2\phi, \qquad (1.5)$$

where $E_c = e^2/2C$ is the charging energy, which satisfies the Schrodinger equation,

$$i\frac{\partial \Psi}{\partial t} = H\Psi, \qquad (1.6)$$

where $\Psi$ is the wavefunction of the junction. The junction itself behaves like a quantum mechanical object when the 'quantum mechanical' kinetic energy term $-4E_c\partial^2/\partial\phi^2$ that acts on the wavefunction $\Psi$ dominates over the 'classical' potential energy term $-E_J \cos 2\phi$. This introduces a dimensionless parameter $E_J/E_c$ which governs the dual physics of the large and small Josephson junctions.

Solving the time-dependent Schrodinger equation ($i\partial\Psi/\partial t = E$) requires solving the well-known Mathieu's equation, which describes a fictitious particle in a periodic cosine potential. The solution to this Eigenvalue problem is given by,

$$\Psi_n(\phi) = u_n(\phi)\exp(\phi q/2e). \tag{1.7}$$

Here, $u_n(\phi)$ is a periodic function satisfying the boundary condition $u_n(\phi) = u_n(\phi + 2\pi)$ and $q$ is known as the quasi-charge. Plugging this solution back into the Schrodinger equation gives the energy spectrum $E_n$ of the Josephson junction. The spectrum comprises Cooper pair $2e$-periodic functions in the energy and quasi-charge space (Brillouin zones) where the energy band structure is entirely determined by the ratio of $E_J/E_c$ as depicted in Fig. 1.4. The first Brillouin zone extends from $-e \leq q \leq +e$. In the limit, $E_J/E_c \ll 1$, the energy bands can be approximated as charge parabolas, $E_n \simeq q^2/2C$ with gaps of amplitude given by $\Delta E_n = 4E_c(E_J/4E_c)^{n/n^{n-1}}$. Conversely, in the limit $E_J/E_c \ll 1$, the ground state of the Josephson junction system takes the form $E_0 \simeq -E_c^* \cos(2\pi q/2e) \equiv U(q)$ where $E_c^* = 16\left(\omega_p^2\left[E_J/8E_c\right]^{1/2}/2\pi\right)^{1/2}\exp\left(-4E_J/\omega_p\right)$ where $\omega_p = 8E_JE_c$ is the plasma frequency. This leads to the dual equations,

$$V = \frac{d}{dq}U(q) = \frac{2\pi}{2e}E_c^* \sin(2\pi q/2e), \tag{1.8}$$

$$\frac{\partial q}{\partial t} = I. \tag{1.9}$$

Due to the dependence of the critical voltage $E_c^*/2e$ on $E_J$ in the exponent, applying a magnetic field diminishes the superconducting gap, which in turn diminishes $E_J$ and by extension also the critical voltage. Further introduction to the dynamics of the quasi-charge $q$ in the Brillouin zone can be found in refs. [1, 2, 68, 69].

According to these equations, no current flows within the region $V \leq E_c^*/2e$, the hallmark of Coulomb blockade. Thus, Coulomb blockade is dual to the dc Josephson effect. Moreover, when the current, $I$ exceeds this critical value, the voltage oscillates. This is dual to the ac Josephson effect. Numerical calculations considering a bias current $I_b$ as well as an interaction term in eq. (1.4) given by $H_{\text{int}} = \frac{1}{C^2R}\int Q^2 dt$ for the effect of the environment on the junction, have found a periodic time solution resulting from an oscillatory voltage across the Josephson junction with a frequency (dual to $\Omega_J$)

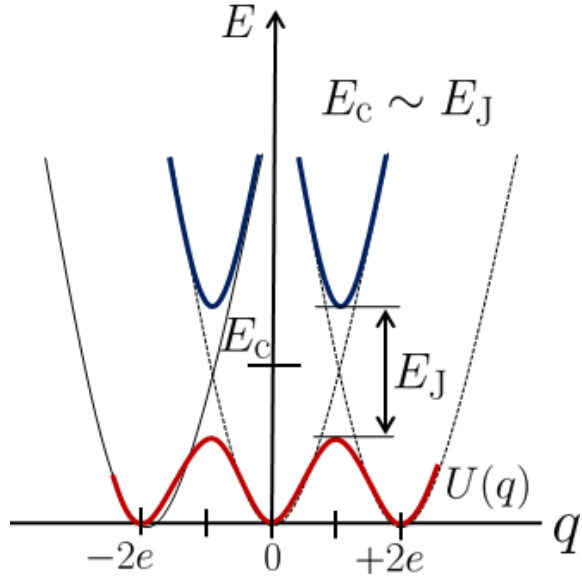

**Figure 1.4.:** The energy spectrum of the Josephson junction for $E_c \simeq E_J$ calculated from eq. 1.6, centered at the first Brillouin zone ($-e \leq q \leq +e$) where the Josephson coupling energy $E_J$ is the energy gap between the $E_0 \equiv U(q)$ and $E_1$ state at $q = \pm e$.

given by,

$$\omega_B = \frac{1}{2e}I = \frac{1}{2e}(I_b - \langle V \rangle/R), \qquad (1.10)$$

where $R \geq 1/2e^2$ is the environmental resistance of the junction representing dissipation effects due to the environment and $\langle V \rangle$ is the averaged voltage across the junction. These oscillations are analogous to the Bloch oscillations in spatially periodic crystals.[1, 2, 6, 12, 39]

## Observation of dual effects by tuning the electromagnetic environment

Bloch oscillations in small Josephson junctions represent the coherent tunneling of Cooper pairs, whereby a region of negative differential resistance in the $I$–$V$ characteristics of the Josephson junction is observed. The intermediate state between Coulomb blockade and Bloch oscillations is a nose structure in the $I$–$V$ characteristics. In analogy with the Josephson effect, a dc and ac bias current $I = I + I_{ac} \cos \Omega t$ phase lock at integer frequency $\omega_B = n\Omega$. In fact, this biasing current can phase-lock either with the basic frequency of the bloch oscillations, $\omega_B$ or one of its harmonics or subharmonics $\Omega_B/m$, thus generally leading to a fractional dual Shapiro step given by $\omega_B = n\Omega/m = I_n/2e$. To date, these have not been exclusive observed experimentally. This is in part due to the difficulty in observing the block nose in experiments with single junctions since their environmental impedance $R$ is not large enough but dominates over $I_b$ the term in eq. (1.10). On the other hand, Josephson junction arrays offer an alternative at least for the clear observation of Coulomb blockade effects since the environmental impedance scales with the number of junctions, $R \propto N_0 R_j$ where $R_j$ is the impedance of each junction and $N_0$ as shown in Fig. 1.6.

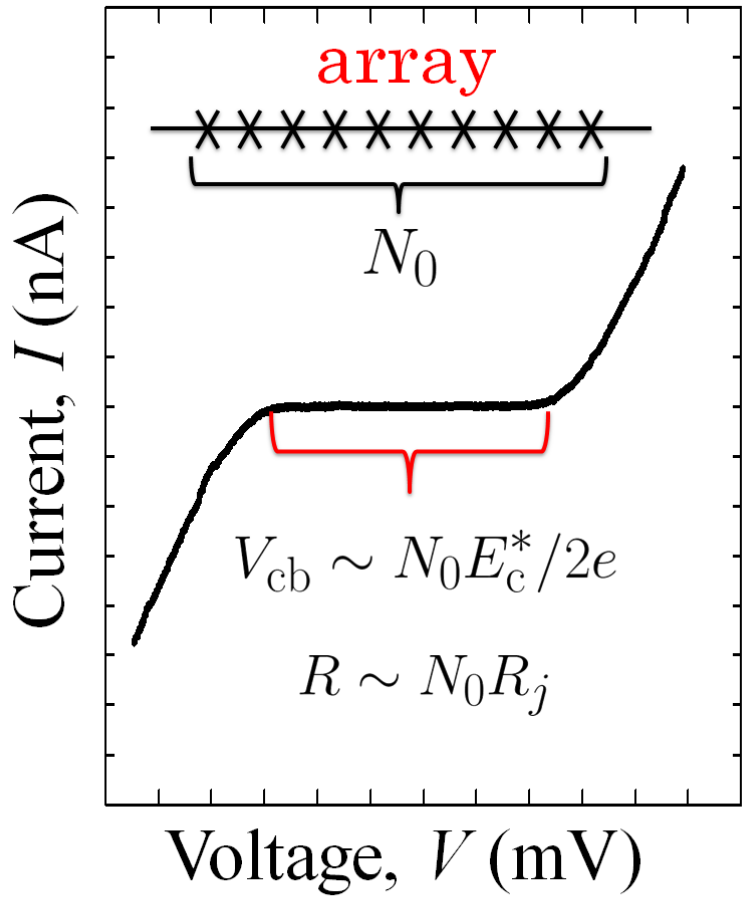

**Figure 1.6.:** The Coulomb blockade characteristics of an array of $N_0$ number of junctions. The Coulomb blockade voltage, $V_{cb}$ and the environmental impedance $R$ both scale with the number of junctions, $N_0$ making the array favorable for Coulomb blockade experiments over the single junction.

A solution for the single junction was suggested by Watanabe et al.[36, 37], where a method using a linear array of SQUIDs an environment with very weak coupling to dissipation is introduced, which allows for the measurement of a well-defined charge quantum state. By measuring the I–V characteristics of a single Josephson junction placed in the high impedance environment, the existence of the well-defined charge state which is a manifest feature of the Coulomb blockade of Cooper pair tunneling, was ascertained. The transfer of Cooper pairs through the junction is thus

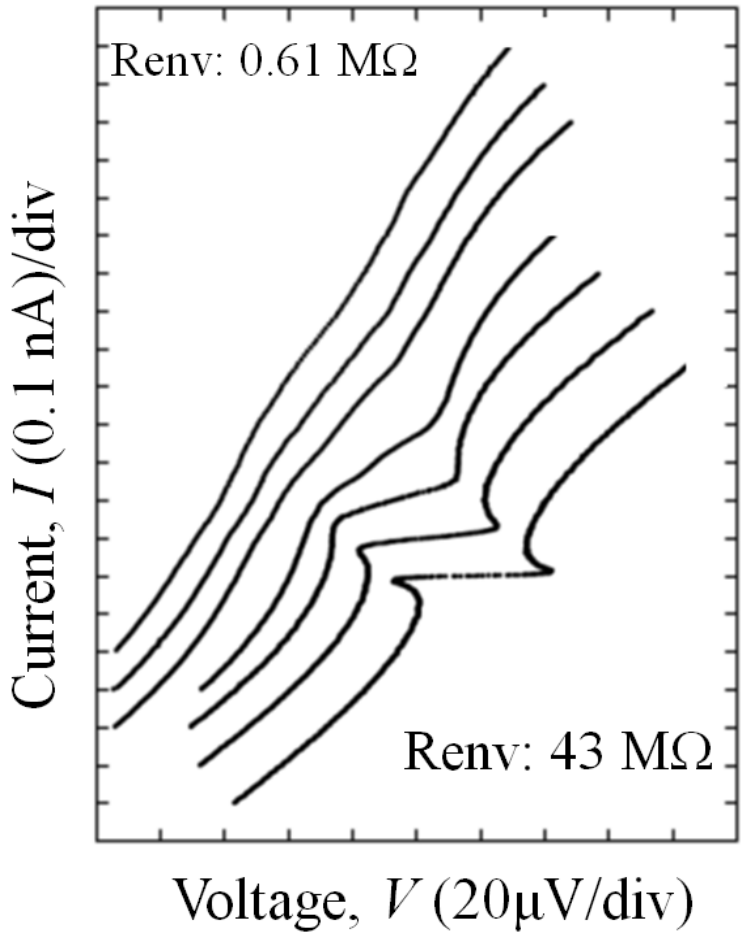

**Figure 1.8.:** The characteristics of a single junction embedded in a high impedance environment of SQUID arrays as shown in Fig. 1.7. The characteristics are tuned from low impedance ($R_{env}$ = 0.61 MΩ), where they exhibit near ohmic characteristics, to high impedance ($R_{env}$ = 43 Mω), where they exhibit Coulomb blockade characteristics with a back-bending structure referred to as a Bloch nose by an external magnetic field.[36, 37]). Since the environmental resistance $R$ (and hence the Coulomb blockade voltage) scales with the number of junctions for the array; and is maximum when $eBA = n\pi$ for the single junction embedded within an environment of SQUIDs, both techniques can be employed for a bigger Coulomb blockade voltage and larger Bloch nose.[12] (Figure reproduced from ref. [36])

governed by overdamped quasi-charge dynamics, leading to Coulomb blockade and Bloch oscillations as shown in Fig. 1.6 and 1.8 respectively. This experiment confirmed exact duality between the standard overdamped phase dynamics of a Josephson junction, resulting in a dual shape of the current-voltage characteristic, with current and voltage exchanging their roles. Subsequently, F. Maibaum et al. designed an experiment and performed extensive simulations and preliminary measurements to identify a set of realistic circuit parameters that should allow for the observation of constant-current steps in short arrays of small Josephson junctions under external ac drive of frequency.[70] Indeed, observation of these steps demonstrating phase locking of the Bloch oscillations with the external drive requires a high-impedance environment for the array. They concluded through their simulation results that the width of the dual Shapiro steps is proportional to the number of junctions in the sample measured. Subsequently, Shimada et al. used arrays of dc-SQUIDs as leads to a linear array of 20 small

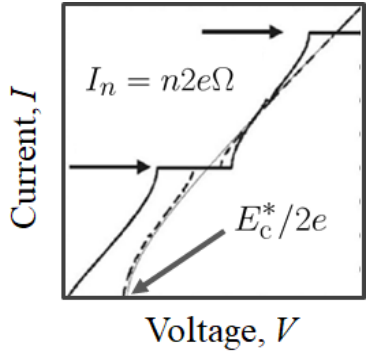

**Figure 1.5.:** The Shapiro steps predicted in small Josephson junctions with a Coulomb blockade voltage, $V_{cb} = E_c^*/2e$. The dc voltages occur at current values $I_n = 2en\Omega$, where $\Omega$ is the frequency of the ac current through the junction. Thus, the duality between the tunneling current phenomena in the large and small Josephson junctions is apparent from comparing Fig. 1.2 and Fig. 1.5.

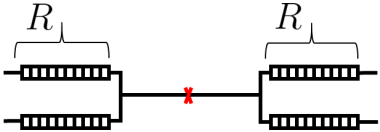

**Figure 1.7.:** A single junction (red) embedded in an environment of four linear arrays of SQUIDs. This technique was used in ref. [12, 36, 37] to increase the environmental impedance of the junction and observe distinct dual effects in small Josephson junctions, as shown in Fig. 1.8

2: Maximal frustration corresponds to the largest value for the effective environmental resistance.

Josephson junctions to tune its environmental impedance with an external magnetic field, thus observing a distinct Bloch nose with a negative differential resistance as large as 14.3 MΩ at maximal frustration of the SQUIDs[2].

# 3. Electromagnetic environment in small junctions

The aforementioned results place overcoming the contribution of the electromagnetic environment at the heart of successfully observing the phase locking of Bloch oscillations, which holds promise in high precision current standard metrology applications. In this regard however, no such experiments have yielded convincing results to date. This is, in part, due to the lack of comprehensive understanding of the effects of the electromagnetic environment especially in Josephson junction arrays.

## Experiments with microwave irradiated small Josephson junctions

### Phase diffusion

Thus far, we have merely considered the duality of the large and small junctions by checking the ratio, $E_J/E_c$ which defines the phase or charge regime of the junction. However, the Josephson junction has another energy parameter, $K_BT$, which determines the thermal fluctuations of Josephson phase. For the small junction, where $E_J/E_c < 1$, the $P(E)$ function, which is temperature dependent is generally used to accurately determine the tunneling rate for $E_J, E_c \ll k_BT$, since in this regime, phase quantum fluctuations are well described by the phase-phase correlation function. However, in the opposite regime when $E_J/E_c > 1$, and $E_J, E_c \ll k_BT$, quantum fluctuations are considerably suppressed while thermal fluctuations are enhanced. In this case, the average in the phase-phase correlation function is carried out over the Boltzmann distribution $\rho_B \propto \exp(-\beta H'(\phi, Q, t))$. The thermal phase-phase correlation $\langle \phi(t)\phi(0) \rangle \propto t$ displays a diffusive behaviour. In particular, the potential energy term of eq. (1.17) given by $U(\phi) = 2eE_J - 2eE_J \cos(2\phi) + I/2e\phi$ gives a washboard profile over which the phase fluctuates in accordance to $\rho_\mathrm{B}$. Large thermal fluctuations lead to a higher fluctuation rate $\Gamma$, which on average leads to a finite voltage of the supercurrent even for $I < 2eE_\mathrm{J}$, as depicted in Fig. 1.9.[73]

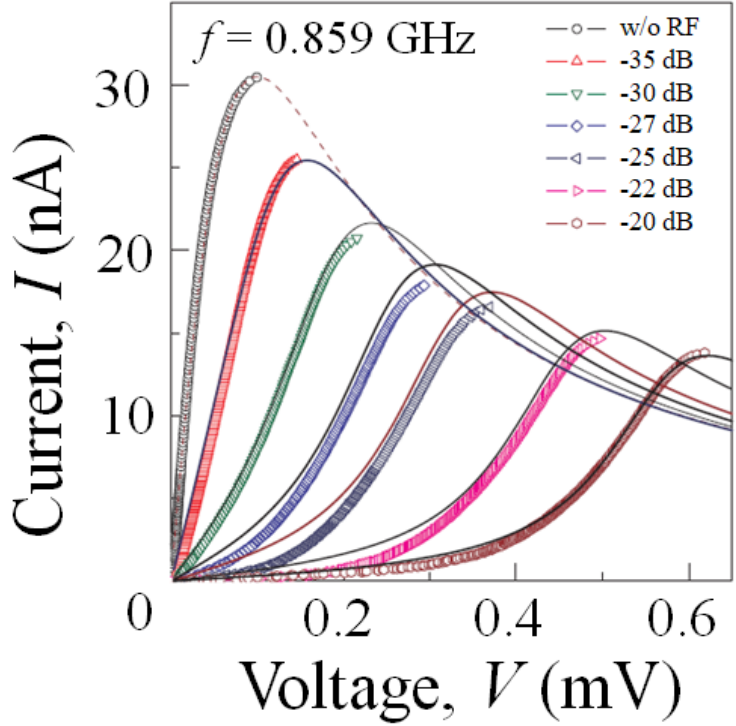

**Figure 1.10.:** The simulated low frequency ($f \ll 2eV_\mathrm{m}$), low temperature $K_\mathrm{B}T \ll 2eV_\mathrm{m}$ characteristics of a Josephson junction (where $V_\mathrm{m}$ is defined as the voltage value of the current peak in the absence of microwaves) with a phase diffusion branch irradiated by microwaves of frequency 0.859 GHz. The increase of microwave power results in larger phase diffusion corresponding to an increase in $V_\mathrm{m}$. Figure reproduced from ref. [73]

Microwave irradiation of junctions satisfying $E_\mathrm{J} > E_\mathrm{c}$ will exhibit similar behaviour, as the phase gains enough energy to 'roll' down the $U(\phi)$ potential for $I < 2eE_\mathrm{J}$. Defining the corresponding voltage of the maximum current $I_\mathrm{m}$ as $V_\mathrm{m}$, phase diffusion causes $V_\mathrm{m}$ to increase in value. Koval et al. [73] demonstrated that incoherent multi-photon absorption by the junction best explains the features observed in the I–V characteristics. They studied phase diffusion in a single Nb/AlO$_\mathrm{x}$/Nb Josephson junctions with different cross-sectional areas, at varied thermal temperatures and microwave power and frequency. Experimental results showed that, $V_\mathrm{m}$ decreases ($I_\mathrm{m}$ increases) with increase in junction size. This is due to enhanced Cooper-pair tunneling currents which dominate over quasi-particle currents thus suppressing phase diffusion.

Thus, using Josephson junctions with large cross-sectional areas (small normal resistances 0.1 ∼ 0.6 kΩ) irradiated with microwaves, they showed that phase diffusion results were enhanced by microwave power at high temperatures (5 K) and low microwave frequency (5 GHz). Conversely, microwave irradiation at low temperatures (0.3 K) and high frequencies (40 GHz) showed Shapiro like negative resistance peaks at zero bias current. All results excellent agreement with incoherent photon absorption in the phase regime ($E_J > E_c$), with the unirradiated I–V characteristics is given by,

$$I_S^0 = \frac{2eE_J}{\alpha} \frac{\nu}{\nu^2 + \delta^2}, \qquad (1.11)$$

where $\nu = V/\sqrt{8E_JE_c}$ is the normalized voltage, $\delta = V_m/\sqrt{8E_JE_c}$ and $\alpha \propto 1/R$ is a damping parameter, and the characteristics for during incoherent multi-photon absorption by,

$$I(V) = \sum_{n=-\infty}^{\infty} J_n^2\left(\frac{2eV_{\text{ac}}}{\alpha\Omega}\right) I_S^0(V - n\Omega/2e), \qquad (1.12)$$

where $\Omega$, $V_{\text{ac}}$ are the microwave frequency and amplitude. TYpical results (simulation) for $f = 0.859 GHz$ have been reproduced in Fig. 1.10.

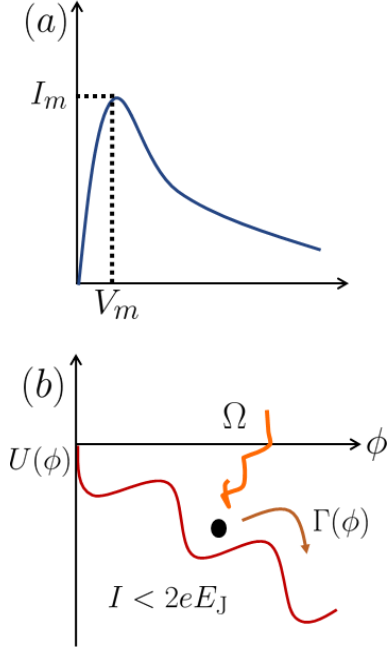

**Figure 1.9.:** A schematic showing, (a) the supercurrent when the junction is undergoing phase diffusion where it shows the voltage $V_m$ corresponding to the maximum supercurrent $I_m$; (b) The energy profile (washboard potential[71]) of the Josephson junction undergoing phase diffusion. A photon of energy $\Omega$ can be absorbed by the junction and lead to diffusive behaviour of the supercurrent, where $\Gamma(\phi)$ is Kramer's rate for such a process to occur.[71, 72]

Consequently, Liou et al. [41] studied microwave phase diffusion and the cross-over into the Coulomb blockade regime using one dimensional arrays of Josephson junctions. Similar results to Koval et al. were obtained in the phase diffusion regime, where the group considered instead $P(E)$ theory.[16] The resulting equation from $P(E)$ theory is essentially eq. (1.12), but with the damping factor $\alpha = 1$, thus suggesting a detector application. In addition, oscillations of the zero bias resistance as the microwave power is increased were observed, with an oscillatory behaviour comparable to the superconducting gap.

### Coulomb blockade regime

Early work with Josephson junction arrays irradiated with microwaves in the Coulomb blockade regime focused on photon-assisted tunneling and thermal tunneling effects of quasi-particles. Experiments by Delsing et al. [32] concentrated on the deep Coulomb blockade regime $E_c \gg E_J$ and $2\pi/e^2 < R_T$. Here, the tunneling current is dominated by quasi-particles. The results showed that the quasi-particle Coulomb blockade characteristics were diminished for high temperatures as well as high microwave powers. Moreover, the work reported electron tunneling oscillations[39] in the differential conductance at particular current value, $I_n = ne\Omega$, where $n$ is an integer and $\Omega$ is the microwave frequency.

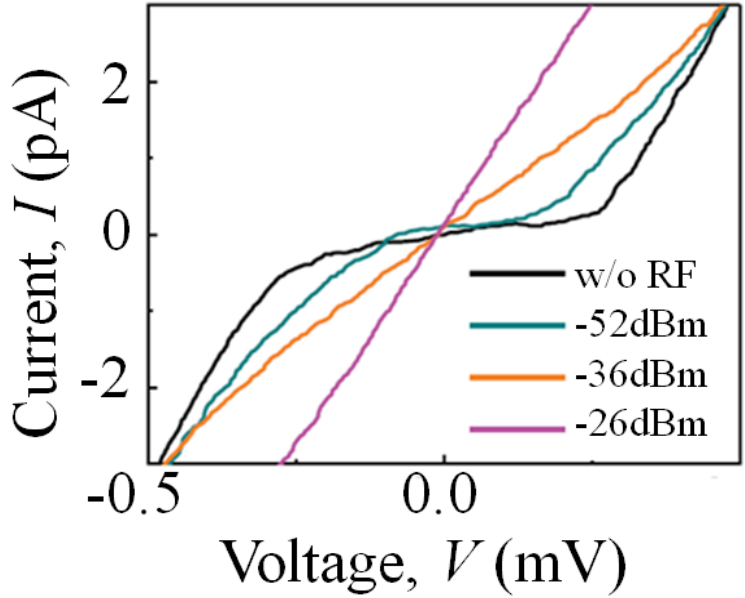

**Figure 1.11.:** The diminishing of Cooper pair Coulomb blockade characteristics by microwave power. The device measured was a SQUID array tuned by an external magnetic field into the deep Coulomb blockade regime, $E_J \ll E_c$. The explored frequency range was 3 GHz to 26 GHz. The figure was lifted from ref. [33]

Using a linear array of SQUIDs, Liou et al. [33] studied high frequency regime of Cooper pair tunneling $E_J \geq E_c$ and the deep Coulomb blockade $E_J \ll E_c$ photon assisted tunneling, reproducing the diminishing of Coulomb blockade characteristics with microwave power as shown in Fig. 1.11. However, the report lacks environmental impedance analysis e.g. the Coulomb blockade power characteristics. Nonetheless, it is suggested that linear arrays in the Coulomb blockade regime are suitable for microwave detection.

### P(E) theory: complexity in determining the environmental impedance

The standard theory that incorporates the effect of the electromagnetic environment in small Josephson junctions is the $P(E)$ theory, formulated in terms of the phase-phase correlation function,

$$P(E) = \frac{1}{2\pi} \int dt \, \exp(iEt) \langle \exp[i\phi(t)] \exp[-i\phi(0)] \rangle, \quad (1.13)$$

where the tunneling rate of the Cooper-pair across the Josephson junction is given by,

$$\Gamma(V) = \frac{\pi}{2} E_J P(2eV). \tag{1.14}$$

The information of the electromagnetic environment, given by a complex-valued impedance function $Z_{\text{eff}}(\omega) = 1/(R^{-1} + i\omega C)$, is contained in the term $\mathcal{J}(t)$ in phase-phase correlation term,

$$\langle \exp i[\phi(t)] \exp[-i\phi(0)] \rangle = \exp(-4\mathcal{J}(t)), \tag{1.15}$$

where, $\mathcal{J}(t) = \langle [\phi(t) - \phi(0)]\phi(0) \rangle$ and the phase-phase correlation function is given by,

$$\langle \phi(t)\phi(0) \rangle = \frac{e^2}{2\pi} \int \frac{d\omega}{\omega} \frac{[Z_{\text{eff}}(\omega) + Z_{\text{eff}}(-\omega)]}{1 - \exp(-\beta\omega)} \exp(-i\omega t), \tag{1.16}$$

with $\beta = k_B^{-1} T^{-1}$ the inverse temperature. The form of the effective impedance function $Z_{\text{eff}}(\omega)$ corresponds to the case for the effective Josephson Hamiltonian ($H' = H + H_{\text{int}}(t) + H_{\text{bias}}$) where $E_J = 0$,

$$H'(Q, \phi, t) = \frac{Q^2}{2C} - E_J \cos 2\phi + \int dt \frac{Q^2}{RC^2} + I\phi/2e \tag{1.17}$$

where $H_{\text{bias}} = I\phi/2e$ is the energy due to a bias current through the junction and $H_{\text{int}} = \frac{1}{2C^2 R} \int dt Q^2$ is the interaction term between the Josephson junction and the environment, effectively introducing dissipation since $\partial H / \partial t \neq 0$. Due to the non-linearity of the Josephson coupling term, it is difficult to incorporate the effect of $E_J$ on the effective impedance $Z_{\text{eff}}(\omega)$ of the $P(E)$ function even for the single junction.[3] In the case of the array, it is extremely difficult to accurately determine $Z_{\text{eff}}(\omega)$ even for $E_J = 0$ since the complexity of the theory scales with the number of junctions in the array.[4]

3: It can trivially be considered only when the Josephson coupling term is linearized by taking $\phi \ll 1$ and writing $E_J \cos 2\phi \simeq E_J(1 - 2\phi^2)$, which leads to an inductive contribution to the effective impedance given by $Z'_{\text{eff}}(\omega) \simeq 1/(R^{-1} + i\omega C + 1/i\omega L_J)$, where $L_J = 1/2e^2 E_J$.

4: Thus, this problem also applies for arrays of normal junctions. However, since normal junctions do not possess the Josephson coupling term, the problem is merely one-fold.

### Recourse for complexity in $P(E)$ theory

Thus, the non-linearity of the Josephson junction, coupled with the complexity exhibited by Josephson junction arrays serve as an unexpected bottleneck towards the range of applicability of the $P(E)$ theory. Nonetheless, $P(E)$ theory has successfully been tested to a great degree of accuracy in a myriad of experiments leading to its widespread application in describing progressively complex tunneling processes such as dynamical Coulomb blockade in small Josephson junctions and quantum dots.[17, 22] However, in most instances, creative theoretical models that essentially are meant to approximate $Z_{\text{eff}}$ have been proposed over the years. For instance, the $P(E)$ theory of dynamical Coulomb blockade in single small Josephson junctions is formulated on the basis of phase correlation functions[15, 16] where tunneling across the barrier is influenced by a high impedance environment treated within the Caldeira-Leggett model [10]. However, the Caldeira-Leggett form of the environmental impedance neglects the back-action of the Josephson junctions on the environment (with the bath and the junction becoming entangled) which has been reported to dramatically change the predictions of the $P(E)$ theory.[31, 34, 35] This back-action manifests through a non-linear inductive response $L_J^* \simeq 1/2e^2 E_J \langle \cos 2\phi \rangle = 1/2e^2 E_J^*$ of the junction where the renormalized Josephson coupling energy $E_J^*$ appears in the renormalized effective impedance function $Z'_{\text{eff}}(\omega) = 1/(R^{-1} + i\omega C + 1/i\omega L_J^*)$ thus altering the insulator-superconductor phase transition conditions for the single Josephson junction.[35][5]

5: A method to determine the impedance of an arbitrary environment of a single Josephson junction has been proposed in ref. [34]

Moreover, owing to significant improvement in microwave precision measurement technology such as near-quantum-limited amplification[23, 24] and progress in theory, recently published works suggest novel features in the $P(E)$ framework ranging from time reversal symmetry violation[25] to the renormalization of electromagnetic quantities appearing in the $P(E)$ function[27–31]. Despite this progress, aspects of the theory remain elusive especially in the case of one dimensional arrays. Useful techniques for Josephson junction arrays involve mapping the dynamics of the quantum phase of the Josephson junction onto dynamics of a system with known solutions e.g. charge solitons,[32, 50, 51] Tomonaga-Luttinger Liquid[26] (TLL) and/or depinning models.[74, 75] However, these techniques apply to specific domains where $P(E)$ theory is not well-understood.

## 4. Motivation and Aim

Due to the aforementioned difficulties, $P(E)$ theory has not been comprehensively tested with Josephson junction arrays. This is because of the aforementioned complexity due to the strong coupling of junctions with the electromagnetic environment, which scales with the number of junctions. For instance, $P(E)$ theory cannot account for all photon-assisted tunneling features of Cooper pairs and quasi-particles, even for the simplest case of two small Josephson junctions in series forming a superconducting single electron transistor (SSET).[42, 66] Here, Cooper-pair tunneling is incoherent while tunneling events are expected to be uncorrelated[43] leading to radiative and photon assisted tunneling processes such as the Josephson quasi-particle cycle.[44, 76] This has broad implications for the interaction of arrays with the electromagnetic field. Moreover, a voltage biased array with no special coupling, fabricated adjacent to another unbiased array with a similar structure, has been shown to induce a strongly correlated current through the latter.[45, 46] The current has the characteristic that reversing the polarity of the bias voltage does not reverse the polarity of the induced current. Such effects require the understanding of how the electromagnetic quantities appearing in $P(E)$ theory couple to the electromagnetic field. Thus, this shows the need for further theoretical and experimental studies pertaining the interaction of single Josephson junctions as well as arrays with the electromagnetic environment, serving as motivation for the research herein. Explicitly stated, **the aim of the research presented herein is to experimentally and theoretically study the electromagnetic environment of small Josephson junctions irradiated by microwaves.**

# Experiment

# Microwave Irradiation of small Josephson junction arrays

# 2.

In this chapter, we conducted an experiment to examine the effect of RF electromagnetic fields on an array of small Josephson junctions satisfying $0.1 < E_J/E_c < 1$ and $R_T > R_Q = 2\pi/4e^2 \simeq 6.45$ k$\Omega$ by measuring its current, $I$–voltage, $V$ characteristics. Under these conditions, the tunneling of charges at small voltages is dominated by Cooper pairs, and the characteristics exhibited are in the charge regime, dual to the phase regime. However, Cooper-pair tunneling can easily be precluded by the electromagnetic environment of the array, leading to Coulomb blockade. Thus, the array offers great utility over the single junction which requires the tuning to observe Coulomb blockade.[38] In our experiment, the Coulomb blockade of tunneling Cooper-pairs was steadily diminished when radio-frequency electromagnetic radiation was applied, independent of frequency $f = \Omega/2\pi$ in the sub-gigahertz band 1 MHz $\leq f \leq$ 1000 MHz with $\Omega \leq k_B T$.[1] The observed diminishing of Coulomb blockade with microwave radiation is dual to the phase diffusion effect reported by Liou et al in ref. [41] for a linear Josephson junction array in the regime, $E_J/E_c > 1$.

Moreover, the environment of the superconducting array is susceptible to an externally-applied magnetic field $H$ through the quotient $E_J(H)/E_c$ that governs the dynamics of the quasi-charge of each Josephson junction in the array within their respective Brillouin zone of the Bloch energy band. In particular, the energy band gap, which is comparable to $E_J(H)$, is diminished by applying a magnetic field $H \leq H_{\max}$ where $H_{\max}$ is the magnetic field that leads to the most enhanced Coulomb blockade of Cooper-pairs in the sample.[2] [2]

In the experiment, a substantial non-varying magnetic field, $H = 500$ Oe is perpendicularly applied to the unirradiated array in order to raise the value of the Coulomb blockade (threshold) voltage $V_{cb}$ to its maximum. This corresponds to a factor of approximately 1.4 its original value for $H = 0$ Oe.[38, 47] Nonetheless, the $V_{cb}$ versus $V_{ac}$ characteristics of the irradiated array when $H = 500$ Oe coincide with those

1: $k_B$ and $T$ are the Boltzmann constant and temperature respectively.

2: In the studied array, $H_{\max} \simeq 500$ Oe.

for $H = 0$ Oe when both axes of the $V_{\text{cb}}$–$V_{\text{ac}}$ plots are rescaled by the aforementioned factor of $\simeq 1.4$.

To analyse the experimental results, we simulate the characteristics of the irradiated linear array using well-known equations[3] for photon-assisted tunneling[48, 49] within the $P(E)$ theory.[15][4] Comparing the simulated curves with the experimental results by plotting $V_{\text{cb}}$ versus $V_{\text{ac}}$ curves, we discover that, a mismatch of a factor, 0.87 persists between the values of the absorbed microwave power by the array in the experiment and the values corresponding the simulated curves with the same Coulomb blockade threshold voltage even after calibration of the microwave line.[5]

We discuss other possible origin of this mismatch by considering the uncertainties relating to the microwave generator,[6] transmission line calibration procedure,[7] the influence of electron heating at the islands of the array[8] and a possible voltage division effect that leads to the renormalization of the microwave amplitude by a factor, $\Xi_A \sim \exp(-\Lambda^{-1}) \simeq 0.89$, where $\Lambda$ is the length over which the applied microwave is damped from the edge.[9]

These results demonstrate pristine Josephson junction arrays are poised for microwave detection applications in a wide range of environments[10] due to their high sensitivity to low-power, of order $10^6$ V/W.

## 1. Experimental Method

### Design and fabrication of the sample

Typically, the Coulomb blockade voltage, $V_{\text{cb}}$ is roughly proportional to the number of Josephson junctions in a linear array.[70] Since the response of the array to irradiation by microwaves depends on $V_{\text{cb}}$, a greater response is exhibited for linear arrays with many junctions as well as for applied magnetic fields below $H_{\text{max}}$. This is the basis for choosing the array over the single junction.[11]

Here, an array of $N_0 = 10$ Josephson junctions aranged in series (linear array) is fabricated such that its soliton length in the semi-infinite model of the array is approximately of the same length. This is based on our analysis in chapter 5 and

---

3: eq. (2.2) and eq. (2.3)

4: Photon-assisted tunneling in the classical regime where $\Omega \ll 2eV_{\text{ac}}$.

5: This factor is neither dependent on frequency nor the applied magnetic field after rescaling the $V_{\text{cb}}$–$V_{\text{ac}}$ axes by 1.4

6: Total uncertainty amounts to $\sim 4\%$ whereas the total mismatch would correspond to a large uncertainty of $\sim 0.13$

7: Gross failure of calibration unlikely due to the clear frequency independence of the $V_{\text{cb}}$–$V_{\text{ac}}$ plots; Fig. 2.6

8: The quasi-particle number is negligibly small to account for the mismatch for the microwave amplitude range considered.

9: The so-called soliton length of the array.[51]

10: e.g. on-chip detection schemes[54]

11: However, unlike the single junction, the array lacks a comprehensive ($P(E)$) theory[15, 16] for its response to microwave irradiation.

Table 2.1.: Parameters of the array (per junction): the tunnel resistance $R_T$, charging energy $E_c$, capacitance $C$, the aluminium electrode superconducting gap given by $\Delta$, Josephson coupling energy $E_J$, and $E_J$-to-$E_c$ ratio. The parameters per junction when magnetic field $H = 0$, 500 Oe is applied, with 500 Oe $\equiv H_{max}$ the value of the magnetic field that leads to the largest Coulomb blockade voltage in the sample.

| $R_T$ / k$\Omega$ | $C$ / fF | $E_c/\mu$eV | $\Delta/\mu$eV | $E_J/\mu$eV | $E_J/E_c$ | $H$ / Oe |
|---|---|---|---|---|---|---|
| 35.1 | 0.72 | 110 | 165 | 30.3 | 0.27 | 0 |
|  |  |  | 133 | 24.5 | 0.22 | 500 |

appendix E that such an array can be treated within the $P(E)$ framework as an effective single junction using the soliton model of a semi-infinite array.[32, 51] The array, with 10, 100 × 200 nm² Al/AlO$_x$/Al junctions and island electrodes of length $l = 1\,\mu$m, base and counter electrode thickness of 25nm and 40 nm respectively, is designed by creation of an evaporation mask using Electron Beam Lithography (EBL), aluminium deposition and subsequent lift-off of the electron-resist. In this process, the chip used was fabricated on a 7 × 7 mm² silicon (Si) wafer engulfed by a silicon dioxide (SiO$_2$) layer. Optical lithography is used to design a pad on the chip with 16 leads that converge to the center of the chip leaving 200 × 200 $\mu$m² at the center, over which the present sample is fabricated; The resist used to fabricate the sample is industry standard PMMA (6% and 2% respectively); Aluminium metal evaporation is conducted in an electron-beam evaporator with a base vacuum pressure of $10^{-7}$ torr using the shadow evaporation technique (evaporation 1st and 2nd angles given by −35° and +25°), whilst oxidation performed with pure oxygen at $10^{-2}$ torr.

**Sample parameters**

The desired single junction parameters ($0.1 < E_J/E_c < 1$ and $R_T > R_Q$) can be chosen by during electron-beam lithography and shadow evaporation during oxidation. The tunnel resistance, $R_T$ and $E_c$ are determined from the offset voltage and the differential conductance d$I_0(V)$/d$V$ respectively, of the linear array characteristics, $I_0(V)$ when $H = 0$ Oe.[32, 77] The differential conductance for $H = 0, 500$ Oe, together with measured $\Delta$–$H$ dependence determine the superconducting gap $\Delta$. The Josephson coupling energy, $E_J$ is then determined by the Ambegaokar-Baratoff relation.[60] The values determined above are displayed in Table 2.1.

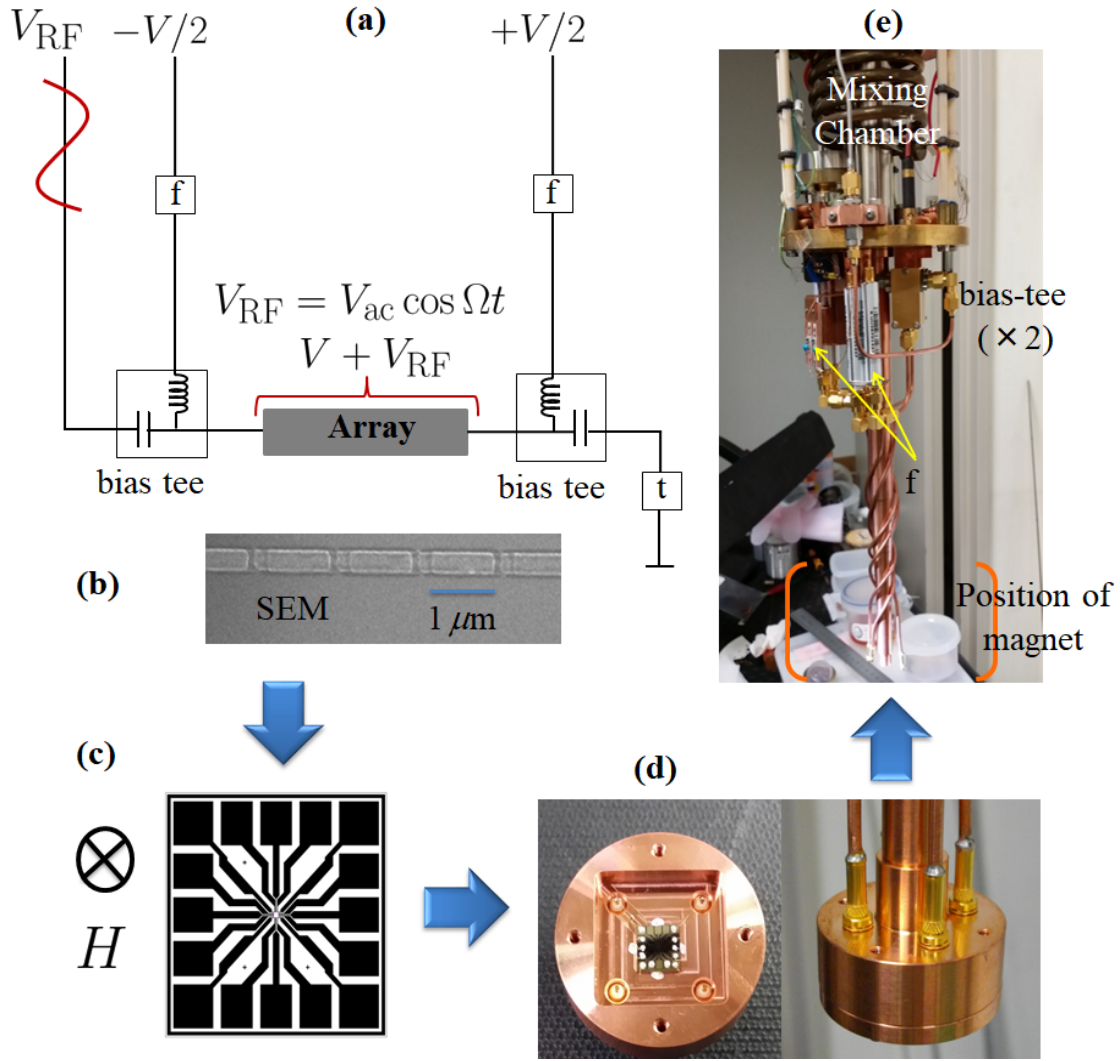

**Figure 2.1.:** Measurement set-up for the linear array of 10 small Josephson junctions. The signal from the RF generator, $V_{RF}$ is combined by the pair of bias tees with the dc signal $\pm V/2$, $V + V_{RF}$ where $V_{RF} = V_{ac} \cos \Omega t$ is the ac voltage or the applied microwaves. The bias tee on the right has a terminator"t" of 50 $\Omega$ at one of its terminals. The magnetic field indicated by $H$ and a circled-cross is applied are right angles to the sample: (a) The diagram of the circuit and array used in experiment; (b) The array as seen by a scanning electron microscope (SEM), displaying 4 of the $N_0 = 10$ fabricated junctions; (c) Schematic of the sample holder containing the fabricated chip; (d) Sample holder containing the fabricated chip; (e) The base of the dilution refrigerator showing the positions of the two bias tees, low-pass filters denoted by "f", the magnet and the sample holder.

### Experimental setup

A standard dilution refrigerator with a stable operational temperature of 40 mK is used for the low temperature measurements. The semi-rigid circuitry incorporated in the refrigerator has a terminator"t" of 50 $\Omega$ at one of its terminals.

A coaxial cable made from CuNi is fitted from room temperature to still plate where thermal anchors are used to fix cryogenic attenuators (10 dB, 20 dB) *, as well as in the vicinity of the 1 K pot. A coaxial cable made from stainless-steel extended from the still plate to the mixing chamber plate is joined to a coaxial cable made of copper through another cryogenic attenuator (20 dB) anchored thermally (Fig. 2.2 (a), (b) and (c)). The copper coaxial cable transmits the microwave signal to the sample through a commercial bias tee, which is fixed at the mixing chamber plate. Similarly, a second bias tee at the mixing chamber plate is connected to the sample chamber and its port fitted with a 50 Omega terminator as shown in Fig. 2.1. Thus, the constant (dc) and alternating (ac) voltage signals ($V + V_{RF}$) are combined by the bias tees, where the high frequency noise signal in the dc line is cut off by a 3.4 kHz low pass filter at the mixing chamber plate before the signal is directed to the ac line through the bias tees.

A copper sample holder with MMCX connectors for the ac signal was fitted at the mixing chamber plate with the sample, with the MMCX connectors connected directly to the gold (Au) pads of the chip with Au wires of length 4mm and 3 mm to the pad and from the pad to the array respectively. All measurements were conducted in an environment shielded from electromagnetic fields. A typical radio-frequency generator (Agilent 8753ES) capable of supplying a signal of 1 MHz $\leq f \leq$ 1000 MHz was used to irradiate the sample via the described circuitry.

The well-known $r$-bias method was applied when measuring the characteristics ($I$–$V$) of the array.[78] It entails incorporating a fixed resistance given by $r$ (1 MΩ, 5 MΩ) serially connected to the array and biasing both (the resistor and the array) with a voltage, and measuring the current and voltage values employing differential amplifiers with high input impedance. Noise reduction was achieved by applying half the dc voltage in each terminal with opposite polarity ($-V/2$ and $V/2$) relative to the ground (Fig. 2.1). Finally, a superconducting coil was used to generate a sufficient magnetic field $H = 500$ Oe, applied at right angle to the sample.[12]

12: The applied magnetic field reduces the superconducting gap per junction of the array, hence increasing the coulomb blockade voltage[38] by a factor of approximately 1.4, as shown in Table 2.1

### Ac input Calibration

We calibrated the transmission characteristics of the line prior to the commencement of the experiment. The transmission coefficient was obtained by fitting a pair of identical transmission lines to the cryostat each spanning from the room temperature environment at the top to the sample chamber at the bottom, with appropriate attenuation and bias tees. The two lines were then shorted at their terminals with a single semi-rigid cable made from copper instead of the sample to create a double line (main line + auxiliary line) as shown in Fig. 2.2 (d) and then the characteristics of the RF line at Room temperature (RT), Liquid Nitrogen temperature (85

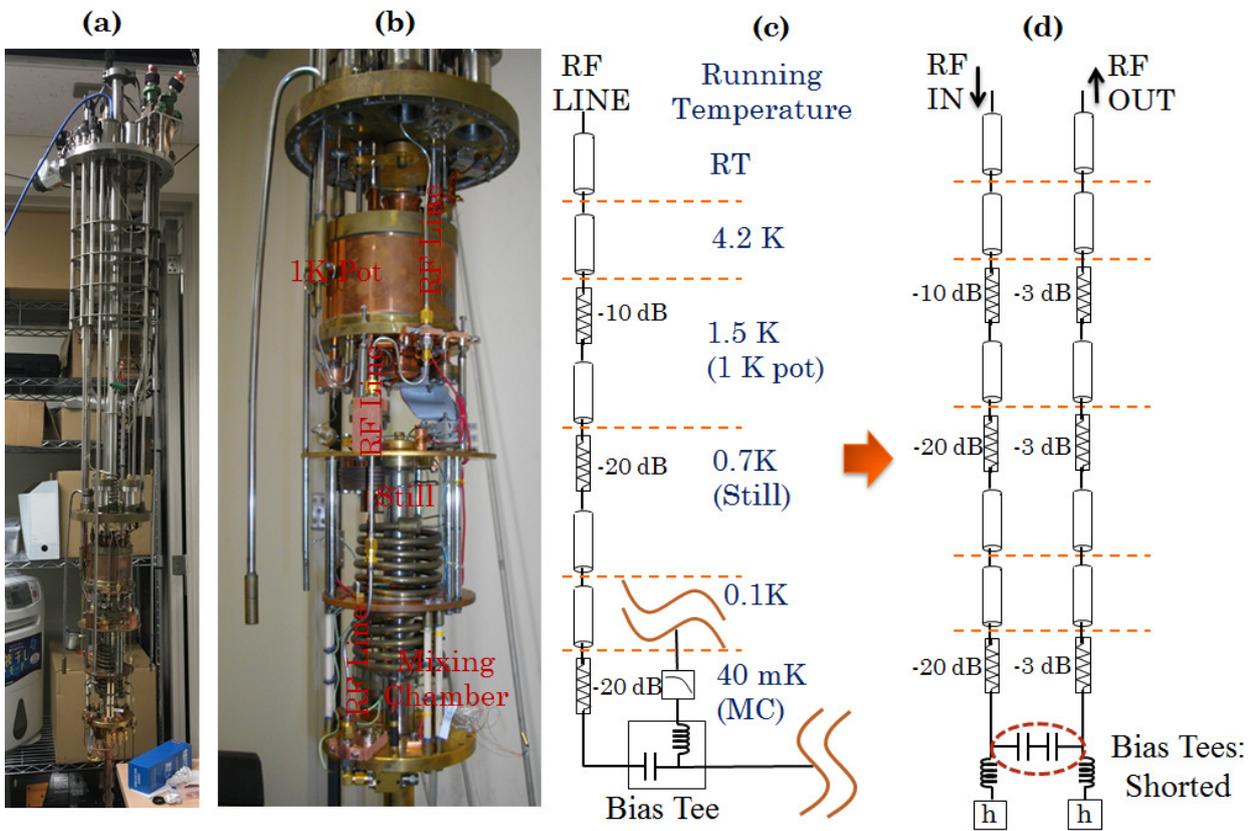

**Figure 2.2.:** The standard dilution refrigerator used in the measurement of the sample and the RF circuitry. (a) The typical dilution refrigerator with a single RF cable attached with the application of RF signal in the measurement set-up depicted in Fig. 2.1 (a) at room temperature (RT); (b) 1 K pot, Still and Mixing Chamber of the dilution refrigerator; (c) The RF line equivalent circuit showing the position of the attenuators (-10 dB, -20 dB and -20 dB) and bias tee in the measurement set-up; (d) Two identical transmission lines in the cryostat, each extending from the room temperature terminal to the sample chamber with their bias tees shorted, replacing the sample with a Cu semi-rigid cable. The total attenuation of the RF line is 59 dB = (10 + 20 + 20 = 50 dB) + (3 + 3 + 3 = 9 dB) (where we have dropped the minus, "-" signs).

---

* minus sign, "-" is implied by the word "attenuation".

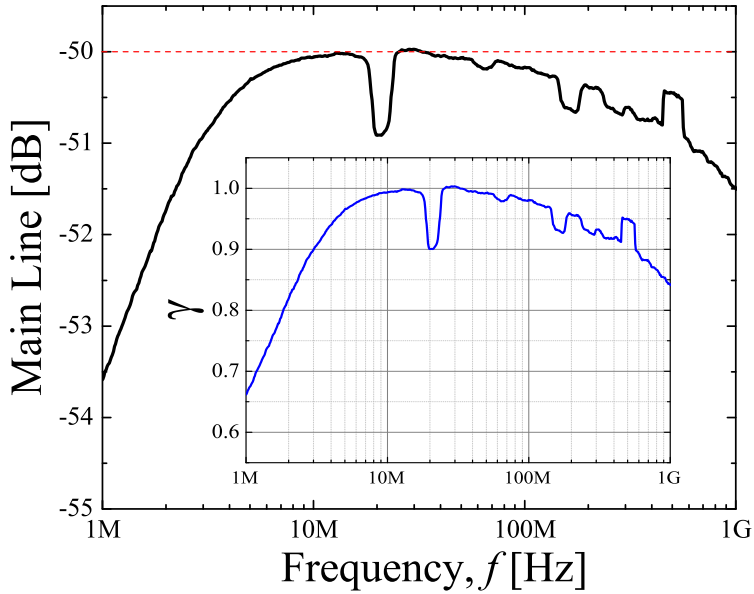

**Figure 2.3.:** The calculated transmission characteristics of the main line using the measured characteristics of the set-up in Fig. 2.2 (d) at approximately the cryostat running temperature ($\lesssim 150$ mK). The red line marks the attenuation value (- 50 dB) of the main line. Inset: The transmission coefficient, $\gamma(\Omega)$ of the main line calculated under the assumption of identical lines, $\gamma = \sqrt{\gamma_{\text{double}}}$.

K), Liquid Helium temperature (4.2 K) and cryostat base temperature ($\sim 0.15$ K) were measured. Since there is no way of measuring the main line transmission coefficient $\gamma$ at the operational temperature of the dilution refrigerator ($\sim 100$ mK) without connecting it to the auxiliary line, we measured the double line transmission coefficient $\gamma_{\text{double}}(\Omega)$ and calculated $\gamma(\Omega)$ from it (Fig. 2.3). Assuming that the main and auxiliary lines are identical except for the attenuation, we have $\sqrt{\gamma_{\text{double}}(\Omega)} \simeq \gamma(\Omega)$. We measured $\gamma_{\text{double}}(\Omega)$ at different equilibrium temperatures and discovered that it was temperature independent at sub-liquid helium temperatures ($T < 4.2K$) for all sub-giga hertz frequencies. The length difference of the two transmission lines at room temperature was taken into account by estimating the corresponding error of the estimation $\sqrt{\gamma_{\text{double}}} \simeq \gamma$ to be 2%.

We determined the error in setting $\sqrt{\gamma_{\text{double}}(\Omega)} \simeq \gamma(\Omega)$ due to the small length difference of the two lines at room temperature to be 2%. Defining the input power to the main line as $P_0^{\text{input}}$, its attenuated power becomes $P_0 = P_0^{\text{input}} - 50$ dB. Thus, the desired incident power on the sample is given by $P \simeq \gamma(\Omega)P_0(\Omega)$ where $P_0$ is the attenuated power of the main line connected to the sample.

Finally, using the fact that the line impedance, $Z_0 = 50$ $\Omega$ for the sub-gigahertz frequency range is much smaller than the sample impedance $Z$, the reflection coefficient of the applied microwave at the input terminal of the linear array is given by $\Gamma = (Z - Z_0)/(Z + Z_0) \simeq 1$. This means the incident voltage gets twice the chance to be absorbed by the array. In particular, the total magnitude of the ac voltage is given by,

$$V_{ac} = 2\sqrt{2PZ_0}. \tag{2.1}$$

**Table 2.2.:** A summary of the uncertainties related to the determination of the applied microwave amplitude, $V_{ac}$.

|  | Uncertainty |
|---|---|
| RF generator | 1% |
| Transmission line characteristics estimation | 2% |
| Output impedance of the line | 5% |
| Impedance mismatch at the connection | Negligible ($\leq$ 100 MHz) |
|  | 1 $\sim$ 10% (1 GHz) |
| Combined | 4% ($\leq$ 100 MHz) |
|  | < 10% (1 GHz) |

The uncertainty of applying the aforementioned procedure to calculate $V_{ac}$ is estimated to be 4%, calculated from the uncertainties of the determined $Z_0$ and $P$ values. The summary of all the relevant uncertainties is given in Table 2.2.

## 2. Experimental Results

The $I$–$V$ characteristics of the array were measured for varied microwave power $P$ values between -115 dBm and -60 dBm at the refrigerator's lowest stable running temperature (40 mK). The $I$–$V$ characteristics for $f = 100$ MHz and $H = 0$ Oe are given in Fig. 2.4a. Distinct Coulomb blockade characteristics were observed for $P = 0$ (represented by the dotted line). The coulomb blockade characteristics are diminished with increase in $P$ to near-ohmic characteristics at $P \geq -63$ dBm.

The Coulomb blockade voltage $V_{cb}(H, V_{ac})$ is defined at the current value $I_{th} = 1$pA for all $I$–$V$ curves. The $I$–$V$ characteristics for varied microwave power values, $-115\,\text{dBm} \leq P \leq -60\,\text{dBm}$ were measured and the microwave amplitude $V_{ac}$ subsequently determined by eq. (2.1). The characteristics for $f = 100$ MHz at $H = 0$ Oe are plotted in Fig. (2.4a) whilst for $f = 100$ MHz at $H = 500$ Oe in Fig. 2.4b.

Evidently, distinct Coulomb blockade characteristics were observed when $P = 0$, as indicated by the dotted line. The curves become increasingly linear for large $P$ values, nearly satisfying ohm's law at $P$ = -63 dBm. Thus, Coulomb blockade characteristics are increasingly diminished as the microwave power is increased. To clearly show this trend, we proceed to plot the Coulomb blockade voltage $V_{th}(H, V_{ac})$ versus the microwave power $V_{ac}$ for the experimental results in Fig. (2.4a) and Fig. (2.4a).

Consider the plot for $f = 100$ MHz in the absence of magnetic field, $H = 0$ Oe. The coulomb blockade voltage was decreased from $V_{th}(H = 0, 0) = 0.25$ mV to $V_{th}(H = 0, V_{ac}) < 0.05$ mV in the presence of maximum microwave $V_{ac} = 0.42$ mV ($P = -63.5$ dBm) as displayed in Fig. 2.4(a) and Fig. 2.5(a). Likewise, $V_{th}(H = 500\,\text{Oe}, 0) = 0.4$ mV was decreased to $V_{th}(H = 500\text{Oe}, V_{ac}) < 0.05$ mV in the presence of maximum microwave $V_{ac} = 0.42$ mV ($P = -63.5$ dBm) as displayed in Fig. 2.4(b) and Fig. 2.5(b) for $H = 500$ Oe where $V_{th}(H = 500$

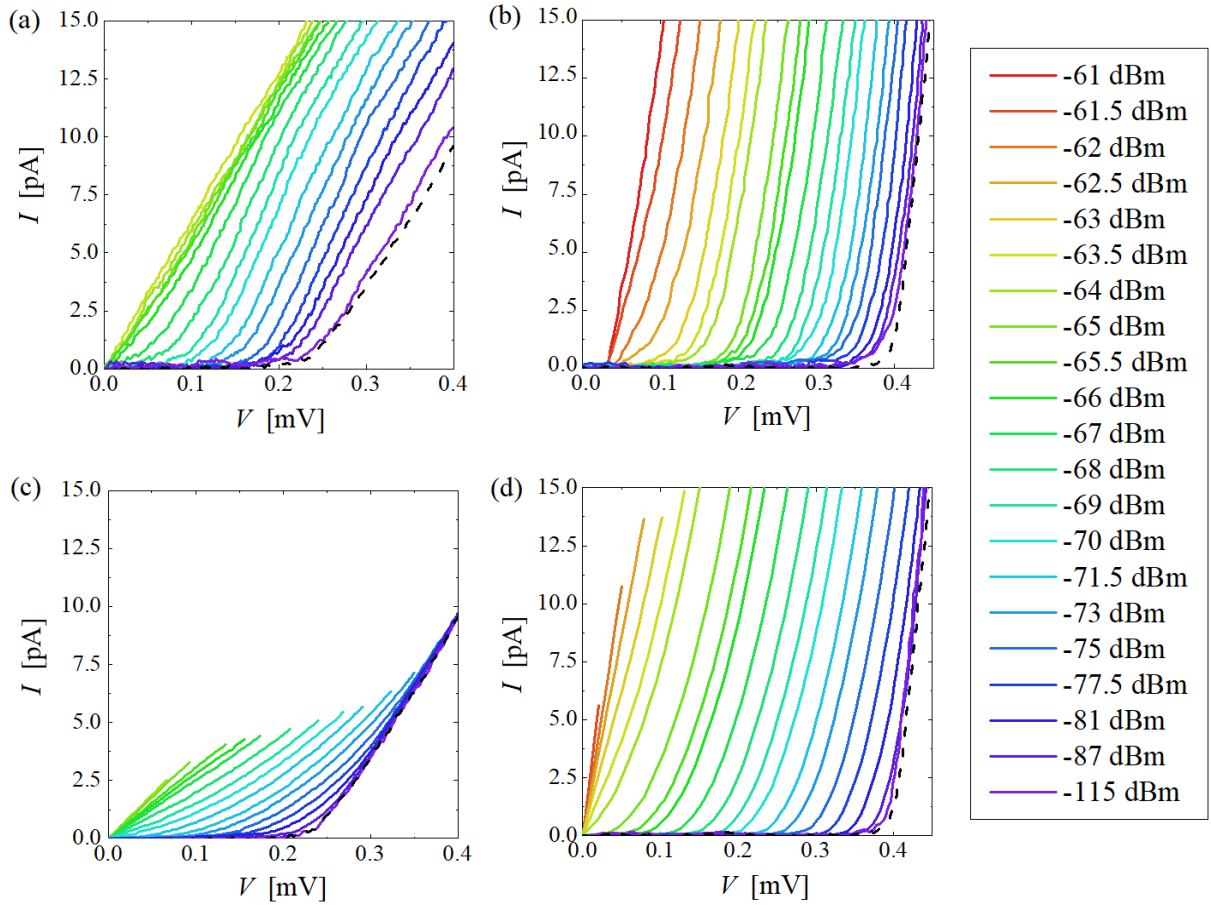

**Figure 2.4.:** The array *I–V* characteristics measured at 40 mK for (a) $H = 0$ and (b) $H = 500$ Oe. Different curves correspond to different values of applied microwave power $P$ for microwave frequency $f = 100$ MHz. The *I-V* curves were calculated using eq. (2.3) for (c) $H = 0$ Oe and (d) $H = 500$ Oe. The Coulomb blockade characteristics of the unirradiated array, $I_0(V)$ for $H = 0$ Oe and $H = 500$ Oe are displayed as dashed curves. The microwave amplitude, $V_{ac}$ is obtained from $P$ in eq. (2.1). The Coulomb blockade voltage $V_{cb}$ for the characteristics in (a), (b), (c) and (d) is defined at $I_{th} = 1$ pA. Figure reproduced from ref. [61] with permission from the journal.

Oe, 0) = 0.34 mV = 1.4 × $V_{th}(H = 0$ Oe, 0). The Josephson coupling energy for each junction in the array was decreased from $E_J(H = 0) = 30.3$ $\mu$eV to $E_J(H = 500$ Oe$) = 24.5$ $\mu$eV by applying the magnetic field $H = 500$ Oe (Table 2.1), where 500 Oe is the value of the magnetic field that leads to the largest measured Coulomb blockade of Cooper-pairs in the array. The above results were reproduced for 1 MHz $\leq f \leq$ 1000 MHz microwave frequency range, and representative results displayed in Fig. 2.5 for $f = 1, 10, 100, 1000$ MHz.

Finally, the simulated characteristics for $H = 0$ and $H = 500$ Oe in Fig. 2.4(c) and Fig. 2.4(d) respectively are calculated using their corresponding unirradiated array characteristics [†], $I_0(V)$ in eq. 2.3 and plotted alongside the aforementioned experimental results, in Fig. 2.5 and Fig. 2.5. Further discussion on the simulation is given in the next the next section.

---

[†] given by the dotted curves in Fig. 2.4

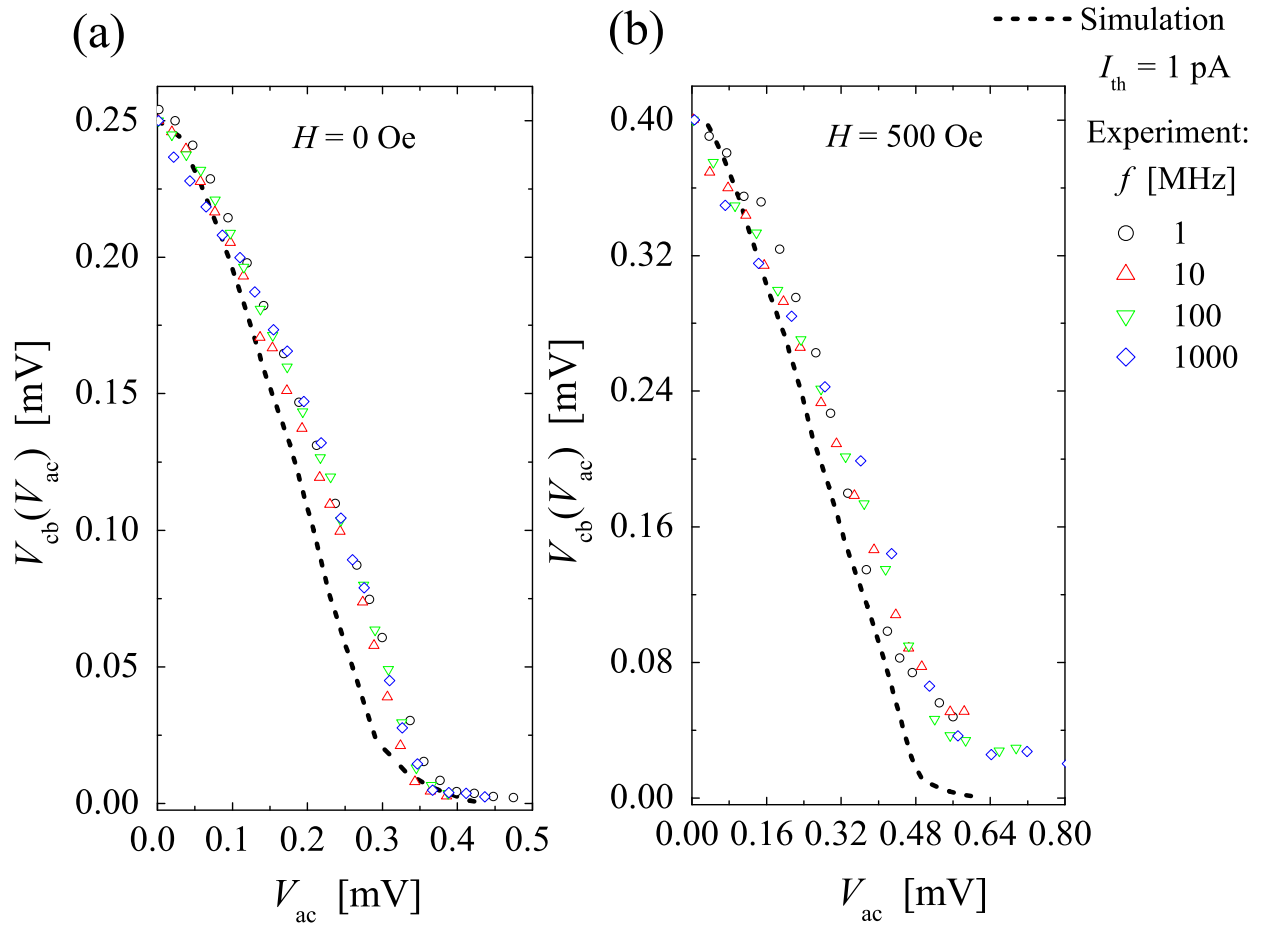

**Figure 2.5.:** The dependence of Coulomb blockade voltage, $V_{cb}(H = 0, 500\,\text{Oe}, V_{ac})$ to microwave amplitude, $V_{ac}$ plotted from the measured I–V characteristics for magnetic field $H = 0$ Oe and $H = 500$ Oe and frequencies $f = 1$ MHz, 10 MHz, 100 MHz and 1000 MHz. The results from the simulated I–V characteristics are plotted as dashed curves. The coulomb blockade voltage is determined at the current value, $I_{th} = 1$ pA. (Figure partially reproduced from ref. [61] with permission from the journal.)

## 3. Discussion

### Irradiation effects due to electron heating in the array

It is important to consider the effects of electron heating of the sample by the applied microwaves.[79, 80] Electron heating occurs via the excitation of quasi-particles in the islands and electrodes leading to dissipation effects. Since these quasi-particles are thermal, their average number can be approximated by the Boltmann distribution given by $\exp(-\epsilon_p/k_B T)$, where $\epsilon_p = \sqrt{p^2/2m + \Delta^2}$ is the BCS quasi-particle energy formula, $p$ the momentum of the quasi-particle measured from the Fermi surface, $m$ the effective mass of the quasi-particle and $\Delta$ the BCS energy gap.[58] Typically, these quasi-particles will not propagate in the sample, leading to $p = 0$.

By roughly estimating the differential resistance near $I_{\text{th}} = 1$ pA to be 10 MΩ, we can estimate the dissipation for the maximum power ($P$ = -61 dBm) incident on the array to be $\approx$ 10 fW. Using the volume 0.12 $\mu$m$^3$ of the array and the electron-phonon coupling constant 2 nW K$^{-5}$ $\mu$m$^{-3}$ in Aluminium, we can calculate[‡] the temperature of the electrons to be $T_e \approx 130$ mK, which is greater than the substrate temperature of 40 mK. However, the corresponding decrease in the Coulomb blockade voltage, $V_{\text{cb}}$ will be small ($\leq$ 10%) up to 130 mK, as was reported in ref. [81] for such a Al/AlO$_x$/Al-junction with similar geometry and parameters. In other words, electron heating via thermally excited quasi-particle excitations will only take for large $V_{\text{ac}}$ values in the curves reported in Fig. 2.5.

In particular, the quasi-particle number per electrode at the electron temperature, $T_e \approx 130$ mK can be estimated to be around 0.01 for $H = 0$ Oe and 0.2 for $H = 500$ Oe, using the Boltzmann factor, $\exp(-\Delta/k_B T)$ and the formular for the effective number of quasi-particle states $N_{\text{eff}} \sim \mathcal{V}\rho(0)\sqrt{2\pi k_B T \Delta}$ with $\rho(0) = 1.45 \times 10^{47}$ m$^{-3}$J$^{-1}$ the Aluminium density of states, $\mathcal{V} = 0.013$ $\mu$m$^3$ the electrode volume and $\Delta$ the superconducting gaop given in Table 2.1 [81]. This leads to a negligibly small quasi-particle number at the electrode for the microwave amplitude range considered in this experiment.

### Cooper pair photon-assisted tunneling

Photon assisted tunneling of cooper-pairs in a Josephson junction is successfully described within the context of $P(E)$ theory[15, 16] by the following formula,

$$I(V) = \sum_{n=-\infty}^{\infty} J_n^2(2eV_{\text{ac}}/\Omega)\, I_0(V - n\Omega/2e), \quad (2.2)$$

containing the *I–V* characteristics of the unirradiated junction ($V_{\text{ac}} = 0$), $I_0$ where $J_n(x)$ are the Bessel functions of the first kind.[13] When the frequency $f$ is infinitismal compared to microwave amplitude ($\Omega \ll 2eV_{ac}$), eq. (2.2) can be approximated with a classical expression[§] describing the microwave absorption by the junction,[49]

$$I(V) = \frac{1}{\pi} \int_{-\pi/2}^{\pi/2} d\theta\, I_0(V - V_{ac} \sin \theta). \qquad (2.3)$$

---

[‡] by the discussion given in refs. [79, 80]

[13]: This is the well-known Tien-Gordon formula and corresponds to the photon-assisted tunneling of Cooper pairs through the junction.[48]

This implies that photon assisted tunneling is restricted to the effect of a large number of photons leading to the smearing out of photon assisted tunneling effect in the characteristics of the junction by steady state current averages for each tunneling event.[49]

The simulated curves in Fig. 2.4 were numerically produced using eq. (2.3) by approximating the integral with Simpson's rule,

$$I \simeq \frac{\Delta\theta}{3\pi}\left[I_0(\theta_0) + I_0(\theta_{2l}) + 2\sum_{m=1}^{l} I_0(\theta_{2m}) + 4\sum_{m=1}^{l} I_0(\theta_{2m+1})\right]$$

where $\theta_m = -\pi/2 + m\pi/2l$ defined between the entire integration interval $\pi = 2l\Delta\theta$ for $2l + 1$ values bounding $2l$ equally spaced intervals of width $\Delta\theta$. The sum was then carried out using spline interpolation[82] with the code appended in G.

Using the aforementioned procedure and substituting $I_0(V)$ with the experimental curves for the irradiated array (given by the dashed curves) for $H = 0$ Oe and $H = 500$ Oe in Fig. 2.4(a) and 2.4(b), we simulate numerically the I–V curves given in Fig. 2.4(c) and 2.4(d). The simulation reproduces the diminishing of Coulomb blockade characteristics via irradiation by microwaves. To qualitatively compare and contrast the simulated results with the experimental curves, we plotted the $V_{cb}$–$V_{ac}$ curves of the simulated characteristics as dashed curves alongside the experimental characteristics in Fig. 2.5 in a similar fashion as before ($I_{th} = 1$ pA). This confirms the general trend of the Coulomb blockade voltage diminishing for the simulated curves, albeit with a larger gradient compared to the characteristics of the irradiated array.

---

[§] Classical here means frequency independent

**Comparing two characteristics with different Coulomb blockade voltage values**

There is need to compare and contrast the $H = 0$ Oe to $H = 500$ Oe results. This can be carried out under by assuming the characteristics for the unirradiated array at zero temperature, $I_0(V)$ can be approximated as,

$$I_0(V) = \frac{V - V_{cb}}{R^*}\Theta(V - V_{cb}), \qquad (2.4)$$

where $R^*$ is the magnetic-field dependent differential resistance above the Coulomb blockade voltage $V_{cb}$ whereas $\Theta(x)$ is the Heaviside function. Note that $I_0(V)$ is slightly smeared near $V_{cb}$ by finite temperature effects but otherwise eq. (2.4) is an excellent approximation for the Coulomb blockade characteristics at near absolute zero temperatures.[16]

To analyzing our results, we employ a scaling form for $I_0(V)$,

$$I_0(V) = \frac{V_{cb}}{R^*} g(V/V_{cb}), \qquad (2.5)$$

with $R^*$ the differential resistance for voltages slightly greater than $V_{cb}$, and $g(x = V/V_{cb})$ a general scaling function that is temperature and magnetic field independent[14]

14: Generally, since $V_{cb}$ and hence the absolute zero temperature characteristics of the unirradiated array depend on applied magnetic fields through $E_J/E_c$, we should write this expression as $I_0(V) = \frac{V_{cb}}{R^*} g(E_J/E_c, V/V_{cb})$ instead. However, since $E_J$ is infinitesimal compared to $E_c$, we can drop this consideration from our analysis.

15: Generally, $g(x)$ asymptotes to zero for $x < 1$, and rapidly increases for $x > 1$.

Thus, using the ideal case given by eq. (2.4), we can write the scaling function as $g(x) = (x - 1)\Theta(x - 1)$.[15] This scaling guarantees the $V_{cb}(V_{ac}, H)$–$V_{ac}$ characteristics fall on a singular line after both axes are normalized by their corresponding $V_{cb}(V_{ac} = 0, H)$ values. As can be seen in Fig. 2.6, the data from experiment conforms to this scaling at least near $V_{cb}$. For such an analysis to be valid, we assume that the Coulomb blockade characteristics of the unirradiated array is such that the *I-V* curves keep a trivial form below a threshold voltage $V_{th}$ and exhibit a steep rise above $V_{th}$ signaling the injection of charge carriers into the array. For the unirradiated array, the this threshold voltage can be determined from the equation,

$$I_{th} = \frac{V_{th}}{R^*} g(V_{th}(V_{ac} = 0)/V_{th}). \qquad (2.6)$$

leading us to equate it to the Coulomb blockade threshold

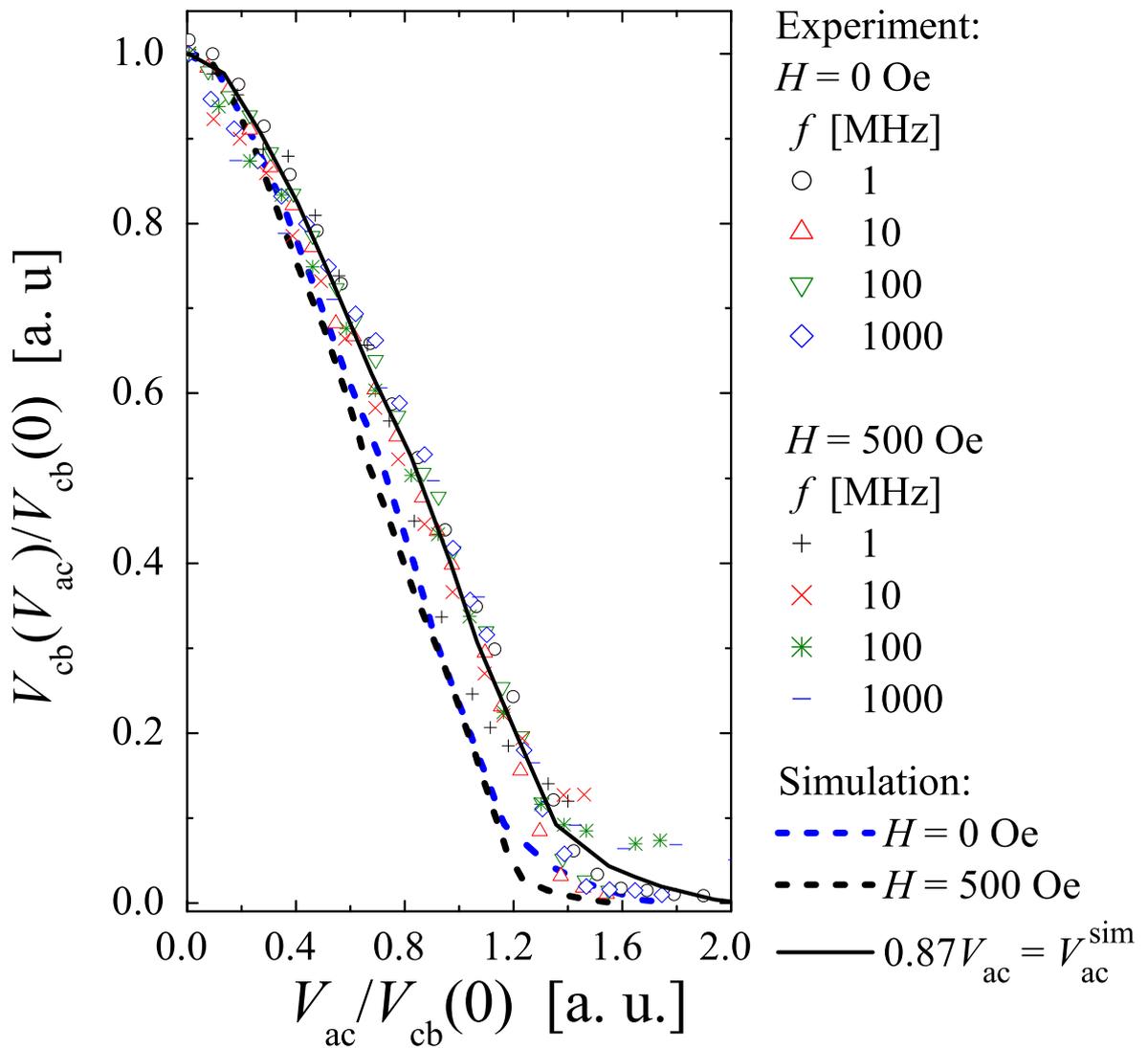

**Figure 2.6.:** The dependence of normalized Coulomb blockade voltage, $V_{cb}(H, V_{ac})/V_{cb}(H, 0)$ to normalized microwave amplitude, $V_{ac}/V_{cb}(H, 0)$ plotted from the measured I–V characteristics for magnetic field $H = 0$ Oe and $H = 500$ Oe and frequencies $f = 1$ MHz, 10 MHz, 100 MHz and 1000 MHz. The coulomb blockade voltage is determined at the current value, $I_{th} = 1$ pA. The normalization is carried out by dividing both axes by their respective $V_{cb}(H, 0)$ values. The normalized plots for the $H = 0$ Oe and $H = 500$ Oe simulated curves, given by the blue and black dashed curves respectively, are presented alongside the experimental results. The two simulated curves coincide within a small margin of error. The two simulated curves labeled by $V_{ac}^{sim}$ in the legend exhibit a steeper gradient than the experimental results labeled by $V_{ac}$ by a factor of $1/0.87$, suggesting the need to consider other effects to successfully explain the experimental results. (Figure partially reproduced from ref. [61] with permission from the journal.)

voltage of the experiment,

$$V_{\text{th}}(V_{\text{ac}} = 0) \simeq V_{\text{cb}}, \tag{2.7}$$

for $I_{\text{th}} = 1$ pA.

Thus applying eq. (2.7) into eq. (2.3), the expression for $I_{\text{th}} R^* / V_{\text{cb}}$ when the array is irradiated by microwaves is,

$$\frac{1}{\pi} \int_{-\pi/2}^{\pi/2} d\theta \, g(V_{\text{cb}}/V_{\text{cb}}(V_{\text{ac}} = 0) - V_{\text{ac}}/V_{\text{cb}}(V_{\text{ac}} = 0) \sin \theta). \tag{2.8}$$

Thus, $V_{\text{cb}}/V_{\text{cb}}(V_{\text{ac}} = 0)$ is effectively a function of $I_{\text{th}} R^* / V_{\text{cb}}$ and $V_{\text{ac}}/V_{\text{cb}}(V_{\text{ac}} = 0)$. This leads to a magnetic field independent approximation for $I_{\text{th}} R^* / V_{\text{cb}}$ at the threshold current, $I_{\text{th}} = 1$ pA.

On the other hand, the scaling for $V_{\text{cb}}$ is obtained by,

$$V_{\text{cb}}/V_{\text{cb}}(V_{\text{ac}} = 0) = h(V_{\text{ac}}/V_{\text{cb}}(V_{\text{ac}} = 0)), \tag{2.9}$$

where $h(x)$ is a magnetic field, microwave amplitude independent scaling function. The plot for measured $y = V_{\text{cb}}/V_{\text{cb}}(V_{\text{ac}} = 0)$ versus $V_{\text{ac}}/V_{\text{cb}}(V_{\text{ac}} = 0)$ for $H = 0$ Oe and $H = 500$ Oe at varied frequencies of the microwave radiation is given in Fig. 2.6. Consequently, the characteristics fall approximately on a singular line independent of the applied magnetic field $H$ and frequency $f$. All the characteristics fall on the same line irrespective of $H$ or $f$ as expected, which demonstrates the validity of this scaling.[16]

16: This demonstrates a singular relevant energy scale given by $2eV_{\text{cb}}$ for photon assisted tunneling of Cooper-pairs in the classical limit and the lifting of Coulomb blockade in Josephson junction arrays satisfying $E_J \ll E_c$.

We proceed to plot, in Fig. 2.6, the curve for $H = 0$ Oe from numerical simulation with eq. (2.3) alongside the aforementioned curves from experiment. The simulated curves show approximately the same characteristics except for the gradient which is steeper, differing from the curves from experiment by a factor of $1/0.87$. We define $\Xi_A = 0.87$ for the array to represent this factor whose origin is yet unexplained. Such a factor less than unity ($\Xi_A$) implies the response of the array to irradiation is suppressed relative to the approach to photon-assisted assisted tunneling encapsulated by eq. (2.2) and eq. (2.3). In this chapter, we shall analyze the origin of such as factor by considering the difference in the response between applied dc and ac voltages.[17]

17: Later in subsequent chapters, we consider this effect to arise from renormalization of the amplitude of the microwave radiation applied to the array, as comprehensively discussed in chapter 5.

## Dependence on magnetic field

In the previous section, we have argued that the magnetic field does not alter the scaling $h(x)$. However, the effect of $H$ will still appear in the Coulomb blockade voltage $V_{cb}$. In the experiment, a substantial non-varying magnetic field, $H = 500$ Oe is applied perpendicular to the unirradiated array in order to raise the value of the Coulomb blockade (threshold) voltage $V_{cb}$ to its maximum. In the experiment, this corresponds to a factor of approximately 1.4 its original value for $H = 0$ Oe.[38, 47] Nonetheless, the $V_{cb}$ versus $V_{ac}$ characteristics of the irradiated array when $H = 500$ Oe coincide with those for $H = 0$ Oe when both axes of the $V_{cb}$–$V_{ac}$ plots are rescaled by the aforementioned factor,

$$\frac{V_{cb}(H = 500 \text{ Oe}, V_{ac} = 0)}{V_{cb}(H = 0, V_{ac} = 0)} \simeq 1.4, \quad (2.10)$$

as discussed in the previous sections.

Heuristically, this can be explained within the context of the dynamics of the quasi-charge of each Josephson junction in the array within their respective Brillouin zone of the Bloch energy band. In particular, the energy band gap, which is comparable to $E_J(H)$, is diminished[18] by applying a magnetic field $H \leq H_{max}$ where $H_{max} = 500$ Oe is the value of the magnetic field that leads to the largest Coulomb blockade of Cooper-pairs in the sample.[2] However, it is rather unwieldy to calculate the $E_J$-dependence of $V_{cb}(V_{ac} = 0)$, since many-body effects for the tunneling Cooper-pairs in the array have to be considered in detail. Such a calculation has been conducted in within the context of a depinning potential in ref. [83]. Using the parameters of the array in our experiment[19] yields,

$$\frac{V_{cb}(H = 500 \text{ Oe}, V_{ac} = 0)}{V_{cb}(H = 0, V_{ac} = 0)} = [U(0.22)/U(0.27)]^{2/3} \simeq 1.05, \quad (2.11)$$

with $U(E_J/E_c)$ the depinning potential. A more accurate calculation, incorporating comprehensive measurements of the array parameters and simulation of the characteristics of the array under $H \neq 0$ is beyond the scope of this thesis.

18: As is apparent in Table 2.1

19: and Fig. 4 of Ref. [83]

**Figure 2.7.:** (a) A diagram depicting the symmetric dc-biasing of the array of $N_0$ Josephson junctions $(-V/2, +V/2)$ and asymmetric ac bias $(V_{RF} = V_{ac} \cos \Omega t)$ from the left corresponding to the effect of microwave irradiation; (b) The equivalent circuit of the array showing the positions of relevant circuit elements where $R_j$, $C$, $C_0$ correspond to the resistance due to the environment of each junction, the junction capacitance and the stray capacitance of adjacent islands respectively. ; (c) A simplified circuit of half the array depicted in (b) and (a). The environment is now given by the sum of resistances $R = N_0 R_j/2$ where $J_{HA}$ indicates half the array (HA). The total capacitor of the half array can be calculated in the semi-infinite approximation ($N_0 \gg 1$) as $C_{HA} = \left(C_0 + \sqrt{C_0^2 + 4CC_0}\right)/2$; (d) The equivalent circuit of (c). [Figures (a), (b) and (c) have been reproduced from ref. [61] with permission from the Journal.]

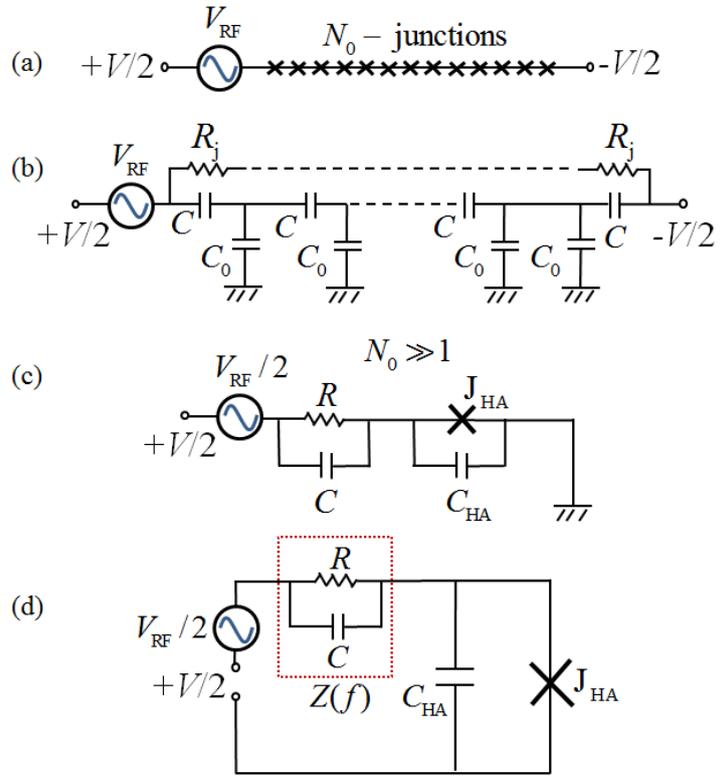

20: Note here that, 'renormalization' means the rescaling of the amplitude of the incident microwaves relative to the absorbed amplitude by the array. The justification of using 'renormalization' instead of 'rescaling' will be tackled in subsequent chapters.

21: Inductive effects can dramatically alter the characteristics of the array by inducing a superconducting phase form the Coulomb blockade phase. [35]

### Renormalization effect

Here, we shall discuss the origin of $\Xi_A = 0.87$ as arising from the 'renormalization' of the amplitude of the applied microwave.[20] The approach considers the impedance of the equivalent circuit of the array consistent with ref. [84]. The impedance analysis is simplified by setting $E_J = 0$, which is justified by the condition $E_J \ll E_c$, that neglects inductive effects of the array.[21]

Proceeding step by step, we first consider the equivalent the circuit of the array biased symmetrically with a dc voltage $(-V/2, +V/2)$ and asymmetrically biased by an ac voltage, $V_{RF} = V_{ac} \cos \Omega t$ from the left corresponding to the effect of microwaves on the array depicted in Fig. 2.7(a). Such a circuit can be analyzed by the equivalent circuit depicted in Fig. 2.7(b). This equivalent circuit shows the positions of relevant circuit elements where $R_j$, $C$, $C_0$ correspond to the resistance due to the environment of each junction, the junction capacitance and the stray capacitance of adjacent islands respectively. The symmetric biasing renders the voltage drop at the center of the array zero. This observation can be exploited to transform Fig. 2.7(b) into the circuit of half the array shown in Fig. 2.7(c) which we shall use to compute the effective impedance of the circuit. One approach to achieve this is by employing the theory of continued fractions.[85] However, such an approach yields unwieldy insoluble expressions for the impedance. An alternative approach is to substitute the circuit elements in series in half the array with effective ones. Thus, we substitute the series of resistances, $R_j$ with the sum $R = N_0 R_j/2$ representing the total environmental resistance, the series of junction elements with a single element for the half array $J_{HA}$, and the series of the capacitances, $C_0$ and $C$ with a unified capacitance given by the analytic expression for capacitance of the half array,

$$C_{HA} = \frac{1}{2}\left(C_0 + \sqrt{C_0^2 + 4CC_0}\right), \quad (2.12)$$

valid for a large number of junctions $N_0 \gg 1$. This leads to the equivalent circuit diagram given in Fig. 2.7 (d), typical for studying the Coulomb blockade within the context of $P(E)$ theory.[15, 16, 27]

Thus, the effective impedance for the half array in Fig. 2.7 (d) becomes,

$$Z_{\text{eff}}(\Omega) = \frac{1}{R^{-1} + i\Omega C}. \quad (2.13)$$

In the absence of applied microwaves, ($V_{ac} = 0$), the effective impedance $Z_{\text{eff}}(\Omega)$ becomes $R$ and the final steady state is characterized by the direct biasing of $J_{HA}$ parallel to $C_{HA}$ by the dc voltage $V$. However, when an ac voltage is applied ($V_{ac} \neq 0$), the effective impedance explicitly depends on the capacitance. For large frequencies compared to the time constant of the circuit, $\Omega \gg 1/RC$, the effective impedance becomes, $Z_{\text{eff}}(\Omega) \simeq 1/i\Omega C$.[22]

22: This approximation is valid if we take $R > 10^9\ \Omega$ to be always greater than the zero bias resistance of the Coulomb blockade characteristics of the array, which yields $1/RC < 6.28$ MHz.

Therefore, the effective ac voltage applied to $J_{HA}$ is renormalized by the impedance ratio,

$$\lim_{RC\Omega \to +\infty} \frac{(i\Omega C_{HA})^{-1}}{Z_{\text{eff}}(\Omega) + (i\Omega C_{HA})^{-1}} = \frac{C}{C + C_{HA}}. \quad (2.14)$$

This ratio merely corresponds to the varied response of the dc voltage form the ac voltage across the array to the center junction producing the factor,

$$\Xi_A = \frac{C}{C + C_{HA}} = \exp(-\Lambda^{-1}), \qquad (2.15)$$

where $\Lambda$ is the characteristic decay length of the electric potential along the array,[41, 51, 52]

$$\Lambda = \left[\cosh^{-1}\left(1 + \frac{C_0}{2C}\right)\right]^{-1} \simeq \sqrt{\frac{C}{C_0}}, \qquad (2.16)$$

known as the soliton length of the array. We measured an array of the same structure using the gate effect[86] [23] to yield $\Lambda \simeq \sqrt{C/C_0} \simeq 9$. From this result, we can determine the renormalization factor of the array to be $\Xi_A \simeq 0.89$, which is within the margin of error of the experimental results.[24]

**Detector application**

An obvious application of the diminishing of Coulomb blockade is the detection of microwaves. We can estimate the sensitivity of such a detector from our results in Fig. 2.5, where a slight change in the Coulomb blockade voltage of order 10 $\mu$V, corresponds to a microwave amplitude (power) of order 40 $\mu$V (4 pW, using eq. (2.1)). Consequently, the sensitivity to small signals becomes greater than $10^6$ V/W. This value far surpasses the sensitivity of standard microwave detection schemes using diodes by about $5.0 \times 10^3$. Such high sensitivities enable Josephson junction arrays to respond to microwave power even in situations where only small coupling schemes to the microwave source is available.

Since a clear Coulomb blockade characteristic is indispensible for such high sensitivity detection schemes, single small Josephson junctions and single Cooper pair transistors are inadequate, since they need to be embedded within a high impedance environment to exhibit clear Coulomb blockade characteristics.[36, 37] Thus, the linear array offers a non-complex and extremely tractable device for microwave detection. An exemplary method of use is to bias the array with a current of order pA, and observe the Coulomb blockade characteristics at around 100 $\mu$V level diminish. This results in extremely low power dissipation of ranging from 0.1 to 1 fW, suitable for detection schemes in dilution refrigerators. Such a detection scheme has successfully been

---

23: We applied a voltage bias $U + V/2$ and $U - V/2$ at each ends of the array respectively corresponding to a total bias of $V$ across the array and an additional offset voltage $U$. The additional potential produces responses in the differential conductance of the array, $dI/dV$ that correspond to maximal disorder where the noise is nearly periodic in $U$. Hence, this periodicity is used to extract the stray capacitance $C_0 = 9$ aF when the array is subjected to a magnetic field higher than the superconducting critical value of aluminium, $H > H_c$, where the conductance structure is exactly periodic with $U = e/C_0$.

24: A different sample was measured prior to the sample reported here at threshold current $I_{th} = 3$pA, frequency range 1MHz $\leq f \leq$ 3GHz and no applied magnetgic field ($H = 0$ Oe), finding $\Xi_A \simeq 0.80$ as can be seen in appendix F. However, despite the large frequency range, it has a large scatter and fewer measured points.

carried out by detecting microwave emission from a single Cooper pair transistor fabricated on-site 2 $\mu$m adjacent to and decoupled from the array applied as the detector.[54] This detection scheme differs from the one presented in ref. [40], where the SCPT is coupled to the detector. Since the detector registers a strong signal even at large separation distances of order 30 $\mu$m between the microwave source and detector, it offers a tractable microwave detection scheme without coupling it to a microwave source.

## 4. Conclusion

In this chapter, we have described the irradiation of a linear array of small aluminium Josephson junctions in the charge dominant regime ($E_c > E_J$) exhibiting distinct Cooper-pair Coulomb blockade characteristics, by microwaves in the sub-gigahertz frequency range. Irradiation has the effect of diminishing the Coulomb blockade characteristics. This effect is independent of the microwave frequency in accordance with the standard theory of Coulomb blockade ($P(E)$ theory) and the classical expression for the photon-assisted tunneling of Cooper pairs. [15, 16, 48, 49] However, our results differ from expected results by a microwave amplitude renormalization factor of approximately 0.87. We analyze such a factor by considering the effective impedance of a semi-infinite model circuit of the array, where the factor can be attributed to the difference in response of the circuit to dc and ac voltages. This connects the factor to the soliton length $\Lambda$ of the array[41, 50–52] by the expression, $\Xi_A = \exp(-\Lambda^{-1}) \simeq 0.89$. Further deliberation on this connection has been carried out in subsequent chapters.

# Theoretical Basis For Renormalization

# Introduction 3.

The interaction of matter with radiation has been the cornerstone of quantum mechanics since the successful theoretical description of the black-body spectrum by Max Planck,[87] where the spectral radiance of a black-body for a mode of frequency $\Omega$ and at absolute temperature $T = \beta^{-1} k_B^{-1}$ is given by $\rho(\Omega, T) = \Omega^3/\pi^3 c^3 \left[\exp(\beta\Omega) - 1\right]$. The mantle was taken up by Einstein, Bohr, Schrödinger, Dirac and others[88] in the early development of quantum theory, later leading to the fully fledged theory of quantum electrodynamics (QED) by Tomonaga, Swinger, Feynmann, Tomonaga and Dyson.[89] Einstein introduced the concept of photons in a seminal paper on the photon electric effect[90] and years later proceeded to describe spontaneous emission, stimulated emission and stimulated absorption of matter at equilibrium with thermal radiation by introducing his famous $A_{21}(\Omega)$, $B_{21}(\Omega)$ and $B_{12}(\Omega)$ coefficients[91, 92] representing the transition rates per particle for the radiation processes respectively (Fig. 3.1). In this introductory chapter, we establish how these coefficients manifest in the Josephson junction, and their relation to the dc and ac effect of renormalization described in chapter 2. This chapter serves as a conduit between the experiment part of the thesis, and the subsequent chapters detailing the theoretical work pertaining renormalization.

## 1. Einstein's *A* and *B* coefficients

Einstein showed, through a detailed balance calculation, [91]

$$0 = -A_{21} N_2 - B_{21} \rho(\Omega) N_2 + B_{12} \rho(\Omega) N_1, \quad (3.1)$$

$$\rho(\Omega) = \frac{A_{21}(\Omega)/B_{12}(\Omega)}{N_1/N_2 - B_{21}(\Omega)/B_{12}(\Omega)}, \quad (3.2)$$

that the necessary condition for Planck's law of black-body radiation to be satisfied is $B_{12}(\Omega) = B_{21}(\Omega) \equiv B(\Omega)$ and $A_{21}(\Omega)/B_{21}(\Omega) = \Omega^3/\pi^3 c^3$, where the number of particles populating energy states $\mathcal{E}_2 > \mathcal{E}_1$ (assuming no degeneracy)

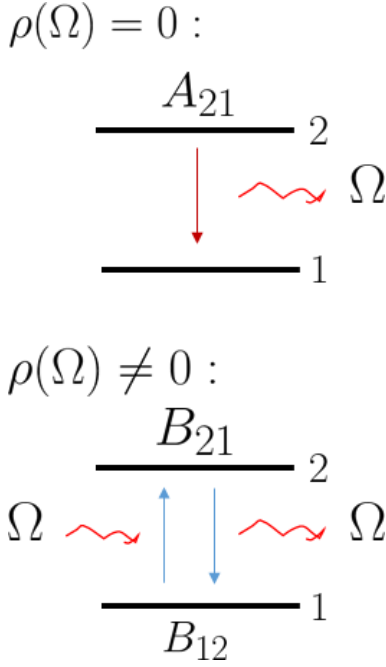

**Figure 3.1.:** A schematic depicting a two level system undergoing spontaneous emission $\rho(\Omega) = 0$ and stimulated emission, $\rho(\Omega) \neq 0$

**Table 3.1.:** Transforming the sum for infinite modes $\omega_i$ of oscillation of the dipole to an integral:

$$\sum_i E_i^2 \to 8\pi \int d\omega \rho(\omega)$$

in the two level system considered is $N_1 = N \exp(-\beta \mathcal{E}_1)$ and $N_2 = N \exp(-\beta \mathcal{E}_2)$ respectively with $\mathcal{E}_2 - \mathcal{E}_1 = \Omega$ and $N$ the total number.

## 2. Determining $B$ (Fermi's golden rule)

These coefficients govern the quantum mechanical transition rates per particle, which are intrinsic to the radiating system, and thus can actually be independent of thermal equilibrium. For instance, the derivation of the stimulated emission rate per particle, $\Gamma_{2\to1} = B_{21}(\Omega)\rho(\Omega)$ between the two quantum mechanical energy states can effectively be carried our in quantum mechanics by applying Fermi's golden rule,[93]

$$\Gamma_{2\to1}(\Omega) = 2\pi \sum_i |\langle \mathcal{E}_2 | H_i | \mathcal{E}_1 \rangle|^2 \delta(\mathcal{E}_2 - \mathcal{E}_1 - \omega_i), \quad (3.3)$$

Here, $\mathcal{E}_2 - \mathcal{E}_1 = \Omega$ as before, $H_i = (\vec{d}^* \cdot \vec{n})E_i |\mathcal{E}_2\rangle \langle \mathcal{E}_1| + (\vec{d} \cdot \vec{n})E_i |\mathcal{E}_1\rangle \langle \mathcal{E}_2|$ is the transition Hamiltonian for the two level system, $E_i$ is the amplitude of an oscillating electric dipole moment of energy, $H_d = (\vec{d} \cdot \vec{n})E_i \exp(-i\omega_i t)$ where $\vec{n}$ is the unit vector, $\vec{d} = q\vec{r}$ is the electric dipole moment, $q$ the dipole charge and $\vec{r}$ the displacement vector. The transformations in Table 3.1 leading to $B_{21}(\Omega) = 4\pi^2 |\vec{d} \cdot \vec{n}|^2$. Then $A_{21}(\Omega)$ can be determined indirectly using $A_{21}(\Omega)/B_{21}(\Omega) = \Omega^3/2\pi^3 c^2$.

The calculation of $A_{12}(\Omega) \equiv A(\Omega)$ from first principles of quantum theory required the formulation of QED, where the radiation field as well as the particle states are both treated quantum mechanically, since spontaneous emission is due to fluctuations of the QED ground state.[94, 95] In contrast, standard quantum mechanics is incapable of calculating $A_{12}(\Omega)$ directly since the radiation field is treated classically by exploiting the gauge invariance of the Schrödinger equation. Instead, spontaneous emission calculations are undertaken within the context of Weisskopf-Wigner theory of spontaneous emission.[96]

The quantum mechanical understanding of the radiation processes has found a wide range of applications particularly in the field of photonics in microwave amplification by stimulated emission of radiation (maser) and light amplification by stimulated emission of radiation (laser) technologies.[97]

# 3. Fluctuation-dissipation theorem

Essentially any quantum mechanical system that couples to radiation can be exploited for amplification of radiation similar to the laser, provided the electrodynamics governed by the dynamics of the system in question is known. In particular, Callen and Welton, demonstrated that the real part of the admittance (kernel) for a system driven by oscillating electromagnetic fields appears as the coefficient of the black-body spectrum,

$$\frac{A(f)}{B(f)} \sim q^2 K^{-1}(f) f^n, \quad (3.4)$$

where $K(f)$ is the kernel of the system, $n$ is the number of space-time dimensions of the system in consideration and $f = \Omega/2\pi$ is the oscillation frequency.[1] This is simply the manifestation of the well-known fluctuation-dissipation theorem.[9] This theorem is universal, for any system whose dynamics are highly dependent on the fluctuations of the environment. Such exotic systems ranging from black holes[98, 99] to small Josephson junctions[16, 100] satisfy the fluctuation-dissipation theorem.[9]

1: The ~ is used to mean not always satisfied by an equal sign, since one has to sometimes consider negative frequencies, as is the case in the Caldeira-Leggett model tackled below.

# 4. Tunnel junctions

Within the $P(E)$ theory, Cooper pair tunneling events are governed by the probability per unit energy, $P(E)$ for the junction to absorb energy $E$ from the environment, given by the function,

$$P(E) = \frac{1}{2\pi} \int d(t - t') \exp(-4\mathcal{J}(t - t')) \exp(-iE[t - t']), \quad (3.5)$$

where, $\mathcal{J}(t - t') = \langle \phi(t)\phi(t') \rangle - \langle \phi(0)\phi(0) \rangle$ is given by the phase-phase correlation function,

$$\langle \phi(t)\phi(t') \rangle = \frac{e^2}{2\pi} \int \frac{d\omega}{\omega} [Z(\omega) + Z(-\omega)] \frac{\exp(-i\omega[t - t'])}{1 - \exp(-\beta\omega)}, \quad (3.6)$$

where $\phi$ is the phase difference of the junction and $Z(\omega)$ is the impedance related to the environment.[16] The spectral radiance of the junction is simply the $n = 1$ (space-time) dimensional spectral radiance, and is related to eq. (3.6) by the Fourier transform of,

$$\frac{\langle \partial \phi(t) \partial \phi(t') \rangle}{\partial t \partial t'} \equiv \rho(t - t') =$$
$$\frac{e^2}{2\pi} \int d\omega\, \omega \left[ Z(\omega) + Z(-\omega) \right] \frac{\exp(-i\omega[t - t'])}{1 - \exp(-\beta\omega)},$$

given by,

$$\rho(-\Omega) = \frac{1}{2\pi} \int d(t - t') \rho(t - t') \exp(-i\Omega[t - t']) =$$
$$\frac{e^2}{2\pi} \left[ Z(\Omega) + Z(-\Omega) \right] \frac{\Omega}{\exp(\beta\Omega) - 1},$$

where we can read-off the Kernel using eq. (3.4) to yield, $K^{-1}(\Omega) = \left[ Z(\Omega) + Z(-\Omega) \right] / 2\pi$.

The form of $Z(\Omega)$ depends on the quantum circuit considered, and is generally difficult to ascertain. This is due to the strong coupling of the small junction to the electromagnetic environment, which significantly modifies the *I–V* characteristics compared to the large junction.

### Large Josephson junctions

In particular, the supercurrent density $J_S$ through the large junction[59] can be written as,

$$\vec{J}_S(t, \vec{r}_\perp) = -\vec{n} \frac{2eE_J}{\mathcal{A}} \sin 2\phi(t, \vec{r}_\perp), \quad (3.7)$$

where $E_J$ is the Josephson coupling energy, $e$ is the electron charge, $\mathcal{A}$ is the cross-sectional area of the tunnel barrier, $\vec{n}$ is the unit vector normal to the tunnel barrier pointing in the tunneling direction and $\phi(t, \vec{r}_\perp)$ is the phase difference across the junction that depends on time, $t$ and the spatial coordinates, $\vec{r}_\perp$ perpendicular to the tunneling direction. The phase couples to the electric field $\vec{E}$ and magnetic field $\vec{B}$ through its space-time derivatives,

$$\frac{\partial \phi(t, \vec{r}_\perp)}{\partial t} = -e d_{\text{eff}} \vec{n} \cdot \vec{E}(t, \vec{r}_\perp), \quad (3.8a)$$
$$\vec{\nabla}_\perp \phi(t, \vec{r}_\perp) = e d_{\text{eff}} \vec{n} \times \vec{B}(t, \vec{r}_\perp), \quad (3.8b)$$

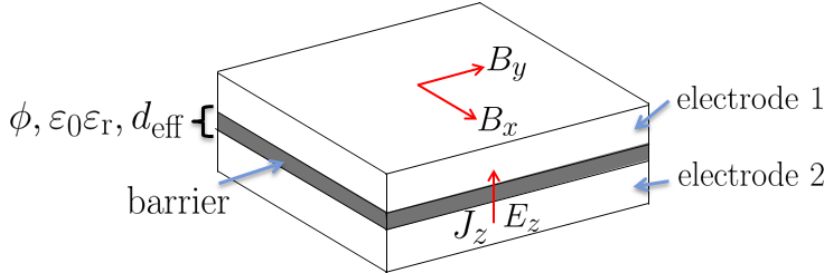

**Figure 3.2.:** The schematic of a large junction showing the electric field $\vec{E} = (0, 0, E_z)$ and magnetic field $\vec{B} = (B_x, B_y, 0)$, as well as the tunneling current $\vec{J} = \vec{J}_S + \vec{J}_N = (0, 0, J_z)$ where $\vec{n} = (0, 0, 1)$ points in the $z$ direction. The quantum phase difference of the electrodes, the permittivity of the barrier and the effective barrier thickness are given by $\phi$, $\varepsilon_0 \varepsilon_r$ and $d_{\text{eff}} \simeq d_0 + \lambda_1 + \lambda_2$ respectively where $\varepsilon_0$ is the permittivity of the vacuum, $\varepsilon_r$ is the relative permittivity of the barrier, $d_0$ is the thickness of the barrier and $\lambda_1, \lambda_2$ is the London penetration depth of electrode 1, 2.

as shown in fig. 3.2, where $\vec{\nabla}_\perp$ is the divergence operator and $d_{\text{eff}}$ is the effective thickness of the tunnel barrier. Note that the above equations have to be consistent with electromagnetism (Maxwell's equations),

$$\vec{\nabla}_\perp \times \vec{B}(t, \vec{r}_\perp) = \frac{1}{\varepsilon_0 \varepsilon_r} \left[ \vec{J}_S(t, \vec{r}_\perp) + \vec{J}_N(t, \vec{r}_\perp) \right] + \frac{\partial \vec{E}(t, \vec{r}_\perp)}{\partial t},$$

where $\varepsilon_0$ and $\varepsilon_r$ are the permittivity of the vacuum and the relative permittivity of the barrier, $\vec{J}_N(t, \vec{r}_\perp) = \vec{n} \int dt' \sigma(t - t') \vec{E}(t', \vec{r}_\perp)$ is the normal current through the system. Hence, $\sigma(\omega) = \int dt \sigma(t) \exp(-i\omega t)$ is the conductance, related to the kernel by $K(\omega) = \mathcal{A} \sigma(\omega) / d_{\text{eff}}$. Plugging in eq. (3.7) and eq. (3.8) into Maxwell's equations yields the field equation for the phase,

$$\frac{\partial^2 \phi(t, \vec{r}_\perp)}{\partial t^2} - \frac{1}{C} \int dt' K(t - t') \frac{\partial \phi(t', \vec{r}_\perp)}{\partial t} - \vec{\nabla}_\perp^2 \phi(t, \vec{r}_\perp) = -4 E_c E_J \sin 2\phi(t, \vec{r}_\perp), \quad (3.9)$$

where we have used the capacitance of the junction, $C = \varepsilon_0 \varepsilon_r \mathcal{A} / d_{\text{eff}}$, the charging energy, $E_c = e^2 / 2C$ and $\mathcal{A} \sigma(t - t') / d_{\text{eff}} = K(t - t')$.

## Small Josephson junctions: Caldeira-Leggett model

In the case of the small junction, the term $\vec{\nabla}_\perp^2 \phi = 0$ can be approximated to vanish due to the small cross-sectional area (small capacitance) of the junction. Such an approximation is convenient since one can apply a simple Hamiltonian formalism known as the Caldeira-Leggett model[10, 101, 102] to arrive at eq. (3.9). This entails employing the Hamiltonian,

$$H_{\text{CL}} = \frac{Q^2}{2C} - E_J \cos 2\phi + \sum_j \left\{ \frac{Q_j^2}{2C_j} + \frac{1}{2e^2 L_j}(\phi_j - \phi)^2 \right\} \tag{3.10}$$

2: Actually, the charge stored by the Josephson junction is $2Q$.

where $Q = CV$ is the charge stored[2] by the small junction of capacitance $C$ biased by voltage $V$, and the small junction couples to the electromagnetic environment represented by $L_j C_j$ circuit elements, where $\omega_j = 1/L_j C_j$ is the characteristic frequency of each mode and $Q_j = C_j V_j$ is the charge stored by each element in the circuit with a voltage drop $V_j$ and $\phi_j(t) = e \int_{-\infty}^{t} dt\, V_j(t)$ is the phase difference across each element. The Hamilton's equations of motion, $\partial H_{\text{CL}}/\partial Q = e^{-1}\partial \phi/\partial t$ and $-\partial H_{\text{CL}}/\partial \phi = e^{-1}\partial Q/\partial t$ for the phase of the junction $\phi$,[3] are,

3: $\partial H_{\text{CL}}/\partial Q_j = e^{-1}\partial \phi_j/\partial t$ and $-\partial H_{\text{CL}}/\partial \phi_j = e^{-1}\partial Q_j/\partial t$ for the environment degrees of freedom, $\phi_j$ yield, $\partial \phi_j(t)/\partial t = eQ_j(t)/C$, and $\partial Q_j(t)/\partial t = -e^{-2}L_j^{-1}\left[\phi_j(t) - \phi(t)\right]$ respectively.

$$\frac{\partial \phi(t)}{\partial t} = \frac{eQ(t)}{C}, \tag{3.11a}$$

$$\frac{\partial Q(t)}{\partial t} = -2eE_J \sin 2\phi(t) + \sum_j \frac{1}{e^2 L_j}\left[\phi_j(t) - \phi(t)\right]. \tag{3.11b}$$

When the time dependence of the phase $\phi(t)$ is known, the equations of motion for the environment can be solved in terms of their initial values, $Q_j(0)$, $\phi_j(0)$ and their influence on $\phi(t)$,[101]

$$\frac{1}{e^2 L_j^2}\left[\phi_j(0) - \phi(0)\right] =$$

$$\frac{1}{eL_j}\left[\phi_j(0) - \phi(0)\right]\cos \omega_j t + \omega_j Q_j(0) \sin \omega_j t$$

$$- \frac{1}{C}\int dt'\, \frac{Q(t')}{L_j} \cos \omega_j(t - t'). \tag{3.12}$$

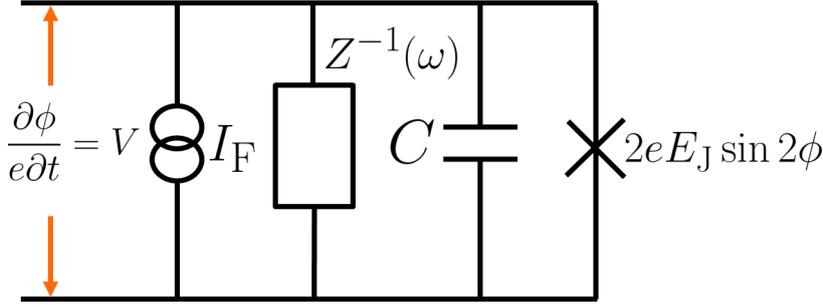

**Figure 3.3.:** A circuit depicting the terms in eq. (3.13) as a fluctuation current, a displacement current through a capacitance $C$, a dissipative current through an admittance $Z^{-1}(\omega)$ and a supercurrent through a Josephson junction. Thus, eq. (3.13) simply corresponds to Kirchhoff's Current Law for the circuit.

Plugging this back to eq. (3.11b) and using eq. (3.11a) yields eq. (3.9),

$$\frac{\partial^2 \phi(t)}{\partial t^2} - \frac{1}{C} \int dt' K(t-t') \frac{\partial \phi(t')}{\partial t'}$$
$$= -4E_c E_J \sin 2\phi(t) + 2e^{-1} E_c I_F(t), \quad (3.13)$$

where,

$$I_F(t) = \sum_j e^{-1} L_j^{-1} \left[\phi_j(0) - \phi(0)\right] \cos \omega_j t + \sum_j \omega_j Q_j(0) \sin \omega_j t,$$

is a fluctuation current due to the environmental degrees of freedom resulting in the normal current[4],

$$I_N = C^{-1} \int dt' Q(t') K(t-t')$$

4: Fluctuation-dissipation theorem[9, 10, 101]

and $K(t) = \sum_j L_j^{-1} \cos \omega_j t$. The sum can be transformed into an integral by $K(t) = \sum_j L_j^{-1} \cos \omega_j t \to \frac{1}{2\pi} \int d\omega Z^{-1}(\omega) \exp i\omega t$ where $Z(\omega)$ is the impedance of the environment. [5]

5: The kernel is given by $K(\omega) = \left[Z^{-1}(\omega) + Z^{-1}(-\omega)\right]/2\pi$. However, we shall first consider positive frequency modes.

## 5. Renormalization

### Different responses between dc and ac voltages

Since the junction biased with a voltage $V$ satisfies $V = -d_{\text{eff}} \vec{n} \cdot \vec{E}$, the coupling of the electric field in $\partial \phi/\partial t = eV$ allows for a swift comparison with the aforementioned case of the oscillating electric dipole. (For ease of expression, we set $E_J \simeq 0$, which is generally satisfied in small junctions.) In particular, we note the response of the junction biased via its electromagnetic environment by a constant voltage $\partial V/\partial t =$ 0 (dc response) versus the response for an oscillating bias voltage $\partial V_{\text{RF}}/\partial t \neq 0$ (ac response) will differ based on eq. (3.13). In particular, taking the Fourier transform of eq. (3.13), we find,

$$i\omega Z^{-1}(\omega)\phi(\omega) = eI_{\text{F}}(\omega), \qquad (3.14)$$
$$i\omega[i\omega C + Z^{-1}(\omega)]\phi_{\text{RF}}(\omega) = eI_{\text{F}}(\omega), \qquad (3.15)$$

for the dc and ac responses respectively. For negative frequency, we simply have to invert the frequency variable $\omega \to -\omega$ in the equations above.

### The quantum significance of different ac and dc responses (renormalization)

For convenience, we define two functions,[6]

$$G(\omega) = -e\phi(\omega)/I_{\text{F}}(\omega), \quad G_{\text{eff}}(\omega) = -e\phi_{\text{RF}}(\omega)/I_{\text{F}}(\omega). \quad (3.16)$$

6: $G(\omega)$ and $G_{\text{eff}}(\omega)$ are simply the Green's functions of the quantum circuit. See chapter 4

Moreover, since we can express the ratio, $A(\omega)/B(\omega)$ using eq. (3.15) as, $i\omega^2[G(\omega) - G(-\omega)]/2\pi = A(\omega)/B(\omega)$, we plug in eq. (3.16) to find,

$$\frac{A(\omega)}{B(\omega)} = \frac{ei\omega^2}{2\pi}\frac{I_{\text{F}}(\omega)\phi(-\omega) - \phi(\omega)I_{\text{F}}(-\omega)}{I_{\text{F}}(\omega)I_{\text{F}}(-\omega)}. \qquad (3.17a)$$

for the dc response. The same procedure with $i\omega^2[G_{\text{eff}}(\omega) - G_{\text{eff}}(-\omega)]/2\pi = A_{\text{eff}}(\omega)/B(\omega)$ yields,

$$\frac{A_{\text{eff}}(\omega)}{B(\omega)} = \frac{ei\omega^2}{2\pi}\frac{I_{\text{F}}(-\omega)\phi_{\text{RF}}(\omega) - \phi_{\text{RF}}(-\omega)I_{\text{F}}(\omega)}{I_{\text{F}}(\omega)I_{\text{F}}(-\omega)}, \qquad (3.17b)$$

where we have unwittingly introduced the concept of renormalization of the spontaneous emission rate by the junction, i.e. $A(\omega) \to A_{\text{eff}}(\omega)$. Consequently, due to the difference in the dc and ac response of the junction, the spontaneous emission rate for the ac voltage is rescaled relative to the case for the dc response. Since this rescaling is traced to the rescaling of the Green's function of the quantum circuit, $G(\omega)$ (due to the difference in the ac response from dc response), we refer to it as Lehmann/wavefunction renormalization.[103]

The wavefunction, $\psi(t)$ of a quantum system satisfies the time dependent Schrodinger equation, $i\partial\psi/\partial t = \mathcal{E}\psi$, where $\mathcal{E}$ is the energy of the system. The Green's function of the system is given by the Fourier transform of the wavefunction, $G(\omega) = -i \int dt \psi(t) \exp(i\omega t)$.[102] Thus, rescaling the Green's function implies renormalizing the wavefunction. Typically, finite wavefunction renormalization factors appear calculating quasi-particle energies when the spectrum of excitations is not in the exact eigenstate of the system.[104] In this case, the various techniques used to diagonalize quantum mechanical operators and the Hamiltonian include Bogoliubov transformations and Feynman diagrams. [104, 105]

# 6. Discussion

### Significance of renormalization

In the case considered here, this renormalization implies that the junction will spontaneously emit photons at different rates for the dc and the ac voltage. In fact, this observation is reminiscent of the rescaling of $A(\Omega)/B(\Omega) = \Omega^3/2\pi^3c^2 \to \Omega^3(1 + a^2/\Omega^2)/2\pi^3c^2 = A_{\text{eff}}(\Omega)/B(\Omega)$ in Unruh-Hawking radiation[98, 99] predicted for a non-inertial observer under uniform acceleration $a$ relative to an inertial observer.[106, 107] Such a renormalization factor is related to the nature of the quantum vacuum in QED and the fact that quantum operators of the inertial observer versus the accelerated observer can be related by a Bogoliubov transformation.[105, 108] In fact, Unruh-Hawking radiation is analogous to the dynamical Casimir effect where an oscillating mirror relative to a static one, spontaneously radiates photons from the quantum vacuum.[7] For a system at its ground state, $A(\omega)$ is zero. However, due to this driving of the mirror, the $A(\omega)$ coefficient is rendered finite.

Note that we could have easily chosen to renormalize $B(\omega)$ instead of $A(\omega)$, or to renormalize both. The motivation to choose to renormalize $A(\omega)$ instead is in analogy with the above argument of the casimir effect. Moreover, observe that $B(\omega) \propto I_F(\omega)I_F(-\omega)/e^2$ has the right dimensions the rate of stimulated emission.[8] Defining the unknown proportionality constant for $B(\omega)$ as a dimensionless coefficient $\gamma$, we write $A(\omega) = i\omega^2\gamma\left[I_F(-\omega)\phi(\omega) - \phi(-\omega)I_F(\omega)\right]/2\pi e$.

7: Dynamical Casimir effect has been observed with a SQUID array attached to the terminal of a transmission line, which when a rapidly oscillating magnetic field is applied, it acts as a rapidly oscillating mirror changing the boundary condition of the transmission line.[109]

8: Since $\rho(-\omega)$ has dimensions of energy [1/s] and $I_F(\omega)/e$ is dimensionless in natural units, $\hbar = 1$.

Recall that in chapter 2 we found that the applied microwave $V_{\text{RF}}$ was rescaled by a factor $\Xi_A(\omega)$ related to the difference of dc and ac response of the equivalent circuit of the array. This also implies the renormalization of $A(\omega)$ as exemplified by our discussion in this chapter. Although the arguments presented in this chapter for the renormalization of the $A(\omega)$ coefficient by considering dc versus ac response may appear new, the results herein are familiar. In fact, the impedance in the $P(E)$ function for the single small junction is always effective $Z_{\text{eff}} = 1/[i\omega C + Z^{-1}(\omega)]$.[16, 100] The rescaling of $V_{\text{RF}}(\omega)$ is therefore by the factor $\Xi(\omega) = Z_{\text{eff}}(\omega)/Z(\omega)$ which in turn is related to the rate which the junction spontaneously emits a photon back to the environment.[9]

### Determining *A* and *B* for small Josephson junctions

In eq. (3.3), we used Fermi's golden rule to calculate the $B(\omega)$ coefficient. From $P(E)$ theory, we know that stimulated emission processes in small junctions must be accompanied by emission or absorption of photons. The $A(\omega)$ and $B(\omega)$ coefficients we have derived do not capture this intuitive observation. Nonetheless, we know they are valid due to fluctuation-dissipation requirements: we have expressed the coefficients using the phase $\phi(\omega)$ and the fluctuation current $I_{\text{F}}(\omega)$. In chapter 5, we shall see that $B(\omega) \propto I_{\text{F}}(\omega)I_{\text{F}}(-\omega)$ and $A(\omega) \propto I_{\text{F}}(-\omega)\phi(\omega)$ appear in the Caldeira-Leggett action as the Coulomb interaction and coupling terms respectively suggesting a deeper connection than is offered in the thesis.

Nonetheless, we can go farther to determine the rate of stimulated emission $\Gamma_{2\to 1}(\omega) = B(\Omega)\rho(\Omega)$ using Fermi's golden rule by following closely the dipole example. In fact, the transition Hamiltonian in eq. (3.3) should be replaced with the tunneling Hamiltonian of Cooper-pairs $H_i = a_i^\dagger E_J \exp(-i2\phi)/2 + a_i E_J \exp(i2\phi)/2$ with the slight modification where we include the photon creation $a_i^\dagger$ and annihilation $a_i$ operators for the infinite $\omega_i$ modes satisfying $[a_i, a_j^\dagger] = \delta_{ij}$, analogous to $E_i$ in the dipole case.[10] Thus, $\Gamma_{2\to 1}(\omega) = B(\omega)\rho(\omega)$ simply becomes the standard tunneling rate calculation for Cooper-pairs using Fermi's golden rule, e.g. the calculation, for the tunneling rate, given e.g. in eq. (135) in ref. [16] where $\mathcal{E}_2 - \mathcal{E}_1 = 2eV + E_R' - E_R$ and $E_R', E_R$ are energies related to the environment. The

---

9: Such a factor has been reported describing the effect of excited environmental modes by an alternating voltage in single normal junctions[27] in an extended $P(E)$ theory approach. Such a factor that may lead to amplification effects has also been reported in ref. [28, 29] and ref. [53] Moreover, the Caldeira-Leggett form of the environmental impedance neglects the back-action of the Josephson junctions on the environment (with the bath and the junction becoming entangled) which has been reported to dramatically change the predictions of the $P(E)$ theory.[31, 34, 35] This back-action manifests through the non-linear inductive response of the junction where the Josephson coupling energy is renormalized and the insulator-superconductor phase transition conditions for the single Josephson junction are altered.[35] Novel features not yet observed experimentally with arrays for Josephson junction arrays include the renormalization of the radio frequency (RF) power absorbed by the array junction as well as higher harmonic modifications of the time-averaged current[32, 33]. We thus view the observation of the renormalization effect detailed in chapter 2 as a crucial first step extending $P(E)$ theory to Josephson junction arrays.

**Table 3.2.:** Transforming the sum of infinite modes $\omega_i$ in the Josephson junction Cooper-pair tunneling rate.

$$\sum_i \to \int d\omega\, \delta(\Omega - \omega),$$
$$a_i \to eV(\omega),$$
$$a_i^\dagger \to eV(-\omega).$$

10: $a_i$ and $a_i^\dagger$ account for tunneling of a Cooper-pair accompanied by stimulated photon emission or absorption processes.

calculation yields, $\Gamma_{2\to1} = \pi \sum_i \langle a_i^\dagger a_i \rangle E_J^2 P(2eV - \omega_i)/2$. Here, $P(2eV - \omega_i)$ is the $P(E_i)$ function given in eq. (3.5) with $E_i = 2eV - \omega_i$. Finally, changing the sum to an integral by the transformation in Table 3.2, we arrive at,

$$\Gamma_{2\to1}(\Omega) = \pi E_J^2 P(2eV - \Omega)\rho(\Omega)/2\Omega, \quad (3.18)$$

where we have used $\rho(\Omega) = e^2\Omega\langle V(\Omega)V(-\Omega)\rangle$. Recalling that $\Gamma_{2\to1} = B(\Omega)\rho(\Omega)$, we discover that $B(\Omega) = \pi E_J^2 P(2eV - \Omega)/2\Omega = \Gamma(2eV - \Omega)/\Omega$, which is proportional to the Cooper-pair tunneling rate $\Gamma$ in eq. (1.14), albeit with the voltage shifted by $\Omega/2e$. Moreover, using $A(\Omega)/B(\Omega) = (2e)^2\Omega[Z(\Omega) + Z(-\Omega)]/2\pi$, we find that,

$$A(f) = 2r(f)\Gamma(2eV - 2\pi f) \quad (3.19)$$

with the definition $r(f) \equiv 4e^2[Z(f) + Z(-f)]/4\pi$ and $f = \Omega/2\pi$. Thus, the rate of spontaneous emission, $A(f)$ is related to the spectral density of photon emission $\gamma(f)$ in the dynamical Coulomb blockade regime provided in ref. [110] by $A(f) = \gamma(f)/f$. This is an interesting result since it implies that, whenever the impedance is renormalized ($Z(\Omega) \to Z_{\text{eff}}(\Omega)$), only the $A(\Omega)$ coefficient (and not the $B(\Omega)$ coefficient) is trivially renormalized, where the non-trivial renormalization entails the rescaling of $\rho(\Omega)$ and/or the $P(E)$ function. The succeeding chapters toil towards formulating in detail the renormalization effect using path integral formalism.[102]

# Electromagnetic Environment in Josephson junctions | 4.

A Josephson junction is a Superconductor(S)-Insulator(I)-Superconductor(S) tunnel junction (Fig. 4.1) that admits a supercurrent across it by the coherent tunneling of Cooper pairs[58, 59] even in the absence of applied voltages. Due to its non-linear response to applied electromagnetic fields, Josephson junctions have found varied metrology[4] and detector applications.[5] Moreover, the junction size determines the current-voltage characteristics observed in experiments. In particular, mesoscopic (small) junctions are known to be greatly susceptible to quantum fluctuations and changes in the electromagnetic environment compared to large junctions. This leads to complex theoretical considerations albeit richer physics such as circuit-Quantum Electro-Dynamics in the small junction whose features are often captured by the so-called $P(E)$ theory.[16]

Despite the existence of excellent reviews on the subject and techniques[15, 16, 100], the authors found much of the techniques and prior concepts useful in following the arguments in this thesis scattered in various literature[10, 11, 89, 101, 111, 112]. In particular, the techniques used in the subsequent chapters include path integral formalism[102] and Green's functions[11, 89] to calculate phase-phase correlation functions and four-vector notation[113] where Maxwell's equations appear for compactness. Thus, we include this section as a preamble for completeness and/or compactness. Hopefully it offers a more nuanced understanding of the Caldeira-Leggett model and $P(E)$ theory in the context of Green's functions and generally a path integral framework.

## Quantum Phase Dynamics in Large Josephson Junctions (The Josephson Effect)

The physics of Josephson junctions (schematic shown in Fig. 4.1) is described by the well known Josephson equations[59],

$$I_S = 2eE_J \sin 2\phi_x(t) \tag{4.1a}$$

$$\frac{\partial \phi_x(t)}{\partial t} = eV_x \tag{4.1b}$$

Here, $2\phi_x(t)$ denotes the phase difference across the junction [1] and $E_J$ is the Josephson coupling energy[60]. The simplest derivation of eq. (4.1) follows from the real and imaginary parts of these two coupled Schrodinger equations

[1]: where the subscript $x$ distinguishes it from quantum phases of other circuit elements defined later in the chapter.

$$i\frac{\partial \psi_1}{\partial t} = \mu_1 \psi_1 + m_0 \psi_2, \tag{4.2a}$$

$$i\frac{\partial \psi_2}{\partial t} = \mu_2 \psi_2 + m_0 \psi_1. \tag{4.2b}$$

Here, $\mu_1$ and $\mu_2$ and the chemical potentials of the left (1) and right (2) junction respectively, $\psi_1$ and $\psi_2$ are the Cooper-pair wavefunctions of the left and right superconductors respectively and $m_0$ is a coupling energy term characterizing magnitude of overlap for the two wavefunctions across the insulator. When a potential difference (voltage) $V_x$ is applied across the junction, the two chemical potentials shift relative to each other in order to accomodate this change. This means that we can set $\mu_1 - \mu_2 = 2eV_x$, where $2e$ is the Cooper pair charge. Based on this, it is instructive to define an average chemical potential, $\mu \equiv (\mu_1 + \mu_2)/2$, and solve for $\mu_1$ and $\mu_2$ in terms of $\mu$. This yields, $\mu_1 = \mu + eV_x$ and $\mu_2 = \mu - eV_x$. Plugging this back to eq. (4.2) yields,

$$i\frac{\partial \psi_1}{\partial t} = (\mu + eV_x)\psi_1 + m_0 \psi_2, \tag{4.3a}$$

$$i\frac{\partial \psi_2}{\partial t} = (\mu - eV_x)\psi_2 + m_0 \psi_1. \tag{4.3b}$$

From this, it is clear $\mu$ is simply the common chemical potential relative to which the voltage drop is measured. This observation implies we can set it to zero without loss of generality, $\mu = 0$.

The Cooper pair wavefunctions are defined as $\psi_1 = \sqrt{n_1} \exp(i2\varphi_1)$ and $\psi_2 = \sqrt{n_2} \exp(i2\varphi_2)$ where $n_1$ and $n_2$ are the number of Cooper-pairs in the left (1) and right (2) superconductor respectively, and $2\varphi_1$ and $2\varphi_2$ is their respective macroscopic quantum phases. Plugging these definitions into eq. (4.3), we find,

$$\frac{\partial n_1}{\partial t} = 2m_0\sqrt{n_1 n_2}\sin(2[\varphi_2 - \varphi_1]) \quad (4.4a)$$

$$\frac{\partial n_2}{\partial t} = 2m_0\sqrt{n_2 n_1}\sin(2[\varphi_1 - \varphi_2]) \quad (4.4b)$$

and,

$$2\frac{\partial \varphi_1}{\partial t} = -m_0\sqrt{\frac{n_2}{n_1}}\cos(2[\varphi_2 - \varphi_1]) - eV_x, \quad (4.5a)$$

$$2\frac{\partial \varphi_2}{\partial t} = -m_0\sqrt{\frac{n_2}{n_1}}\cos(2[\varphi_2 - \varphi_1]) + eV_x. \quad (4.5b)$$

Taking the approximation that any tunneling currents that arise have the effect of varying $n_1$ and $n_2$ by only a small amount $(\delta n)^2 \simeq 0$ from an equilibrium value given by $\frac{1}{2}(n_1 + n_2) \equiv n$, i.e. $n_1 = n + \delta n$ and $n_2 = n - \delta n$, we see that the supercurrent across the barrier is given by $I_S \equiv e\partial \delta n/\partial t$ and the voltage drop by $\partial(\varphi_2 - \varphi_1)/\partial t = eV_x$, which yield eq. (4.1) when $E_J = nm_0$ and $\varphi_2 - \varphi_1 = \phi_x$.

There is another advantage of setting $\mu = 0$. In particular, eq. (4.3) becomes a spinor equation,

$$i\frac{\partial \psi}{\partial t} = H_{cp}\psi \quad (4.6a)$$

$$H_{cp} = eV_x\sigma_3 + m_0\sigma_1, \quad (4.6b)$$

$$\psi \equiv \begin{pmatrix} \psi_1 \\ \psi_2 \end{pmatrix}, \quad (4.6c)$$

where $\sigma_1$ and $\sigma_3$ are the Pauli matrices,

$$\sigma_1 = \begin{pmatrix} 0 & 1 \\ 1 & 0 \end{pmatrix}, \quad \sigma_3 = \begin{pmatrix} 1 & 0 \\ 0 & -1 \end{pmatrix}. \quad (4.7a)$$

Since $E_J \propto m_0$, $\sigma_1$ plays the role of the coupling term that results to Cooper-pair tunneling. Using the last Pauli matrix,

$$\sigma_2 = \begin{pmatrix} 0 & -i \\ i & 0 \end{pmatrix}, \quad (4.7b)$$

we can define two operators $\sigma_+ \equiv (\sigma_1 + i\sigma_2)/2$ and $\sigma_- \equiv (\sigma_1 - i\sigma_2)/2$. These operators form tunneling matrix elements with the spinor and a transpose conjugate spinor defined

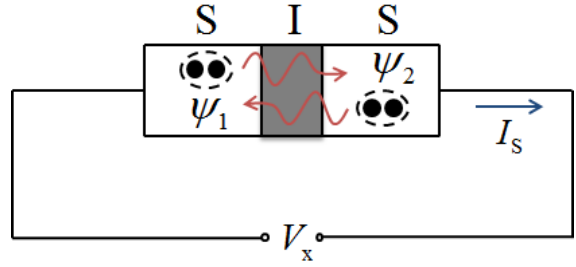

**Figure 4.1.:** A schematic of a Josephson junction (S: superconductor, I: insulator, S: superconductor) depicting a Cooper pair from the left/right electrode tunneling through the insulator to the right/left electrode.

as,

$$\psi^\dagger \equiv \psi^{*T} = (\psi_1^*, \psi_2^*). \tag{4.8}$$

For instance, tunneling from left to right requires replacing $\psi_1$ with $\psi_2$ and annihilating $\psi_1$, which corresponds to $\sigma_+ \psi$. The inverse process process corresponds to $\sigma_- \psi$. These matrices will be useful when calculating Cooper-pair tunneling rates for small junctions (See eq. 4.34).

# The Electromagnetic Environment and Fluctuation-Dissipation in Single Large Josephson Junctions

Equation 4.1 only considers the superconducting current and thus neglects the environment that lead to effects such as Coulomb blockade. The environment consists of all sources of the electromagnetic field (including the field itself) which couple to the Cooper-pair wavefunction via the phase difference thus determining the $I - V$ characteristics satisfying eq. (4.1). Specifically, the environment arises from processes such as the alternating currents and voltages, thermal fluctuations in the form of Johnson-Nyquist noise and coupled high impedance circuit environments[7, 73].

Using eq. (4.1), one can define a conserved energy by treating the junction as a capacitance

$$E = Q_x^2/2C - E_J \cos(2\phi_x) \tag{4.9a}$$
$$Q_x = -CV_x \tag{4.9b}$$
$$\frac{\partial Q_x}{\partial t} = I_S \tag{4.9c}$$

where $C$ is the capacitance of the junction. Modifying the last equation in (4.9) to

$$\frac{\partial Q_x}{\partial t} = \sum_a I_a, \qquad (4.10)$$

one can then include all the environmental sources of energy in the form of currents. In fact, to arrive at eq. (4.10), the phase-difference needs to couple to the electromagnetic field (Maxwell equations) in a straight-forward manner

$$\frac{\partial \phi_x}{\partial x^\mu} = e d_{\text{eff}} N^\nu F_{\mu\nu} \qquad (4.11a)$$

$$N^\mu N_\mu = -\sum_{ij} N_i N_j \delta_{ij} = -1 \qquad (4.11b)$$

$$\frac{\partial F^{\mu\nu}}{\partial x^\mu} = -\frac{1}{\varepsilon_0 \varepsilon_r} \sum_a J_a^\nu \qquad (4.11c)$$

Here, $F_{\mu\nu} = \partial A_\mu/\partial x_\nu - \partial A_\nu/\partial x_\mu = -F_{\nu\mu}$ is the electromagnetic tensor, $d_{\text{eff}}$ is the thickness of the barrier and $N^\mu = (0, N^i)$ points in the direction $N^i$ normal to the tunnel barrier. We have used Einstein notation where only the Greek indices are summed over and the Minkowski space-time signature is diag($\eta_{\mu\nu}$) = (+, -, -, -). Taking the total derivative $\eta^{\mu\nu}\partial/\partial x^\nu = \partial/\partial x_\nu$ of eq. (4.11a) (with $\eta_{\mu\alpha}\eta^{\nu\alpha} = \delta_\mu^\nu$) and using $\partial N^\nu/\partial x^\mu = 0$, we arrive at

$$\eta^{\mu\nu}\frac{\partial^2 \phi_x}{\partial x^\mu \partial x^\nu} = \frac{\partial^2 \phi_x}{\partial x^\mu \partial x_\mu} = -\frac{e d_{\text{eff}}}{\varepsilon_0 \varepsilon_r} \sum_a N_\mu J_a^\mu = -wJ \qquad (4.12)$$

which is eq. (4.10) in disguise. Note that $F_{0i} = \vec{E}$ and $\frac{1}{2}\sum_{i,j} \varepsilon_{ijk} F^{ij} = \vec{B}$ where $\vec{E}$ and $\vec{B}$ are the $x, y, z$ components of the electric and magnetic fields respectively and $\varepsilon_{ijk}$ is the Levi-Civita symbol. eq. (4.12) is the sourced Klein-Gordon equation with $J = \sum_a N_\mu J_a^\mu$ the source and $w = e d_{\text{eff}}/\varepsilon_0 \varepsilon_r$ the coupling constant. The vector $N_\mu$ and the anti-symmetry of the electromagnetic tensor $F_{\mu\nu}$ guarantees that, unlike Maxwell equations, the coupled Klein-Gordon equation lives in 2+1 dimensions instead of 3+1. For instance, when the tunnel barrier is aligned to the y-z direction, $N^\mu = (0, 1, 0, 0)$ and eq. (4.11a) and by extension eq. (4.12) become independent of $x$.

$$\frac{\partial^2 \phi_x}{\partial x^\mu \partial x_\mu} = \frac{\partial^2 \phi_x}{\partial t^2} - \frac{\partial^2 \phi_x}{\partial y^2} - \frac{\partial^2 \phi_x}{\partial z^2} = -wJ \qquad (4.13)$$

Furthermore, taking the limit for small junctions which corresponds to taking the area of the barrier $\mathcal{A}$ to be small such that the phase neither varies with $y$ nor $z$, we arrive at eq. (4.10)

$$\frac{\partial^2 \phi_x}{\partial t^2} = -4E_c E_J \sin(2\phi_x) - \frac{1}{RC}\frac{\partial \phi_x}{\partial t} - 2E_c I_F/e \quad (4.14a)$$

$$\sum_a^3 J_a^1 = J_S^1 + J_N^1 + J_F^1 \quad (4.14b)$$

$$\langle I_F(t) I_F(t') \rangle = \frac{4\beta^{-1}}{R}\delta(t-t') \quad (4.14c)$$

with $J_S^1 = J_{cp}\sin 2\phi_x$ the supercurrent, $J_N^1 = \sigma_{xx} F_{01}$ the normal current and $\sigma_{xx}$ the effective conductivity of the barrier along the $x$ direction. Here, we have used the cross-sectional area of the junction, $\mathcal{A}$ to define $J_a^\mu \mathcal{A} = I_a^\mu$, the junction capacitance $C = \varepsilon_0 \varepsilon_r \mathcal{A}/d_{\text{eff}}$, the charging energy $E_c = e^2/2C$ and the junction conductance $1/R = \sigma_{xx}\mathcal{A}/d_{\text{eff}}$. Finally, $\beta^{-1} = k_B T$ is the inverse temperature and we have assumed the fluctuation current $J_F^1 \mathcal{A} = I_F$ is Gaussian-correlated over a bath (B stands for bath or Boltzmann distribution) with the thermal correlation function given by eq. (4.14c).

It is straight forward to generalize the conductance $1/R$ in eq. (4.14) using a spectral function $K(\omega) = [Z^{-1}(\omega) + Z^{-1}(-\omega)]/(2\pi)$ describing the macroscopic physics of the microscopic degrees of freedom of the system undergoing Brownian motion due to a heat bath comprising $k$ harmonic oscillators,[101]

$$H_B = \sum_{n=1}^{k} \left\{ \frac{Q_n^2}{2C_n} + \frac{(\phi_n - \phi_x)^2}{2e^2 L_n} \right\} \quad (4.15a)$$

$$K(t) = \sum_{n=1}^{k} L_n^{-1} \cos(\omega_n t) = \int_{-\infty}^{+\infty} d\omega K(\omega) \exp -i\omega t \quad (4.15b)$$

where $H_B$ is the Hamiltonian of the heat bath consisting of $L_n C_n$ circuits in parallel where $\omega_n = 1/L_n C_n$, $Q_n$, $\phi_n$ are the charges stored by and the phases of the elements and $K(t)$ is referred to as the Kernel representing the dissipative nature of the circuit.[2] The generalized Lagrangian for the system[3]

---

2: See eq. (3.4) of chapter 3

3: Also the Lagrangian for eq. (3.9) given in chapter 3 with $\phi \to \phi_x$ and $\vec{\nabla}_\perp^2 \phi_x = 0$

is given by

$$\mathcal{L} = \frac{C}{2e^2}\left(\frac{\partial \phi_x(t)}{\partial t}\right)^2 - \frac{1}{2e^2}\int_{-\infty}^{+\infty} \phi_x(t)\frac{\partial K(s-t)}{\partial s}\phi_x(s)ds$$
$$-\frac{1}{e}\int_{-\infty}^{+\infty} I_F(t-s)\phi_x(s)ds + E_J\cos(2\phi_x), \quad (4.16)$$

where the fluctuation current is given by,

$$I_F(t) = \sum_{n=1}^{k}\left\{\omega_n Q_n \sin(\omega_n t) + e^{-1}L_n^{-1}(\phi_n - \phi_x)\cos(\omega_n t)\right\}$$

$$\langle I_F(t)I_F(t')\rangle = \sum_{n=1}^{k} 2L_n^{-1}\langle H_B(\omega_n)\rangle \cos\omega_n(t-t').$$

The average is over the thermal bath degrees of freedom. For the Ohmic conductance above, we have $Z^{-1}(\omega) = 1/R$ and $\langle H_B(\omega)\rangle = \beta^{-1}$ where the continuous, large $k$ limit

$$\lim_{k\to\infty}\sum_{n=1}^{k}L_n^{-1}\times \to \int_{-\infty}^{\infty} d\omega K(\omega)\times \quad (4.17a)$$

is taken in accordance with eq. (4.15b) thus recovering eq. (4.14c). The fluctuation current density certainly satisfies the Green-Kubo relation[111, 112],

$$\sigma_{xx} = \frac{\beta}{4}\int d^4x N^\nu N^\mu \langle J_{F\nu}(t)J_{F\mu}(0)\rangle$$
$$= \frac{\beta}{4}\mathcal{A}^{-1}d_{\text{eff}}\int dt\, \langle I_F(t)I_F(t')\rangle = \frac{d_{\text{eff}}}{R\mathcal{A}}, \quad (4.17b)$$

where we have used eq. (4.14c) in the last line. Note that to obtain the correct equation of motion, integration by parts of the second term in eq. (4.16) should be performed after applying the Euler-Lagrange equations, then the boundary term is dropped

$$\frac{1}{e^2}\int_{-\infty}^{+\infty} \frac{\partial}{\partial s}\left[K(s-t)\phi_x(s)\right]ds = 0 \quad (4.18)$$

# Phase Correlation Functions, *P(E)* function and Coulomb Blockade of Cooper-pairs and Quasi-Particles in Single Small Josephson Junctions

## The Hamiltonian

Consider a mesoscopic tunnel junction with capacitance $C$ driven by a voltage source $V_x$ via an environmental impedance $Z(\omega)$. Each circuit element is characterized by a phase $\phi_a$ related to the voltage drop $V_a$ of the element in the circuit by $\phi_a(t) = \int_{-\infty}^{t} eV_a(\tau)d\tau$, where the subscript a = J, x or z corresponds to the junction, voltage source and environment impedance and $\kappa e = 2e, e$ corresponds to Cooper pair, quasi-particle charge respectively. [4] The voltages $V_z$ and $V_J$ decrease as one moves clockwise along the circuit, whereas the value increases for the voltage source $V_x$ in the same direction. The corresponding charge on the junction is defined as $Q_J = CV_J$ where $C$ is the capacitance of the junction. The circuit can store a topological flux $e\Phi = \phi_J + \phi_z + \phi_x = \sum_a \phi_a$ related to a topological potential $\int_{-\infty}^{t} A(\tau)d\tau = \Phi(t)$, which we will find out, in chapter 5, that it leads to RF power renormalization when present.

4: That the effect of the environmental impedance $Z(\omega)$ can be represented by a single quantum phase $\phi_z$ defined by the voltage drop over $Z(\omega)$ is not at all obvious. At this stage, we treat it as an ansatz. It will not appear in the equations until we impose the topological constraint $\sum_a \phi_a = e\Phi$ on the circuit.

The total Hamiltonian, $\mathcal{H}$ of the circuit (Fig. 4.2) is given by the expression,

$$\mathcal{H} = \sum_{\kappa=1}^{2} H_\kappa + H_J + H_z \tag{4.19}$$

Here, $\sum_{\kappa=1}^{2} H_\kappa = H_1 + H_2$ where the Cooper-pair Hamiltonian $H_2 = \mu \psi^+ \psi = \psi^+ i \left[\partial/\partial t + iH_{cp}\right] \psi$ depends on the chemical potential $\mu$ and the 2-spinor $\psi$ and the quasi-particle Hamiltonian $H_1$ is given by,

$$H_1 = H_L + H_R = \sum_{p,\sigma} \epsilon_{p\sigma} \gamma^\dagger_{p\sigma} \gamma_{p\sigma} + \sum_{q,\sigma} \epsilon_{q\sigma} \gamma^\dagger_{q\sigma} \gamma_{q\sigma} \tag{4.20}$$

where $\gamma_{p\sigma}$ or $\gamma_{q\sigma}$ and $\gamma^\dagger_{p\sigma}$ or $\gamma^\dagger_{q\sigma}$ are the annihilation and creation operators respectively of a quasi-particle state with energy $\epsilon_{p\sigma}$ or $\epsilon_{q\sigma}$, momentum $p$ or $q$ and spin $\sigma$ in the left or right electrode,

$$H_J = \sum_{\kappa=1}^{2} \Theta_\kappa \exp(-i\kappa\phi_J) + h.c. \quad (4.21)$$

is the tunneling Hamiltonian where

$$\Theta_1 = \sum_{p \neq q, \sigma} M_{pq} \gamma_{q\sigma} \gamma_{p\sigma}^\dagger \quad (4.22a)$$

$$\Theta_2 = \frac{E_J}{2}(\sigma_1 - i\sigma_2) \equiv \frac{E_J}{2}\sigma_- \quad (4.22b)$$

$\sigma_{1,2}$ are the $x, y$ Pauli matrices acting on the 2-spinor given in eq. (4.7), $M_{pq}$ is a dimensionful spin-conserving complex-valued quasi-particle tunneling matrix, $p \neq q$ enforces the condition $[H_L, H_R] = 0$ and $E_J$ is the Josephson coupling energy[60],

$$H_z = \frac{(Q_J + CV_x + CA)^2}{2C} + \sum_{n=1} \left\{ \frac{Q_n^2}{2C_n} + e^{-2}\frac{(\phi_n - \phi_J + \phi_x + e\Phi)^2}{2L_n} \right\} \quad (4.23)$$

is the Hamiltonian[5] describing the environmental impedance $Z(\omega)$ and junction capacitance $C$, where $Z(\omega)$ is characterized by an infinite number of parallel $L_n C_n$ circuits coupled serially to the tunnel junction. One can define $Q = Q_J - CV_x$ and $\phi = \phi_J - \phi_x$ as the fluctuation variables of the junction charge $Q_J = CV_J$ and junction phase $\phi_J(t) = e \int_{-\infty}^{t} dt' V_J(t')$ around the mean value determined by the voltage source $V_x$, where $V_J(t')$ is the voltage drop across the junction. (See ref. [16] on page 27. Note that $\phi_J$ and $\phi$ are related by a suitable unitary transformation $\mathcal{U}$ of the Hamiltonian,

$$\mathcal{H}' = i\mathcal{U}^\dagger \frac{\partial}{\partial t}\mathcal{U} + \mathcal{U}\mathcal{H}\mathcal{U}^\dagger, \quad (4.24)$$

where $\mathcal{H}' = H_1' + H_2 + H_J + H_z$, $H_J = \sum_{\kappa=1}^{2} \Theta_\kappa \exp(-i\kappa\phi_J) + h.c.$, $H_1' = \sum_{p \neq q, \sigma} \epsilon_{p\sigma}' \gamma_{p\sigma}^\dagger \gamma_{p\sigma} + \sum_{p \neq q, \sigma} \epsilon_{q\sigma} \gamma_{q\sigma}^\dagger \gamma_{q\sigma}$ and $\epsilon_{p\sigma}' = \epsilon_{p\sigma} + eV_x$. $Q = Q_J - CV_x$, $Q_n$ are the conjugate variables to $\phi = \phi_J - \phi_x$, $\phi_n$ satisfying the charge-phase commutation relation,

$$[\phi_n, Q_m] = i\delta_{mn}e, \quad (4.25)$$
$$[\phi, Q] = ie \quad (4.26)$$

[5]: This Hamiltonian basically corresponds to eq. (21) of ref. [16] with the topological terms, $\Phi = 0$ and $A = 0$.

where $\delta_{ab}$ is the Kroneker delta. Operators, $O(t)$ in the Heisenberg picture are related to the ones in the Schrödinger picture, $O(0)$ by $O(t) = U_0(t)^\dagger O(0) U_0(t)$ with the unitary evolution operator $U_0(t)$ given by $U_0(t) = \exp\left\{-i \sum_{\kappa=1}^{2} H_\kappa t\right\}$ in the absence of tunneling.

6: As before in eq. 4.3.

In what follows, we assume the Cooper-pair ground state energy $\mu = 0$.[6] The tunneling current $I(V)$ at the junction is given by

$$I(V, t') = \text{tr} \left\langle \mathcal{T} \left\{ U^\dagger I_J(0) U \right\} \right\rangle \quad (4.27)$$

Here, $U = U_0 + U_{\text{int}}$ where,

$$U_{\text{int}} = \exp\left(-i \int_{-t'}^{+t'} H_J(t) dt\right)$$

$$= \exp\left(-i \int_{-t'}^{0} H_J(t) dt\right) \exp\left(-i \int_{0}^{t'} H_J(t) dt\right)$$

$$= \exp\left(+i \int_{0}^{-t'} H_J(t) dt\right) \exp\left(-i \int_{0}^{t'} H_J(t) dt\right)$$

$$= U_{\text{int}}^\dagger(-t') U_{\text{int}}(+t')$$

and $\mathcal{T}$ is the time ordering operator with the property given by $\mathcal{T} \Theta_\kappa(t) \Theta_\kappa^\dagger(0) = \Theta_\kappa(t) \Theta_\kappa^\dagger(0)$, $\mathcal{T} \Theta_\kappa(0) \Theta_\kappa^\dagger(t) = \Theta_\kappa^\dagger(t) \Theta_\kappa(0)$. Here, $t'$ is the elapsed time after switching on the interaction term $U_{\text{int}}(t')$ and takes the range $0 \leq t' \leq +\infty$. [Note that $U_{\text{int}}^\dagger(t')$ takes care of $c.c.$ term in eq. (4.29), updating the integral range as discussed: $\int_0^{t'} dt \to \int_{-t'}^{0} dt + \int_0^{t'} dt = \int_{-t'}^{+t'} dt$.] We shall be interested in the current $I(V, t' \to +\infty) = I(V)$ at equilibrium [eq. (4.45)].

The tunneling current operator is

$$I_J(0) = -i[Q_J(0), H_J(0)], \quad (4.28)$$

and the average $\langle \ldots \rangle$ is over, the quasi-particle equilibrium states, whose density matrix is given by $\rho_1 = \rho_L \rho_R = \mathcal{Z}_1^{-1} \exp(-\beta H_1)$ with $\mathcal{Z}_1 = \mathcal{Z}_L \times \mathcal{Z}_R = \prod_p [1 + \exp(-\beta \epsilon_{p\sigma})] \times \prod_q [1 + \exp(-\beta \epsilon_{q\sigma})]$, and the environment $\rho_{\text{env}} = \mathcal{Z}_{\text{env}}^{-1} \exp(-\beta H_z)$ where $\beta = 1/k_B T$ is the inverse temperature while the trace (tr) is over the Pauli matrices.

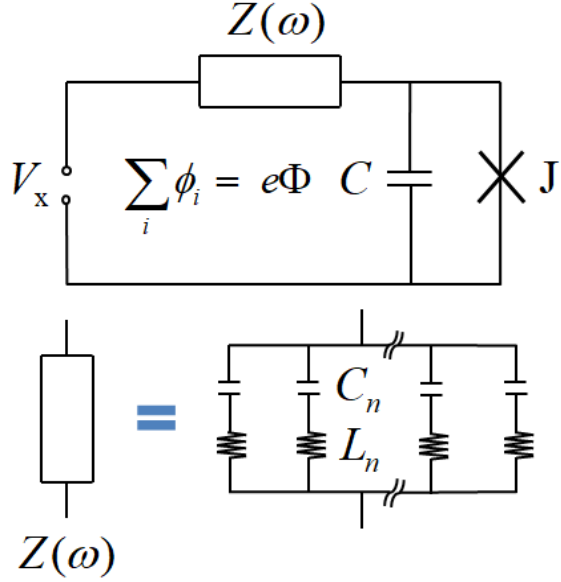

**Figure 4.2.:** A mesoscopic tunnel junction, J with capacitance $C$ driven by a voltage source $V_x$ via an environmental impedance $Z(\omega)$ composed of infinite number of parallel $L_n$ $C_n$ circuits. The circuit stores a flux $e\Phi = \sum_i \phi_i = \phi_J + \phi_x + \phi_z$ related to a topological potential $A(t)$ by $\int_{-\infty}^{t} ds\, A(s) = \Phi(t)$. (This figure corresponds to Fig. 2 of ref. [16] without the topological flux.)

**Perturbation Expansion**

We can then expand eq. (4.27) as a perturbation series in the tunneling Hamiltonian $H_J(t)$ by applying the perturbation expansion formula in Appendix C

$$I = \left\langle I_J(0) - i \int_{-\infty}^{+\infty} [I_J(0), H_J(t)] dt + O(H_J^2) \right\rangle \quad (4.29)$$

Using $\mathcal{T} \Theta_\kappa(t) \Theta_\kappa^\dagger(0) = \Theta_\kappa(t) \Theta_\kappa^\dagger(0), \mathcal{T} \Theta_\kappa(0) \Theta_\kappa^\dagger(t) = \Theta_\kappa^\dagger(t) \Theta_\kappa(0)$,
$\langle \Theta_\kappa(t) \rangle = 0$ and $\langle \Theta_{\kappa'}(t) \Theta_\kappa^\dagger(0) \rangle = \langle \Theta_{\kappa'}^\dagger(t) \Theta_\kappa(0) \rangle = \alpha_\kappa(t) \delta_{\kappa'\kappa}$,
we find that

$$I \simeq \left\langle \int_{-\infty}^{+\infty} [H_J(t), [Q_J(0), H_J(0)]] dt \right\rangle \quad (4.30)$$

$$= ie \sum_{\kappa=1}^{2} \int_{-\infty}^{+\infty} \left( \alpha_\kappa(t) \left\langle \sin[\kappa \Delta \phi_J(t)] \right\rangle_{\phi_J} \right) dt \quad (4.31)$$

where $\Delta \phi_J(t) = \phi_J(t) - \phi_J(0)$. Thus, the particle degrees of freedom $\alpha_\kappa(t)$ and the environment are decoupled and the trace over the environment $\langle \ldots \rho_{env} \rangle$ has been re-written as $\langle \ldots \rangle_{\phi_J}$ in terms of the junction phase $\phi_J$ degree of freedom.

For the quasi-particle current, the kernel $\alpha_1(t)$ scales with the dimensionless tunneling conductance $e^{-2} R_T^{-1}$ but its functional form depends on the gap, reflecting the corresponding structures in the quasi-particle $I - V$ characteristics. It can be computed as,

$$\alpha_1(t) = \left\langle \Theta_1(t)\Theta_1^\dagger(0) \right\rangle = \sum_{p\neq q,\sigma} \sum_{p'\neq q',\sigma} M_{pq} M^*_{q'p'} \times$$

$$\langle R,s|\langle L,s|\gamma_{q\sigma}(t)\gamma^\dagger_{p\sigma}(t)\gamma_{p'\sigma}(0)\gamma^\dagger_{q'\sigma}(0)\rho_1|L,s\rangle|R,s\rangle$$

$$= 2\sum_{p\neq q}\sum_{p'\neq q'} M_{pq}M^*_{q'p'}\langle R|\gamma_q(t)\langle L|\gamma^\dagger_p(t)\gamma_{p'}(0)\rho_L|L\rangle \gamma^\dagger_{q'}(0)\rho_R|R\rangle$$

$$= 2\sum_{p\neq q} f(\epsilon_p)\exp\{-i\epsilon_p t\}\sum_{p'\neq q'} M_{pq}M^*_{q'p'}\delta_{pp'}\langle R|\gamma_q(t)\gamma^\dagger_{q'}(0)\rho_R|R\rangle$$

$$= 2\sum_{p\neq q} f(\epsilon_p)[1-f(\epsilon_q)]\exp\{i(\epsilon_q-\epsilon_p)t\}\sum_{p'\neq q'} M_{pq}M^*_{q'p'}\delta_{qq'}\delta_{pp'}$$

$$= 2\sum_{p\neq q} f(\epsilon_p)[1-f(\epsilon_q)]M_{pq}M^*_{qp}\exp\{i(\epsilon_q-\epsilon_p)t\}$$

$$\rightarrow \frac{1}{\pi e^2 R_T}\int_{-\infty}^{+\infty}\int_{-\infty}^{+\infty} d\epsilon_p d\epsilon_q \frac{\mathcal{N}_L(\Delta)}{\mathcal{N}_L(0)}\frac{\mathcal{N}_R(\Delta)}{\mathcal{N}_R(0)}\times$$
$$f(\epsilon_p)(1-f(\epsilon_q))\exp\{i(\epsilon_q-\epsilon_p)t\}, \quad (4.32)$$

by taking the continuous limit $2\pi e^2 R_T \mathcal{N}_L(0)\mathcal{N}_R(0)M_{pq}M^*_{qp}\rightarrow 1$. Here, $e^2 R_T$ is the dimensionless tunnel resistance, $f(E)=[1+\exp(\beta E)]^{-1}$ is the Fermi-Dirac function and $\mathcal{N}_L(\Delta)$, $\mathcal{N}_R(\Delta)$ is the left, right BCS density of states which reduce to the electron density of states $\mathcal{N}_L(0)$, $\mathcal{N}_R(0)$ when the superconducting gap $\Delta = 0$ vanishes,

$$\frac{dE_p}{d\epsilon_p}\frac{dE_q}{d\epsilon_q} = \frac{\mathcal{N}_L(\Delta)}{\mathcal{N}_L(0)}\frac{\mathcal{N}_R(\Delta)}{\mathcal{N}_R(0)}, \quad (4.33a)$$

$$E_p = \sqrt{\epsilon_p^2 - \Delta^2}, E_q = \sqrt{\epsilon_q^2 - \Delta^2} \quad (4.33b)$$

where $E_p = p^2/2m$, $E_q = q^2/2m$ is the kinetic energy of the electrons above the Fermi sea.

Likewise, calculating $\alpha_2(t)$, we find,

$$\alpha_2 = \left\langle \Theta_2^\dagger(t)\Theta_2(0)\right\rangle = \left\langle \Theta_2^\dagger(0)\Theta_2(0)\right\rangle$$
$$= \left(\frac{E_J}{2}\right)^2 \text{tr}\{(\sigma_1+i\sigma_2)(\sigma_1-i\sigma_2)\}$$
$$= \frac{E_J^2}{4}\text{tr}\{2\sigma_0 + i[\sigma_2,\sigma_1]\} = \frac{E_J^2}{2}\text{tr}\{\sigma_0+\sigma_3\} = E_J^2, \quad (4.34)$$

where $\sigma_0$ is the $2\times 2$ identity matrix. We discover that, unlike $\alpha_1(t)$, the function $\alpha_2(t) = \alpha_2(0) = E_J^2$ is time independent and only depends on the strength of Cooper pair tunneling, $E_J$.

## Path Integrals and Phase Correlations

To calculate the remaining average over $\phi_J$ in eq. (4.31), we work in Minkwoski time at zero temperature (thus by-passing a rigorous but otherwise tedious Wick rotation to Euclidean time) since the finite temperature propagator is trivially related to the zero temperature result [eq. (4.43b) for the trivial relation and Appendix B for the formalism].

In this formalism: Given an observable $O(\phi_J)$, its average at zero temperature is given by the functional/path integral

$$\lim_{\beta \to \infty} \langle O(\phi_J) \rangle_{\phi_J} = \mathcal{Z}^{-1} \prod_{n=1}^{k} \int D\phi_n D\phi_J O(\phi_J) \exp iS_z(\phi_n, \phi_J) \tag{4.35}$$

where $\mathcal{Z} = \prod_{n=1}^{k} \int D\phi_n D\phi_J \exp iS_z(\phi_n, \phi_J)$ is the partition function normalizing eq. (4.35) and the Lagrangian in the action for the environment $S_z(\phi_n, \phi_J)$ is given by the (inverse) Legendre transform of the environment Hamiltonian in eq. (4.23)

$$S_z = \int \mathcal{L}_z dt = \int \left\{ (Q_J + CV_x + CA) \frac{\partial H_z}{\partial Q_J} - H_z \right\} dt, \tag{4.36a}$$

$$C \frac{\partial \phi_x(t)}{\partial t} = eQ_x, \quad C \frac{\partial \phi_J(t)}{\partial t} = eQ_J, \quad \frac{\partial \Phi(t)}{\partial t} = A(t). \tag{4.36b}$$

The effective action $S'_z(\phi_J)$ resulting from performing first the functional integral product over $\phi_n$ is given by

$$S'_z(\phi + e\Phi) = \frac{C}{2e^2} \int_{-\infty}^{+\infty} \left( \frac{\partial [\phi(t) - e\Phi(t)]}{\partial t} \right)^2 dt - \frac{1}{2e^2} \int_{-\infty}^{+\infty} \frac{[\phi(t) - e\Phi(t)]^2}{\sum_n L_n} dt$$

$$- \frac{1}{4\pi e^2} \int_{-\infty}^{+\infty} \int_{-\infty}^{+\infty} [\phi(t) - e\Phi(t)] \frac{\partial Z^{-1}(s-t)}{\partial s} [\phi(t) - e\Phi(t)] ds \, dt$$

$$+ \frac{1}{e} \int_{-\infty}^{+\infty} I_F(t) [\phi(t) - e\Phi(t)] dt \tag{4.37}$$

with a fluctuation current $I_F(t) = 0$ and we have used $\phi = \phi_J - \phi_x$. Here, $Z^{-1}(t)$ is the Fourier transform of a generalized admittance function $Z^{-1}(\omega)$ given by

$$Z^{-1}(\omega) = \sum_{n=1}^{k} \frac{\omega_n}{i\omega L_n} \left\{ \frac{\omega_n}{(\omega + i\varepsilon)^2 - \omega_n^2} \right\} \quad (4.38a)$$

$$= \sum_{n=1}^{k} \frac{\omega_n}{i\omega L_n} \left\{ \frac{1}{\omega - \omega_n + i\varepsilon} - \frac{1}{\omega + \omega_n + i\varepsilon} \right\}, \quad (4.38b)$$

$$\frac{1}{\omega + \omega_n \pm i\varepsilon} = \mp i\pi \delta(\omega + \omega_n) + \text{p.p.}\left( \frac{1}{\omega + \omega_n} \right) \quad (4.38c)$$

where eq. (4.38c) is the Sokhotski-Plemelj formular and p.p. stands for Cauchy principal part. eq. (4.38) is related to the spectral function $K(\omega) = [Z^{-1}(\omega) + Z^{-1}(-\omega)]/(2\pi)$ where $\varepsilon$ is the infinitesimal satisfying $\omega\varepsilon = \varepsilon$ and the nilpotent condition $\varepsilon^2 = 0$. Note, the spectral function is the sum of negative and positive frequency impedance accounting for emission and absorption processes respectively by the circuit. Thus, eq. (4.16) differs slightly from eq. (4.37) where the real-valued spectral function $K(t)$ in the classical Lagrangian gets replaced with the complex valued admittance $Z^{-1}(t)/(2\pi)$ in the quantum case.

Introducing the Dirac delta function $\delta(x)$ for functional integrals with the property

$$\int Dx\, f(x)\delta(x - y) = f(y) \quad (4.39)$$

for any functional $f(x)$, we may proceed to insert $\int D\phi_z \delta(\phi_J + \phi_x + \phi_z - e\Phi) = 1$ into eq. (4.35) thus introducing the constraint $\sum_a \phi_a = \phi_J + \phi_x + \phi_z = e\Phi$ guaranteed by the circuit in Fig. (4.2). Consequently, the average in eq. (4.31) is now taken over both $\phi_J$ and $\phi_z$:

$$\lim_{\beta \to +\infty} \langle \sin[\kappa \Delta \phi_J(t)] \rangle_{\phi_J \phi_z}$$

$$= \mathcal{Z}^{-1} \int D\phi_J \int D\phi_z \delta\left( \sum_a \phi_a - e\Phi \right)$$

$$\times \sin[\kappa \Delta \phi_J(t)] \exp iS'_z(\phi - e\Phi) \quad (4.40)$$

We find,

$$-\langle \sin\left[\kappa\Delta\phi_{\text{J}}(t)\right]\rangle_{\phi_{\text{J}}\phi_{\text{z}}} = \langle \sin[\kappa\Delta\phi_{\text{x}}(t)+\kappa e\int_0^t A(\tau)d\tau+\kappa\Delta\phi_{\text{z}}(t)]\rangle_{\phi_{\text{z}}}$$

$$= \langle \sin\left[\kappa\Delta\phi_{\text{z}}(t)\right]\rangle_{\phi_{\text{z}}} \cos\left[\kappa\Delta\phi_{\text{x}}(t) + \kappa e\int_0^t A(\tau)d\tau\right]$$

$$+ \langle \cos\left[\kappa\Delta\phi_{\text{z}}(t)\right]\rangle_{\phi_{\text{z}}} \sin\left[\kappa\Delta\phi_{\text{x}}(t) + \kappa e\int_0^t A(\tau)d\tau\right] \quad (4.41)$$

with $\Delta\phi_a(t) = e\int_0^t V_a(t')dt'$ and $\Delta\Phi(t) = \int_0^t A(t')dt'$. We have assumed Fubini's theorem for interchange of integration order applies and thus performed first the integral over $\phi_{\text{z}}$. Using the fact that $S'_{\text{z}}$ is quadratic, the resulting functional integral over $\phi_{\text{z}}$ in eq. (4.41) is Gaussian resulting in $\langle \sin\left[\kappa\Delta\phi_{\text{z}}(t)\right]\rangle_{\phi_{\text{z}}} = 0$ term vanishing. Likewise, $\langle \cos\left[\kappa\Delta\phi_{\text{z}}(t)\right]\rangle_{\phi_{\text{z}}}$ satisfies Wick's theorem[16]

$$\langle \cos\left[\kappa\Delta\phi_{\text{z}}(t)\right]\rangle_{\phi_{\text{z}}}$$
$$= \exp\left(\kappa^2 \langle [\phi_{\text{z}}(t) - \phi_{\text{z}}(0)]\phi_{\text{z}}(0)\rangle_{\phi_{\text{z}}}\right), \quad (4.42a)$$

$$\int D\phi_{\text{z}} \exp iS'_{\text{z}}(\phi_{\text{z}}, I_{\text{F}}) = \exp iS''_{\text{z}}(I_{\text{F}}), \quad (4.42b)$$

$$S''_{\text{z}}(I_{\text{F}}) = \frac{2\pi}{2e^2} \int_{-\infty}^{+\infty} I_{\text{F}}(-\omega)G_{\text{eff}}(\omega)I_{\text{F}}(\omega)d\omega \quad (4.42c)$$

$$G_{\text{eff}}(\omega) = -e^2 i\omega^{-1} Z_{\text{eff}}(\omega), \quad (4.42d)$$

$$Z_{\text{eff}}(\omega) = \frac{1}{Z^{-1}(\omega) + y(\omega)} \quad (4.42e)$$

where $y(\omega) = i\omega C - i\omega^{-1}\sum_n L_n^{-1}$.

We introduce the zero temperature propagator $D_{+\infty}(t)$ given by

$$D_{+\infty}(t) = \frac{1}{2\pi}\int_{-\infty}^{+\infty} \frac{d\omega}{\omega} \exp-i\omega t \left\{Z_{\text{eff}}(\omega) + n.f.\right\}, \quad (4.43a)$$

where $n.f.$ stands for negative frequency. The finite temperature propagator is related to $D_{+\infty}(t)$ by a sum over the

photon number states

$$D_{+\infty}(t) \to D_\beta(t) = \sum_{n=0}^{+\infty} D_{+\infty}(t - in\beta)$$

$$= \frac{1}{2\pi} \int_{-\infty}^{+\infty} \frac{d\omega}{\omega} \frac{\exp -i\omega t}{1 - \exp(-\beta\omega)} \{Z_{\text{eff}}(\omega) + n.f.\} \quad (4.43b)$$

Thus, computing the phase–phase correlation function, we find

$$\langle \phi_z(s)\phi_z(t)\rangle_{\phi_z} = \left. \frac{\mathcal{Z}^{-1}\delta^2 \exp iS_z''(I_F)}{e^{-2}\delta I_F(s)\delta I_F(t)} \right|_{I_F=0, \mathcal{Z}=1}$$
$$= e^2 D_{+\infty}(s-t) \to e^2 D_\beta(s-t), \quad (4.44)$$

which satisfies the well-know fluctuation-dissipation theorem.[9]

## Cooper Pair and Quasi-Particle Tunneling Current

Finally, plugging in results (4.41) and (4.43b) in eq. (4.31), and using $\Delta\phi_x(t) = \int_0^t V(t')dt' = Vt$ where $V_x = V$ is a constant external voltage and $\Delta\Phi(t) = 0$, the total $I - V$ characteristics is given by

$$I_0(V) = I_1(V) + I_2(V)$$
$$= e^{-1}R_T^{-1} \int_{-\infty}^{+\infty} d\epsilon_p d\epsilon_q \frac{\mathcal{N}(\epsilon_q)\mathcal{N}(\epsilon_p)}{\mathcal{N}^2(0)} f(\epsilon_p)(1 - f(\epsilon_q))$$
$$\times \{P_1(\epsilon_q - \epsilon_p + eV) - P_1(\epsilon_q - \epsilon_p - eV)\}$$
$$+ e\pi E_J^2 \{P_2(2eV) - P_2(-2eV)\} \quad (4.45)$$

where we have introduced the so called $P(E)$ function[16]

$$P_\kappa(E) = \frac{1}{2\pi} \int_{-\infty}^{+\infty} dt \, \exp \kappa^2 \mathcal{J}(t) \exp iEt, \quad (4.46a)$$
$$e^{-2}\mathcal{J}(t) = D_\beta(t) - D_\beta(0) \quad (4.46b)$$

with $E$ some arbitrary energy. It gives the probability that the junction will absorb energy $E$ from the environment. Note that eq. (4.45) reduces to the normal junction $I - V$

characteristics

$$I(V)|_{\Delta=0} = e^{-1}R_T^{-1} \int_{-\infty}^{+\infty} dE_p dE_q f(E_p)(1 - f(E_q))$$
$$\times \{P_1(E_q - E_p + eV) - P_1(E_q - E_p - eV)\}$$
$$= e^{-1}R_T^{-1} \int_{-\infty}^{+\infty} dE \frac{E}{1 - \exp(-\beta E)}$$
$$\times \{P_1(-E + eV) - P_1(-E - eV)\} \quad (4.47)$$

when the superconducting gap vanishes $\Delta = 0$, since $E_J(\Delta = 0) = 0$, $\mathcal{N}(\Delta = 0)/\mathcal{N}(0) = 1$ and $E_p = \epsilon_p$, $E_q = \epsilon_q$.

Here, we have used path integral formalism and demonstrated how to reproduce the *I–V* characteristics of the single small Josephson junction. Finally, recall we introduced a flux stored by the circuit, $\Phi$ only to set it equal to zero. In the next chapter, we discuss how the microwave amplitude and Green's function are renormalized in a manner analogous to vacuum polarization in QED. We show that this implies that $\Phi \neq 0$ is non-vanishing.

# Renormalization

# Rescaling of microwave amplitude in small Josephson junctions

# 5.

In chapter 4, we introduced the $P(E)$ function in terms of Green's function in eq. (4.42e) for the equivalent circuit of the Josephson junction and the environment given in Fig. 4.2. It is worth noting that the early introduction of $P(E)$ theory and its applications for single junctions is provided in ref. [16, 114] while the case of the single Josephson junction is given in ref. [114]. In particular, $P(E)$ theory accurately predicts the current-voltage characteristics (given by eq. (4.45)) for a single junction biased by a dc voltage ($V$) and strongly coupled to its electromagnetic environment in the form of an impedance function $Z(\omega)$.

Typically, the theory accounts for the effect of an ac bias voltage by considering a shift $V \to V + V_{\text{ac}} \cos \Omega t$ of the $I$–$V$ characteristics, where $V_{\text{ac}}$ is the amplitude and $\Omega$ is the angular frequency of the alternating voltage. Using a unitary transformation method equivalent to eqs. (25 - 31) of ref. [16], entailing the splitting off of the dc component $\phi(t)$ of the quantum phase from the ac component $\phi_{\text{RF}}(t)$, the aforementioned shift was shown not to hold for a normal junction driven via its electromagnetic environment.[27] Instead, the two phases $\phi(t)$ and $\phi_{\text{RF}}(t)$ should satisfy the following constraint,

$$C\frac{\partial^2 \phi_{\text{RF}}}{\partial t^2} - \int dt' Z^{-1}(t-t')\frac{\partial}{\partial t}\left[\phi_{\text{RF}}(t') - \phi(t')\right] = 0, \quad (5.1)$$

equivalent to eq. (4.9) of ref. [27]. Taking its Fourier transform, it is clear that eq. (3.15) satisfies this constraint. The constraint requires that the shift $V \to V + V_{\text{ac}} \cos \Omega t$ be modified to $V \to V + \left|\left[1 + i\Omega C Z^{-1}(\Omega)\right]^{-1}\right| V_{\text{ac}} \cos(\Omega t + \eta)$, where $\eta(\Omega)$ is the argument of $\left|\left[1 + i\Omega C Z^{-1}(\Omega)\right]^{-1}\right|$. Herein, we shall refer to the rescaling $V_{\text{ac}}$ to $|\Xi|V_{\text{ac}}$, where $\Xi$ is a ratio of impedance functions, as *renormalization*.

It is important to note that since the arguments presented in chapter 3, consistent with eq. (5.1) through eq. (3.15), merely require the responses of the dc voltage vary from the ac voltage to be valid, the renormalization result of ref. [27] for 1: A Lehmann weight is a factor which renormalizes the Green's function of a quantum mechanical system. Lehmann weights are also referred to as wavefunction renormalization factors since the Green's function of a quantum mechanical system is often the Fourier transform of the wavefunction.

single normal junctions should also apply for single small Josephson junctions.

Thus, in this chapter, we introduce a method for calculating the renormalization factor using Lehmann weights.[103][1] This method has the advantage of being amenable to varied situations ranging from single normal junctions, single Josephson junctions as well as arrays. Particularly, we consider renormalization effects of applied oscillating voltages due to wavefunction renormalization/Lehmann weights[103] that rescale the environmental impedance of the single junction as well as the array. The array is treated as an infinitely long[51] effective circuit. As is the case for the single junction, we find a Lehmann weight of the general form, $\Xi(\omega) = \exp(-\Lambda^{-1})\exp{-\beta M(\omega)}\exp(i\beta\varepsilon_m)$, where $\Lambda$ is the soliton length of the array[41, 50–52], which also acts as a linear response function for oscillating electromagnetic fields, and can be interpreted as the probability amplitude of exciting a 'particle' of mass $M$ from the junction ground state by the RF field.[115] The quantum statistics of this 'particle' are determined by the argument $\beta\varepsilon_m$, where $\varepsilon_m$ is identified as the Matsubara frequency.[116] In the case of the infinite array, this 'particle' corresponds to a bosonic charge soliton injected into the array. Possible application of these results is in accurately determining the absorbed RF power in dynamical Coulomb blockade experiments especially where long arrays are used as on-site electromagnetic power detectors.[33, 54]

## 1. Rescaling of Oscillating Voltages Applied on Single Junctions

### Introduction

Within the Caldeira-Leggett model[10] introduced in chapter 4, the environment of a dissipative voltage-biased single junction shown in Fig. 4.2 is modeled by the action $S_z = \int dt\, \mathcal{L}_z$, where the Lagrangian is given by,

$$\mathcal{L}_z = \frac{C}{2e^2}\left(\frac{\partial \phi'}{\partial t}\right)^2 + \sum_{n=1}^{k}\left\{\frac{C_n}{2e^2}\left(\frac{\partial \phi_n}{\partial t}\right)^2 - \frac{L_n^{-1}}{2e^2}(\phi_n - \phi')^2\right\},$$

(5.2a)

where $e$ is the electric charge, $C = \varepsilon_0\varepsilon_r\mathcal{A}/d_{\text{eff}}$ is the junction capacitance, $\phi' \equiv \phi_J - \phi_x - e\Phi$ where[2] $\phi_J, \phi_x, \phi_z$ and $e\Phi$ are the phases associated with the the voltage drop at the junction $V_J$, voltage source (external voltage) $V_x$ and the flux stored by the circuit respectively, $\phi_n$ is the bath degrees of freedom represented by $k$ coupled (via $\phi$) $L_n C_n$ oscillators constituting the environment.

[2]: $\phi'$ is the phase defined for convenience, to shorten the expression of the Lagrangian to avoid always writing $\phi_J - \phi_x - e\Phi = \phi - e\Phi$ in the calculations in chapter 4.

The effective action,

$$S'_z = -i \ln \int \prod_{n=1}^{k} D\phi_n \exp iS_z(\phi_n, \phi')$$

$$= \int dt \left\{ \frac{C}{2e^2} \left(\frac{\partial \phi'}{\partial t}\right)^2 - \frac{1}{4\pi e^2} \int ds\, \phi'(s) \left[\frac{\partial Z^{-1}(t-s)}{\partial t}\right] \phi'(t) \right\}$$

$$= \frac{2\pi}{2e^2} \int d\omega\, \phi'(\omega) i\omega Z^{-1}_{\text{eff}}(\omega) \phi'(-\omega) \quad (5.2b)$$

requires the impedance Green's function in the $P(E)$ function to be modified by a wavefunction renormalization (Lehmann) weight[103]),

$$P_\kappa(E) = \frac{1}{2\pi} \int dt \exp\left(\kappa^2 \mathcal{J}(t) + iEt\right), \quad (5.3a)$$

$$\mathcal{J}(t) = \frac{2e^2}{2\pi} \int \frac{d\omega}{\omega} \text{Re}\{Z_{\text{eff}}(\omega)\} \frac{\exp(-i\omega t) - 1}{1 - \exp(-\beta\omega)}, \quad (5.3b)$$

$$Z_{\text{eff}}(\omega) = [Z^{-1}(\omega) + i\omega C]^{-1} = \Xi(\omega) Z(\omega), \quad (5.3c)$$

where $\text{Re}\{Z_{\text{eff}}(\omega)\} = [Z_{\text{eff}}(\omega) + Z_{\text{eff}}(-\omega)]/2$ is the real part of $Z_{\text{eff}}(\omega)$, $\beta$ is the inverse temperature, $\kappa e = 1e, 2e$ is the quasi-particle, Cooper-pair charge, $E$ is the energy exchanged between the junction, $\Xi(\omega)$ is a Lehmann weight[3] and $L_n C_n$ circuits acting as the environment, $\omega$ is the Fourier transform frequency that also plays the role of the thermal photon frequency at finite temperature.

[3]: A Lehmann weight renormalizes a Green's function.

It is known – at least since the work of Callen and Welton[9] – that the (causal) response function[4] $\Xi(\omega) \equiv \int_{-\infty}^{+\infty} dt\, \theta(t)\chi(t) \exp(i\omega t) dt$ for a system driven by oscillating electromagnetic fields appears as the coefficient[5] of the black body spectrum. Consequently, this requires that the response $V'_{\text{RF}}(t)$ as seen by the junction J in Fig. 4.2 be a weighted function of $\chi(t)$ and the applied oscillating voltage $V_{\text{RF}}(t): V'_{\text{RF}}(t) = \int_{-\infty}^{t} ds\, \chi(t-s) V_{\text{RF}}(s)$. Therefore, to accurately describe the I–V characteristics of J driven by an applied oscillating voltage

[4]: Appendix D.

[5]: This co-efficient can be computed using the driven system's equation of motion.

$V_{\text{RF}}(t)$, it is not enough to simply rescale the impedance $Z(\omega)$ in the $P(E)$ function: the amplitude and phase of the applied oscillating voltage $V_{\text{RF}}$ has to be renormalized accordingly. This point of view and its implications has been discussed in detail in chapter 3.

In subsequent sections, we first consider tracing our steps from standard quantum electrodynamics and ease our way into circuit-QED and hence $P(E)$ theory. We then proceed to introduce the finite temperature propagator for the junction and consider how the Lehmann weight arises for the impedance in $P(E)$ theory, and its implications for single junctions and long arrays driven by $V_{\text{RF}}(t)$. We find that, a finite time varying flux $\Phi(t)$ stored by the circuit consistently implements the aforementioned wavefunction renormalization by guaranteeing the circuit responds linearly to $V_{\text{RF}}(t)$.

## Connection of $P(E)$ theory to quantum electrodynamics (QED).

Here, we shall make the connection of the Caldeira-Leggett model to quantum electrodynamics (QED). We shall find out that circuit-QED is merely the 1 dimensional space time version of QED. This also allows us to link the propagator introduced in eq. (4.43), chapter 3 with the photon propagator in QED.

The Fourier transform of the summed terms in Caldeira-Leggett action given in eq. (5.2a) is $S_0 + S_{\text{int}}$ where,

$$S_0 = \frac{2\pi}{2e^2} \sum_n \int d\omega C_n \phi_n(\omega) \left[\omega^2 - \omega_n^2\right] \phi_n(-\omega)$$
$$+ \frac{2\pi}{2e^2} \sum_n \frac{1}{L_n} \int d\omega \phi_n(\omega) \phi'(-\omega) + O(\phi'^2), \quad (5.4)$$

$O(\phi'^2)$ is a term with $\phi'^2$ that we initially neglect, $\phi_n$ are the Caldeira-Leggett phases of $L_n C_n$ circuits in Fig. 4.2, $\omega_n = 1/L_n C_n$ and the interaction term $S_{\text{int}}$ is given by,

$$S_{\text{int}} = \frac{2\pi}{2e^2} \int d\omega \omega^2 C \phi'(\omega) \phi'(-\omega)$$
$$+ \frac{2\pi}{e} \int d\omega I_{\text{F}}(-\omega) \phi'(\omega), \quad (5.5)$$

1. *Rescaling of Oscillating Voltages Applied on Single Junctions* | 89where $I_F$ is the fluctuation current which we shall later set, $I_F = 0$.

Integrating out the fluctuating degrees of freedom, $\phi_n$ as we did in eq. (4.35) of chapter 3,

$$\prod_n \int D\phi_n \exp(iS_0) = \exp(iS_0'), \qquad (5.6a)$$

we arrive at,

$$S_0' = \frac{(2\pi)^2}{4\pi} \int d\omega \phi'(-\omega) G^{-1}(\omega) \phi'(\omega), \qquad (5.6b)$$

$$G^{-1}(\omega) = \frac{1}{e^2} \sum_n \frac{1}{L_n} \frac{\omega_n^2}{\omega^2 - \omega_n^2} \equiv i\omega e^{-2} Z^{-1}(\omega), \qquad (5.6c)$$

where we have transformed the Green's function $G(\omega)$, from the $\phi_n$ degrees of freedom, to the environmental impedance $Z(\omega)$.

Proceeding to combine the two actions yields,

$$S_0' + S_{\text{int}} = \frac{2\pi}{2e^2} \int d\omega \phi'(\omega) \left[ e^2 G^{-1}(\omega) + \omega^2 C \right] \phi'(-\omega)$$
$$+ \frac{2\pi}{e} \int d\omega I_F(-\omega) \phi'(\omega). \qquad (5.7)$$

Note that eq. (5.7) is the action for eq. (3.13) with the Josephson coupling energy $E_J = 0$, already encountered in chapter 3.

Moreover, we saw in chapter 3 that eq. (3.13) is consistent with Maxwell's equations. In fact, by defining the electromagnetic vector potential $A_\mu = (V, \vec{A})$, the electric field $\vec{E} = \partial \vec{A}/\partial t - \vec{\nabla} V$, the magnetic field $\vec{B} = \vec{\nabla} \times \vec{A}$ and the fluctuation current density $J_F^\mu$, it can be seen that the interaction term given by $S_{\text{int}}$ above is actually Maxwell's action in disguise,

$$S_{\text{int}} \propto \int dt d\mathcal{A} dl \left[ \frac{\varepsilon_0 \varepsilon_r}{2} (\vec{E} \cdot \vec{E} - \vec{B} \cdot \vec{B}) + e J_F^\mu A_\mu \right]$$
$$= \frac{1}{2} \int \frac{d^4 k}{(2\pi)^4} \varepsilon_0 \varepsilon_r A^\mu(k) G_{\mu\nu}^{-1} A_\nu(-k) + e \int \frac{d^4 k}{(2\pi)^4} J_F^\mu(-k) A_\mu(k),$$
$$(5.8)$$

with the conditions[6] $\vec{B} = 0$, $\partial \phi'/\partial t = -el\vec{n} \cdot \vec{E}$, $\phi' = e \int dl \, \vec{n} \cdot \vec{A}$, $C = \varepsilon \mathcal{A}/l$ and $\int d\mathcal{A}\vec{n} \cdot \vec{J}_F = I_F$ where $\vec{n} = (1, 0, 0)$ is the normal vector to the junction barrier, $l \equiv d_{\text{eff}}$ the effective barrier thickness and $\mathcal{A}$ the junction area. The last term

[6]: Some of these conditions have been introduced in chapter 3 in eq. (3.8)

7: Actually, the Green's function takes the form,[89]

$$G_{\mu\nu} = \lim_{\varepsilon \to 0} \frac{-\eta_{\mu\nu} + k_\mu k_\nu / \varepsilon^2}{k^2 - \varepsilon^2}.$$

(5.9)

is the Fourier transform of the action where $k \equiv k^\mu = (\omega, \vec{k})$ is the photon energy-momentum satisfying $k^2 \equiv k^\mu k_\mu = \varepsilon^2$ and $G_{\mu\nu}(k) \propto 1/k^2$ is the photon Green's function in $1 + 3$ space-time dimensions.[7]

Integrating out $\phi'(\omega)$ degrees of freedom in $S_0 + S_{\text{int}}$ given by eq. (5.7) as in eq. (5.6a) leads to a circuit-QED term,

$$\frac{\pi}{e^2} \int d\omega I_F(\omega) G_{\text{eff}}(\omega) I_F(-\omega) \qquad (5.10)$$

where $G_{\text{eff}}^{-1} = G_\omega^{-1} + e^{-2}\omega^2 C$ is reminiscent of the famous Coulomb interaction term in QED,

$$\alpha \int \frac{d^4 k}{(2\pi)^3} J_F^\mu(k) G_{\mu\nu}(k) J_F^\nu(-k), \qquad (5.11)$$

8: The difference is the dimensionality of the theory: QED is in $1 + 2$ space-time dimensions while circuit-QED is solely in the time dimension.

where $\alpha = e^2/4\pi\varepsilon_0\varepsilon_r$ is the fine structure constant. The QED term is obtained in a similar fashion by integrating out $A_\mu$ instead of $\phi'$. Nonetheless, both expressions are essentially describing the same process.[8] Thus, we have showed that $1/\omega^2 C$ and hence $G(\omega)$ are the photon propagator in 1 dimensions.

However, a question still remains: is there any significance of this trivial Fourier space transformation given by $G^{-1}(\omega) \to G_{\text{eff}}^{-1}(\omega)$? We notice that we can define a factor $\Xi(\omega) = G(\omega)/G_{\text{eff}}$ and claim that this factor renormalizes the propagator $G(\omega)$ (of the $\phi_n$ degrees of freedom) to $G_{\text{eff}}(\omega)$ due to the presence of the Maxwell term, $S_{\text{int}}$. Since this renormalization takes a photon propagator into a different photon propagator, the Feynman rules to calculate it resemble photon self-energy interactions.[9] Bearing this in mind, we formulate the following Feynman rules for the propagator:

9: In self-energy interactions, the photon polarizes the QED vacuum by creating electron-positron pairs which subsequently annihilate. Such pairs can be created an infinite number of times, thus the contribution to the amplitude of all the processes takes the form: $G_{\text{eff}} = G + GUG + GUGUG\cdots = G/(1 - UG) = 1/(G^{-1} - U)$, where $G$ is the photon propagator and $U$ the vacuum polarization energy (interaction term). e.g. see [89]

1. The photon propagator $G(\omega) = e^2 Z(\omega)/i\omega$ is represented by: 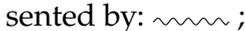 ;
2. The vacuum polarization energy term, $U(\omega) = e^{-2}\omega^2 C$ is represented by: 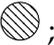 ;
3. Therefore, the leading order interaction term, $G(\omega)U(\omega)G(\omega)$ is drawn as, 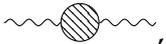,

where time increases from left to right.

Note that diagram 1 reads as follows: A photon of energy $\omega$ is created, propagates with a probability amplitude $G(\omega)$ and annihilates at a later time. Thus the amplitude must be assigned a photon creation operator and an annihilation operator, $a^\dagger$ and $a$ respectively and, in mathematical form, it should be written as $aG(\omega)a^\dagger$.

Likewise, diagram 3 represents an interaction whereby a photon of frequency $\omega$ is produced by acting on the vacuum state with $a^\dagger$, it propagates with an amplitude given by $G(\omega)U(\omega)G(\omega)$ then it annihilates by acting on the vacuum with $a(\omega)$. We emphasize that reversing $\omega$ reverses the aforementioned processes. This implies that the operators themselves should also be defined accordingly as,

$$a(-|\omega|) = a^\dagger, \qquad (5.12a)$$
$$a(|\omega|) = a, \qquad (5.12b)$$
$$[a(\omega), a(\omega')] = [\theta(\omega') - \theta(\omega)]\delta_{\omega,-\omega'}, \qquad (5.12c)$$

where $\theta(\omega)$ is the Heaviside function. This clearly displays the roles of the positive and negative frequencies. Note that negative frequencies are allowed since we are interested in energy differences due to single photon emission and absorption processes. For instance, processes where a photon is created before annihilation are related to processes where a photon is annihilated before creation by reversing the sign of the frequency $\omega$ and re-ordering the $a^\dagger, a$ operators appropriately.[10]

10: This procedure is invalid when arguments for time-reversal asymmetry e.g. discussed in ref. [25] apply.

### Connection to $P(E)$ theory: The Finite Temperature Green's Function and Propagator

Observe that a straightforward regularization procedure verifies $Z_{\text{eff}}(\omega)$ plays the role of effective Green's function of the $P(E)$ function,

$$Z_{\text{eff}}(\omega) = Z(\omega) + Z(\omega)[-i\omega CZ(\omega)] + Z(\omega)[-i\omega CZ(\omega)]^2$$
$$+ \cdots + Z(\omega)[-i\omega CZ(\omega)]^{n \to +\infty} = Z(\omega)\sum_{n=0}^{+\infty}[-i\omega CZ(\omega)]^n$$
$$= \frac{Z(\omega)}{1 + i\omega CZ(\omega)} = \Xi(\omega)Z(\omega), \quad (5.13)$$

analogous to the renormalization of the propagator in QED which often leads to a (Lehmann) factor[103, 117] analogous to $\Xi(\omega)$.

To elucidate this, consider the finite temperature propagator for $S_0'$, given by (∿∿∿),

$$D^\kappa(\omega) = \frac{-\kappa^2}{2\pi i} \{G(\omega)\langle a(\omega)a(-\omega)\rangle + G(-\omega)\langle a(-\omega)a(\omega)\rangle\} \tag{5.14}$$

where $\kappa e = 1e, 2e$ is the Cooper pair, BCS quasi-particle charge.[11][12]

11: The prefactor $\kappa^2/2\pi i$ included for ease of comparison later with the $P(E)$ theory.

12: The processes with $D^{\kappa *}(\omega) = \frac{-\kappa^2}{2\pi i}\{G(-\omega)\langle a(\omega)a(-\omega)\rangle\} + \frac{-\kappa^2}{2\pi i}\{G(\omega)\langle a(\omega)a(-\omega)\rangle\}$ are forbidden since represent *actual* negative frequency photons.

We proceed to include the effect of $S_{\text{int}}$ from a finite tunnel junction impedance by introducing the potential $U(\omega) = e^{-2}(\omega^2 C - \sum_n L_n^{-1}) = -e^{-2}i\omega y(\omega)$ where we have restored the $O(\phi'^2)$ term from eq. (5.4). The effective propagator is given by the sum of all the single photon interactions at the junction due to $U(\omega)$ and is proportional to $\langle a(\omega)a(-\omega)\rangle$ for positive and $\langle a(-\omega)a(\omega)\rangle$ for negative frequencies [eq. (5.14)], we find

$$\begin{aligned}D^\kappa_{\text{eff}}(\omega) &= \frac{-\kappa^2}{2\pi i}\{G(\omega) + G(\omega)U(\omega)G(\omega) + ...\}\langle a(\omega)a(-\omega)\rangle \\ &+ \frac{-\kappa^2}{2\pi i}\{G(-\omega) + G(-\omega)U(-\omega)G(-\omega) + ...\}\langle a(-\omega)a(\omega)\rangle \\ &= \frac{-\kappa^2}{2\pi i}G(\omega)[1 - U(\omega)G(\omega)]^{-1}\langle a(\omega)a(-\omega)\rangle \\ &+ \frac{-\kappa^2}{2\pi i}G(-\omega)[1 - U(-\omega)G(-\omega)]^{-1}\langle a^\dagger a\rangle_{-\omega} \\ &= \frac{-\kappa^2}{2\pi i}\{G_{\text{eff}}(\omega)\langle a(\omega)a(-\omega)\rangle + G_{\text{eff}}(-\omega)\langle a(-\omega)a(\omega)\rangle\}.\end{aligned} \tag{5.15a}$$

The term proportional to $[U(\pm\omega)G(\pm\omega)]^n$ is the finite temperature propagator for the photon interacting $n$ times with the junction impedance and the perturbation series

$$\sum_{n=0}^{+\infty}[U(\pm\omega)G(\pm\omega)]^n = [1 + y(\pm\omega)Z(\pm\omega)]^{-1}$$
$$= 1 - y(\pm\omega)Z_{\text{eff}}(\pm\omega) \equiv \Xi(\pm\omega), \tag{5.15b}$$

is computed by analytic continuation of the series $1 + x + x^2 \cdots + x^n \to [1-x]^{-1}$, where $x$ is given by the diagram,

. Thus, the effective Green's function becomes,

$$G_{\text{eff}}(\omega) = \smile\smile + \smile\!\bigcirc\!\smile + \smile\!\bigcirc\!\smile\!\bigcirc\!\smile + \cdots$$

$$= \smile\smile \times \frac{1}{\left(1 - \bigcirc\!\smile\right)} = \frac{1}{\left([\smile\smile]^{-1} - \bigcirc\right)}.$$

Comparing eq. (5.14) to (5.15a), we find that $G(\omega)$ is rescaled to $G_{\text{eff}}(\omega) = [G^{-1}(\omega) - U(\omega)]^{-1}$. Consequently, the effective action is given by

$$S'_z|_{I_F=0} = S'_0 + S_{\text{int}}|_{I_F=0}$$
$$= \pi \int_{-\infty}^{+\infty} d\omega \, \phi'(\omega) G_{\text{eff}}^{-1}(\omega) \phi'(-\omega), \quad (5.16)$$

which is equivalent to eq. (4.37) in chapter 4 with $I_F = 0$. Indeed we recover $\mathcal{J}(t) = e^2 \left[D_{\text{eff}}^\kappa(t) - D_{\text{eff}}^\kappa(0)\right]$, where $D_{\text{eff}}^\kappa(t) = \int d\omega \, D_{\text{eff}}^\kappa(\omega) \exp(-i\omega t)$ is the Fourier transform of $D_{\text{eff}}^\kappa(\omega)$,[13] by substituting $\langle a(-\omega)a(\omega)\rangle = n(-\omega) = [\exp(-\beta\omega) - 1]^{-1}$ and $\langle a(\omega)a(-\omega)\rangle = n(\omega) + 1 = [1 - \exp(-\beta\omega)]^{-1}$ in eq. (5.14) and (5.16), where the analytic continuation

$$\sum_{n=0}^{+\infty} \exp(\mp\beta n|\omega|) = \frac{1}{1 - \exp(\mp\beta|\omega|)} \quad (5.17)$$

has been used to regularize the divergent sum when calculating averages, $\langle \cdots \rangle$ for negative frequencies.

13: Also see eq. (4.37) and (4.43b).

### Vacuum Excitation and Photon amplitudes

Notice that in eq. (5.15a), the factor $[1 - U(\omega)G(\omega)]^{-1} = \Xi(\omega) \equiv |\Xi(\omega)| \exp\left[i\eta(\omega)\right]$ either rescales $G(\omega)$ or the photon number thermal averages $\langle \cdots \rangle$, implying that the photon number states are modified by the junction impedance. Rear-

ranging, we find

$$\begin{aligned}
G_{\text{eff}}(\omega) \langle a(-\omega)a(\omega) \rangle &= G(\omega)\Xi(\omega) \langle a(-\omega)a(\omega) \rangle \\
&= G(\omega) \langle |\Xi(\omega)| \exp\{i\eta(\omega)\} a(-\omega)a(\omega) \rangle \\
&= G(\omega)\mathcal{Z}_{\text{ph}}^{-1} \sum_{n=0}^{\infty} \langle n|a(-\omega)a(\omega) \times \\
&\quad \exp\left\{-\beta\omega\left[a(-\omega)a(\omega) + \frac{\text{sgn}(\omega)}{2}\right] + \ln[\Xi(\omega)]\right\}|n\rangle \\
&\equiv G(\omega)\mathcal{Z}_{\text{ph}}^{-1} \sum_{n=0}^{\infty} \langle n|a(-\omega)a(\omega) \times \\
&\quad \exp\left\{-\beta\left[\omega[a(-\omega)a(\omega) + \frac{\text{sgn}(\omega)}{2}] + M - i\varepsilon_m\right]\right\}|n\rangle,
\end{aligned}$$
(5.18)

where $\text{sgn}(\omega)$ is the sign function, $M - i\varepsilon_m = -\beta^{-1}\ln\Xi(\omega)$ is a gap in the electromagnetic energy spectrum. Taking $M(\omega)$ to be positive definite leads to $|\Xi(\omega)|^2 \leq 1$.

Thus, $M$ is the energy of a 'particle' excited from the vacuum by the electromagnetic field where the probability that the vacuum will be excited is given by the Boltzmann distribution $|\Xi(\omega)|^2 = \exp(-\beta 2M)$. Since this excitation carries electromagnetic energy, $|\Xi(\omega)|^2$ gives the fraction of electromagnetic power absorbed via excitations at the capacitor $C$ by the junction.[115] We note that the complex nature of $\beta^{-1}\ln\Xi(\omega)$ is not a concern since we are already accustomed to shifting our frequencies or energies by an infinitesimal [e.g. $\omega$ to $\omega + i\varepsilon$ in eq. (4.38)]. Taking in eq. (5.15b) the impedance $Z(\omega) = R$ to be real and $y(\omega) = i\omega C$, corresponding to eq. (4.14), yields

$$\arctan \omega RC = \eta(\omega) = \beta\varepsilon_m. \quad (5.19)$$

We introduce quantum statistics of the 'particle' by identifying $\varepsilon_m = (2m+1)\pi\beta^{-1}$ or $\varepsilon_m = 2\pi m\beta^{-1}$ as the fermionic or bosonic Matsubara frequency[116] respectively where $m \in \mathbb{Z}$. This requires the oscillation period $2\pi/\omega$ of the electromagnetic field to greatly exceed the relaxation time $RC$ of the circuit, $2\pi/\omega \gg 2\pi RC$. Thus, when this condition is not satisfied, it leaves the possibility for 'anyons'[118] with exotic statistics. Consequently, we can take $\Xi(\omega)$ as the amplitude[14] that a photon of frequency $\omega$ is absorbed by the junction creating a 'particle' of mass $M$ and statistics according to the Matsubara frequency $\varepsilon_m$. This realization, together with the

14: Photon amplitude here refers to a wavefunction renormalization (Lehmann) weight, where the photon wavefunction is taken to be the $x$ component of the Riemann – Silberstein vector[119].

fact that the field theory of the Josephson junction given by eq. (4.11) lives in 1 + 2 dimensions, suggests that 'anyonic' excitations cannot be ignored.[118] [15]

## Applied Alternating Voltages

In the previous section, we have established that the fraction of photons absorbed by a tunnel junction is $|\Xi(\omega)|^2 = [1 + y(\omega)Z(\omega)]^{-2}$. A straightforward way to experimentally measure $|\Xi(\omega)|^2$ is applying an external oscillating electric field in the form of ac voltage $V_{RF}(t)$ supplying power $P \propto \int_{-T/2}^{T/2} dt V_{RF}^2(t)$ where $T$ is the oscillation period. Whether the ac power is efficiently transferred to the junction from this ac source ought to depend on $\Xi(\omega)$. We proceed to formally express the explicit form of the effective alternating voltage at the junction.

This can be done by substituting $\Delta\phi_x(t) = \Delta\phi_x(t - 0^+) = \int_{0^+}^{t} V_x(\tau)d\tau$ in eq. (4.10) where $V_x = V + V_{RF}(t)$ and $V_{RF}(t)$ is an alternating voltage corresponding to the effect of the RF field given by

$$V_{RF}(t) = \int_{-\infty}^{+\infty} d\omega V_{RF}(\omega) \exp(-i\omega t) = V_{ac} \cos \Omega t \quad (5.21a)$$

$$V_{RF}(\omega) = \frac{V_{ac}}{2} \{\delta(\Omega - \omega) + \delta(\Omega + \omega)\} \quad (5.21b)$$

where $V_{RF}(\omega)$, $V_{ac}$ and $\Omega$ is the spectrum, the amplitude and frequency of the RF field respectively and the significance of the + sign is that $0^+$ is experimentally, not computationally equal to 0; it is the limit $0^+ \equiv t \to 0$.

Proceeding, the applied power $P$ of the RF field is the mean-square value of $V_{RF}(t)$ given by

$$P \propto \frac{1}{2\pi/\Omega} \int_0^{2\pi/\Omega} [V_{RF}(t)]^2 dt = \frac{V_{ac}^2}{2}, \quad (5.22)$$

where the proportionality factor is the admittance of the junction.

---

15: Notice that the exponent can be re-written conveniently as

$$\mathcal{G}^{-1}(M, i\varepsilon_m) + \frac{1}{2} \coth(\beta\omega/2)$$
$$= \mathcal{G}^{-1}(M, i\varepsilon_m) + \sum_{m=-\infty}^{+\infty} \frac{1}{2\pi m i - \beta\omega}$$
(5.20a)

with $\mathcal{G}^{-1}(M, i\varepsilon_m) = \beta(M - i\varepsilon_m)$ and the statistics of the excitations computed as,

$$\sum_{m=-\infty}^{+\infty} \mathcal{G}(M, i\varepsilon_m) = \sum_{m=-\infty}^{+\infty} \frac{\beta^{-1}}{M - i\varepsilon_m}$$
(5.20b)

where eq. (5.20b) is the inverse thermal Green's function of the 'particle' with mass $M$. Thus, this picture of 'particle' excitation has well defined Green's functions. Ofcourse, there is need to explore this excitation picture further and perhaps get the appropriate structure of the interaction potential, $U(\omega)$ given by . We have made an attempt in ref. [115], for the case of a large junction.

### Lehmann Weight and Linear Response

However, we are interested in the power absorbed by the junction $P_J$ instead since it modifies the $I - V$ characteristics. In the previous sections, we argued that this power $P_J$ is proportional to $|\Xi(\Omega)|^2$ in the presence of renormalization.

Within the context of linear response theory[112], this means that the applied voltage $V_{RF}(t)$ acts as an external force, and the effective voltage as a linear response $V'_{RF}(t)$ of the circuit,

$$V'_{RF}(t) = \int_{-\infty}^{t} \chi(t-s) V_{RF}(s) ds, \quad (5.23a)$$

$$\Xi(\omega) = \int_{0}^{+\infty} \chi(t) \exp(i\omega t) dt, \quad (5.23b)$$

$$V'_{RF}(\omega) = \Xi(\omega) V_{RF}(\omega), \quad (5.23c)$$

where $\chi(t-s)$ is the response function, making $\Xi(\omega)$ the susceptibility. For the special case discussed in eq. (5.19), we have $\chi(t) = (1/RC)\exp(-t/RC)$. Thus, the RF spectrum above gets modified to

$$V'_{RF}(\omega) = \frac{V_{ac}}{2} \Xi(\omega) \left\{ \delta(\Omega - \omega) + \delta(\Omega + \omega) \right\}, \quad (5.24a)$$

$$V'_{RF}(t) = \int_{-\infty}^{+\infty} V_{RF}(\omega) \exp(-i\omega t) d\omega,$$

$$= |\Xi(\Omega)| V_{ac} \cos(\Omega t + \eta), \quad (5.24b)$$

and $P_J$ is proportional to,

$$P_J \propto \frac{1}{2\pi/\Omega} \int_0^{2\pi/\Omega} dt [V'_{RF}(t)]^2 = \frac{V_{ac}^2}{2} |\Xi(\Omega)|^2 \propto P|\Xi(\Omega)|^2, \quad (5.25)$$

where we have used eq. (5.21), (5.22) and (5.24).

### Unitarity and the topological potential

This section displays the unitary nature of the renormalization effect. In particular, we show that the renormalization effect can be split into two: The fraction of ac voltage drop at the junction and the fraction of ac drop at the environment. This entails taking the quantum states of the environment and the junction as two orthornormal states $\psi_{\rm J} = |1\rangle$ and $\psi_{\rm z} = |0\rangle$ respectively (Fig. 5.1) undergoing a time dependent unitary transformation.

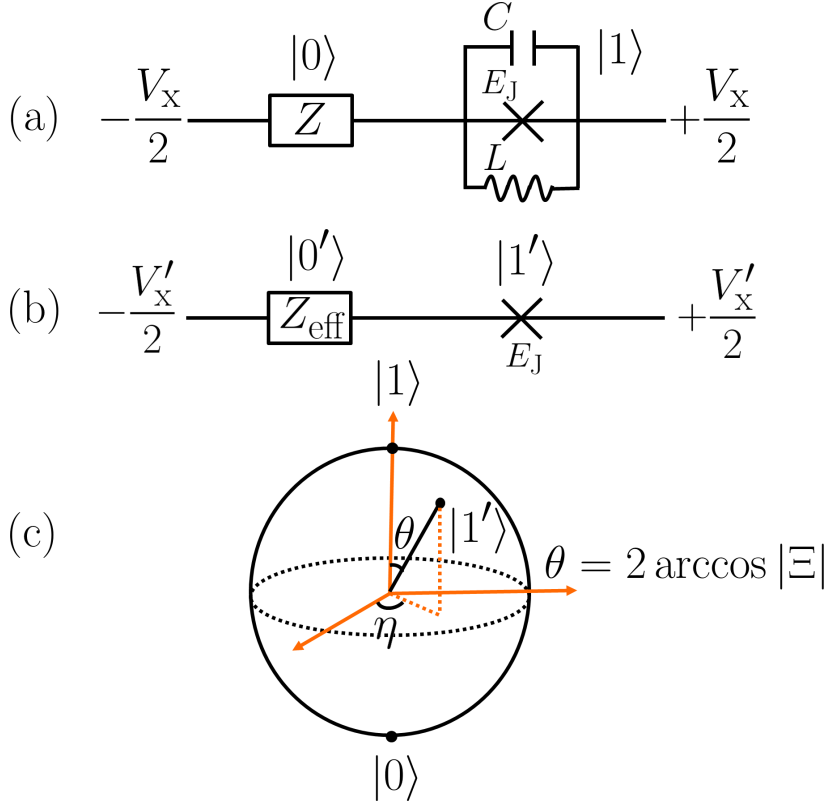

**Figure 5.1.:** Diagrammatic representation of the unitary transformation implemented by the matrix $\mathcal{U}_{\rm RF}(t=0)$ in eq. (5.26). (a) The equivalent circuit of the Josephson junction labeled by the coupling energy $E_{\rm J}$, the capacitance $C$ and inductance $1/L = \sum_n 1/L_n$ where the admittance $y(\omega) = i\omega C + \sum_n 1/i\omega L_n$. The junction is coupled serially to the environmental impedance $Z(\omega)$ and symmetrically biased by an external voltage $V_{\rm x}$ where the quantum states of the environment and the junction can be represented by $|0\rangle$ and $|1\rangle$ respectively; (b) The equivalent circuit of the Josephson junction and its environment. The bias voltage, the quantum states and the environmental impedance are all renormalized by the unitary transformation given by $\mathcal{U}$ in eq. (5.26); (c) A Bloch sphere representing the action of the unitary transformation given by $\mathcal{U}(t=0)$ in eq. (5.26), where $\eta$ is the argument of the renormalization factor $\Xi = |\Xi|\exp(i\eta)$ and $\theta = 2\arccos|\Xi|$.

In particular, defining matrices $\mathcal{V}_{\rm x}$ and $\mathcal{U}_{\rm RF}$ and two quantum states,

$$\mathcal{V}_{\rm x} = \frac{1}{2}\{V\sigma_0 + V_{\rm ac}\mathcal{U}_{\rm RF}\}, \qquad (5.26{\rm a})$$

$$\mathcal{U}_{\rm RF}(t) = \begin{pmatrix} \Xi(\Omega)e^{i\Omega t} & -\sqrt{1-|\Xi(\Omega)|^2}e^{i\Omega t} \\ \sqrt{1-|\Xi(-\Omega)|^2}e^{-i\Omega t} & \Xi(-\Omega)e^{-i\Omega t} \end{pmatrix},$$

$$(5.26{\rm b})$$

such that,

$$\begin{pmatrix} \psi'_J \\ \psi'_Z \end{pmatrix} = \mathcal{U}_{RF} \begin{pmatrix} \psi_J \\ \psi_Z \end{pmatrix}, \quad (5.26c)$$

$$V'_x = \text{tr}\{\mathcal{V}_x\} = V + \int_{-\infty}^{+\infty} V'_{RF}(\omega)\exp(-i\omega t)d\omega, \quad (5.26d)$$

$$\det\{\mathcal{U}_{RF}\} = 1, \quad (5.26e)$$

$$\mathcal{U}^{\dagger}_{RF}\mathcal{U}_{RF} = \mathcal{U}_{RF}\mathcal{U}^{\dagger}_{RF} = 1, \quad (5.26f)$$

where $\sigma_0$ is the $2 \times 2$ identity matrix, we find that these states $\psi_J = |1\rangle$, $\psi'_J = |1'\rangle$ and $\psi_Z = |0\rangle$, $\psi'_Z = |0'\rangle$ are normalized $\langle 0|0\rangle = \langle 0'|0'\rangle = \langle 1|1\rangle = \langle 1'|1'\rangle = 1$ and orthogonal to each other, $\langle 0|1\rangle = \langle 0'|1'\rangle = 0$, while orthonormality is preserved under the unitary transformation $\mathcal{U}_{RF}$ leading to a renormalized external voltage $V_x = V + V_{ac}\cos(\Omega t) \to V'_x = \text{tr}\{\mathcal{V}_x\} = V_x + A(t)$. This requires the topological flux $\Delta\Phi(t) \neq 0$ in eq. (4.41) not vanish in the presence of oscillating electromagnetic fields. Solving for the topological potential $A(t)$, we find

$$A(t) = V'_x - V_x = V_{ac}\{v(\Omega)\sin\Omega t - u(\Omega)\cos\Omega t\} \quad (5.27a)$$

$$\xi(\Omega) = u(\Omega) + iv(\Omega) \quad (5.27b)$$

$$\Xi(\Omega) = 1 - \xi(\Omega) = 1 - u(\Omega) - iv(\Omega) \quad (5.27c)$$

with eq. (5.27c) relating the impedance Lehmann weight $\Xi(\Omega)$ to the topological potential amplitude factors $u(\Omega), v(\Omega)$.

Thus, the purpose of the topological potential is to implement the renormalization scheme above. By eq. (5.27), we find that the topological flux $\Phi(t) = \int_{-\infty}^{t} A(\tau)d\tau$ is ill-defined for $\tau = -\infty$ since $\sin(\omega\infty)$ and $\cos(\omega\infty)$ both oscillate rapidly without converging. Nonetheless, this poses no problem since it is the flux difference $\Delta\Phi(t) = \int_{0}^{t} A(\tau)d\tau$ that appears in the correlation function in eq. (4.41) rendering the I–V characteristics in eq. (4.45) perfectly well-defined.

Moreover, by eq. (5.15b), we find that

$$\Xi(\Omega) = \frac{y^{-1}(\Omega)}{y^{-1}(\Omega) + z(\Omega)} = \frac{1}{1 + z(\Omega)y(\Omega)} \quad (5.28a)$$

$$\xi(\Omega) = \frac{z(\Omega)}{y^{-1}(\Omega) + z(\Omega)} = y(\Omega)Z_{\text{eff}}(\Omega) \quad (5.28b)$$

are ratios of impedances. [16]

16: Thus, in the simple model in eq. (4.14), power renormalization is negligible ($\Xi(\Omega) \simeq 1$) only for extremely low frequencies satisfying $1/RC \gg \Omega$. However, for samples exhibiting Coulomb blockade that satisfy the Lorentzian-delta function approximation $\text{Re}\{Z_{\text{eff}}\} = R/(1 + \Omega^2 C^2 R^2) \sim \pi C^{-1}\delta(\Omega)$, the conductance $1/R$ is extremely small ($1/RC \ll \Omega$) and thus we should expect power renormalization for virtually all applied frequencies.

## Current–Voltage Characteristics with finite RF Field

Now that we have the form of the voltage $V'_x = V + V_{RF} + A = V + |\Xi|V_{ac}\cos(\Omega t + \eta)$, where $|\Xi|V_{ac} = V^{eff}_{ac}$, it should be substituted into eq. (4.45) to determine the Cooper-pair and quasi-particle tunneling current in the presence of microwaves. Thus, substituting $V'_x$ into eq. (4.45) and using the identities, $\sin(x\sin y) = \sum_{n=-\infty}^{\infty} J_n(x)\sin ny$ and $\cos(x\sin y) = \sum_{n=-\infty}^{n=\infty} J_n(x)\cos ny$ where $J_n(x) = (-1)^{-n}J_{-n}(x)$
$= \frac{1}{2\pi}\int ds \exp i(x\sin s - ns)$ is the Bessel function of the first kind, $x, y$ are arbitrary functions and $n \in \mathbb{Z}$ is an integer, the I–V characteristics of the irradiated junction can be expressed in terms of the $I$-$V$ characteristics $I_1$ and $I_2$ of the unirradiated junction,

$$I(V) = \sum_{n=-\infty}^{\infty} J_n^2\left(\frac{eV^{eff}_{ac}}{\Omega}\right) I_1\left(V - \frac{n\Omega}{e}\right)$$
$$+ \sum_{n=-\infty}^{\infty} J_n^2\left(\frac{2eV^{eff}_{ac}}{\Omega}\right) I_2\left(V - \frac{n\Omega}{2e}\right). \quad (5.29)$$

Here, $I_{1,2}$ are the quasi-particle, Cooper pair RF-free $I - V$ characteristics given in eq. (4.45), $J_n(x)$ are Bessel functions of the first kind where the order $n$ is the number of actual photons absorbed by the junction, $V_{ac}$ is the amplitude of the alternating voltage, $\Omega$ is the energy quantum of individual photons, $V$ and $I$ respectively are the applied dc voltage and tunneling current of the junction. Equation 5.29 is the well-known photon-assisted tunneling equation[48] where the total current is shifted by the number of photons reflecting energy conservation and is proportional to the square of the Bessel function reflecting the modification of density of states.[17] However, it differs from the conventional photon-assisted tunneling expression since the applied voltage $V_{ac}$ is renormalized by a factor $\Xi(\Omega)$ to become $V^{eff}_{ac} = \Xi(\Omega)V_{ac}$. Such a factor, including additional higher-order effects that are largely suppressed, was first derived in ref. [27] for a single normal junction using unitary transformations of the environmental degrees of freedom.[18] Our approach differs from [27] since it places $\Xi(\omega)$ within the framework of wavefunction renormalization by treating the environmental impedance of a single normal and the superconducting junction as a Green's function, and $\Xi(\omega)$ as a Lehmann weight,[103] providing a straight forward path to generalization to linear

17: In the case of Shapiro steps, the characteristics depend on a single power of the Bessel function, $J_n$. This can be traced to the fact that the Josephson equations in eq. (1.2) depend on $\phi(t)$ and not $\Delta\phi_x(t) = e\int_{0+}^{t} dt' V_x(t')$. For arbitrary functions $x_{1,2}, y_{1,2}$, the integral produces two oscillatory terms, $\sin(x_1\sin y_1 + x_2\sin y_2)$ while the Josephson current, only one, $\sin(x_1\sin y_1)$

18: The amplitude of the higher (current) harmonics has been shown by Grabert in ref. [27], to correspond to $|\Xi(n\Omega)| = |Z_{eff}(n\Omega)/Z(n\Omega)|$ where $n$ is the $n$-th harmonic (assumed to be positive). Using the impedances $Z(\Omega) = R$ and $Z_{eff}(\Omega) = 1/(R^{-1} + i\Omega C + 1/i\Omega L)$, the renormalization factor $|\Xi(n\Omega)|$ is a rapidly decreasing function of $n$.

arrays. Further discussion is provided in the Conclusion (Section 4).

## 2. Rescaling of Applied Oscillating Voltages in Linear Arrays of Josephson Junctions

We consider the action for $N_0$ number of Josephson junction elements where the action resembles that for a single junction given in eq. (5.2b) where all the admittance elements in the expression are replaced by $(N_0 - 1) \times (N_0 - 1)$ matrices,

$$S_z^A = \sum_{j,k=1}^{N_0-1} \int dt \frac{1}{2e^2} C_{jk} \frac{\partial \phi'_j}{\partial t} \frac{\partial \phi'_k}{\partial t}$$

$$- \frac{1}{4\pi e^2} \int ds dt \, \phi'_j(s) \left[ \frac{\partial (Z^{-1})_{jk}(t-s)}{\partial t} \right] \phi'_k(t)$$

$$- E_J \sum_{i=1}^{N_0-1} \int dt \, \cos(\phi'_i(t) - \phi'_{i+1}(t)), \quad (5.30a)$$

where $C_{jk} = (C_0 + 2C)\delta_{j,k} - C\delta_{j+1,k} - C\delta_{j-1,k}$ is the capacitance matrix of the array with Josephson junctions of equal capacitance $C$, $N_0 - 1$ is the number of islands, $C_0$ is the stray capacitance of each island, $\phi'_j$ and $Q'_j = e^{-1} \sum_k C_{jk} \partial \phi'_k / \partial t$ the phase and charge of each island respectively, $E_J \ll e^2/2C$ the Josephson coupling energy of each island and $(Z^{-1})_{jk}$ is the unspecified environment admittance matrix of the array.[19] The Fourier transform of this action with $E_J \to 0$ as

19: This action has been considered before within the context of one dimensional XY model of topological phase transitions, e.g. in ref. [120] with $(Z^{-1})_{ij} = 0$. This reference can be consulted for introduction on how to approach dynamics and phase transitions in such a linear array.

**Figure 5.2.:** A chain of Josephson junction arrays representing the locations of each element in the array. Here, $C$ is the junction capacitance, $Z$ is the junction environmental impedance and $E_J$ is the Josephson coupling energy. Each island is labeled by an index $j$ and is characterized by a self-capacitance $C_0$. The action for this circuit is given by eq. (5.30a) and eq. (5.30b)

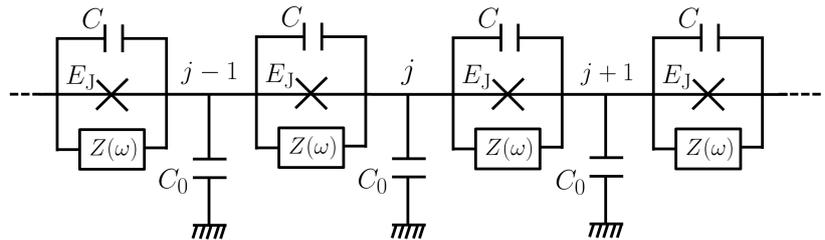

a perturbation parameter takes the general form,

$$S_z^A = \frac{2\pi}{2e^2} \sum_{j,k=1}^{N_0-1} \int d\omega \phi'_j(\omega) i\omega (Z_{\text{eff}}^{-1})_{jk}(\omega) \phi'_k(-\omega), \quad (5.30b)$$

where $(Z_{\text{eff}}^{-1})_{jk} = i\omega C_{jk} + (Z^{-1})_{jk}(\omega)$ is the effective admittance matrix. It is now straightforward to determine the phase-phase correlation function $\langle \phi'_j(t)\phi'_k(t)\rangle$ by recalling that its Fourier transform is equal to the imaginary part of the Green's function of the action read-off from eq. (5.30b) to yield $G_{jk}^A(\omega) = 2e^2 i \omega^{-1}(Z_{\text{eff}})_{jk}(\omega)$ where $(Z_{\text{eff}})_{jk}(\omega)$ is the inverse matrix of $(Z_{\text{eff}}^{-1})_{jk}$. This procedure yields, $\int dt \langle \phi'_j(t)\phi'_k(0)\rangle \exp(i\omega t) = G_{jk}^A(\omega) - G_{jk}^A(-\omega) = 2e^2\omega^{-1}\{(Z_{\text{eff}})_{jk}(\omega) + n.f.\}$. This suggests that the Lehmann/wavefuntion renormalization weight is a matrix of the form, $\sum_l (Z_{\text{eff}})_{jl}(Z^{-1})_{lk} = \Xi_{jk}(\omega)$. The amplitude of the applied ac voltage will be modified by its determinant $\Xi_A = \det(\Xi_{ij}(\omega))$.

For Gaussian correlated phases, $\langle \phi'_j(t)\phi'_k(s)\rangle = 0$ with $k \neq j$, the impedance matrix $(Z_{\text{eff}})_{jk}(\omega)$ has to be diagonalized, with $(Z_{\text{eff}})_{jk}(\omega) = 0$ for $j \neq k$. This is akin to setting all the phase-phase interaction terms to zero. However, this is not the case since the islands will effectively interact when a charge soliton propagates along the array constituting a current. The injection of a soliton/anti-soliton pair into the array depends on the electrostatic potential at the junction at the edge and the one at the center of the array labeled 1 and 2. We shall approximate the array as infinite with $N_0 \gg 1$ junctions, where each junction has a capacitance $C$ and environmental impedance $Z(\omega) = R$. The capacitance of the rest of the array is computed by recognizing that for an infinite array, neglecting the capacitance of the first junction $C$ and the self-capacitance of the first island $C_0$ does not alter the capacitance $C_r$ of the rest of the array, $C_r^{-1} = C^{-1} + (C_0 + C_r)^{-1}$. Solving for $C_r$, we find

$$\frac{1}{(C_0 + C_r + C)C_r} = \frac{1}{(C_0 + C_r)C} \rightarrow C_r^2 + C_0 C_r - C_0 C = 0$$

$$\rightarrow C_r = \frac{1}{2}\left(-C_0 + \sqrt{C_0^2 + 4CC_0}\right).$$

The total capacitance of the infinite array $C_A$ (excluding the first junction) is given by,

$$C_A = C_0 + C_r = \frac{1}{2}\left(C_0 + \sqrt{C_0^2 + 4CC_0}\right) = 2C_3.$$

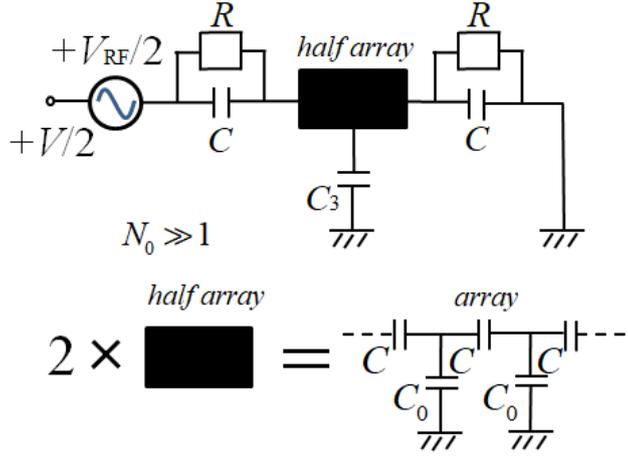

**Figure 5.3.:** A simplified circuit of a symmetrically biased infinite array with $N_0 \gg 1$ showing only half the array as a black box where the effective capacitance of the whole array is given by $2C_3 = \frac{1}{2}\left(C_0 + \sqrt{C^2 + 4CC_0}\right)$.

We thus set the capacitance of half the array as $C_3 = C_A/2$.

Taking the junction phases as $\phi_1$ and $\phi_2$, we can write the leading terms of the effective action as,

$$S_{\text{eff}}^A = \frac{C}{2e^2} \int dt \left\{\left(\frac{\partial \phi_1}{\partial t}\right)^2 + \left(\frac{\partial \phi_2}{\partial t}\right)^2\right\}$$

$$+ \int dt \frac{C_3}{2e^2}\left(\frac{\partial \phi_1}{\partial t} - \frac{\partial \phi_2}{\partial t}\right)^2$$

$$- \frac{1}{4\pi e^2} \int dt\, ds\, \phi_1(t) \frac{\partial Z^{-1}(t-s)}{\partial t} \phi_1(s)$$

$$- \frac{1}{4\pi e^2} \int dt\, ds\, \phi_2(t) \frac{\partial Z^{-1}(t-s)}{\partial t} \phi_2(s). \quad (5.31)$$

This leads to the following $2 \times 2$ capacitance and admittance matrices respectively,

$$i\omega C_{jk} = i\omega \begin{pmatrix} C + C_3 & -C_3 \\ -C_3 & C + C_3 \end{pmatrix}, \quad (Z^{-1})_{jk} = \frac{1}{R}\begin{pmatrix} 1 & 0 \\ 0 & 1 \end{pmatrix}.$$

where $j, k = 1, 2$. The effective admittance matrix is thus given by their sum, $(Z_{\text{eff}}^{-1})_{jk} = (Z^{-1})_{jk} + i\omega C_{jk}$.

Thus, the array via the interaction term with $C_3 = C_A/2$ modifies the phase-phase correlations by destroying their Gaussian nature as the impedance matrix is shifted as $(i\omega C + 1/R)\delta_{ij} \to (Z_{\text{eff}}^{-1})_{jk}$. In turn, this corresponds to a Lehmann weight, $\Xi_{jk} = \sum_{l=1}^{2}(i\omega C + 1/R)\delta_{jl}(Z_{\text{eff}})_{lk} = (i\omega C + 1/R)(Z_{\text{eff}})_{jk}$. This will affect the amplitude of the applied oscillating voltage by a Lehmann weight $\det(\Xi_{jk})$.[20]

Table 5.1.: A summary of the electromagnetic quantities and their rescaled expressions where the junction admittance and the environmental impedance are given by $y(\omega) = i\omega C$ and $Z(\omega)$ respectively.

| Quantity | Expression | Rescaled Expression |
|---|---|---|
| Environmental impedance (Single Junction) | $Z(\omega)$ | $Z_{\text{eff}}(\omega) = \Xi(\omega)Z(\omega)$ |
| Lehmann weight (Single Junction) | 1 | $\Xi(\omega) = [1 + i\omega C Z(\omega)]^{-1}$ |
| Microwave amplitude (single junction) | $V_{\text{ac}}$ | $V_{\text{ac}}^{\text{eff}} = |\Xi(\Omega)|V_{\text{ac}}$ |
| Environmental Impedance (long array: Fig. 5.3) | $(i\omega C + 1/R)\delta_{jk}$ | $(Z_{\text{eff}}^{-1})_{jk} + i\omega C_{jk}$ |
| Lehmann Weight (long array) | $\det[\delta_{jk}] = 1$ | $\Xi_A = \det\left[(i\omega C + 1/R)(Z_{\text{eff}})_{jk}\right]$ $\sim \exp(-\Lambda^{-1})$ |
| Microwave amplitude (long array) | $V_{\text{ac}}$ | $V_{\text{ac}}^{\text{eff}} \sim \exp(-\Lambda^{-1})V_{\text{ac}}$ |

When the angular frequency $\omega$ is much larger than the inverse of the time constant $RC$, we find,

$$\Xi_A = \lim_{\omega RC \to \infty} \det(\Xi_{jk}) = \frac{C}{C + C_A} = \exp(-\Lambda^{-1}), \quad (5.32)$$

where $\Lambda$ is the soliton length of the array. When an alternating voltage is applied across the array, the amplitude of the oscillating voltage will be renormalized by $\Xi_A \sim \exp(-\Lambda^{-1})$. This represents a damping of the applied power of applied oscillating voltage. This damping corresponds to the measured renormalization factor in chapter 2.

Finally, caution should be taken when using this result since the form of the Lehmann weight is valid as long as the effective action given by eq. (5.31) is a valid approximation of the action for the array. In particular, the full eq. (5.30a) should be applied to determine $\Xi_A = \det\left[\sum_l (Z_{\text{eff}})_{jl}(Z^{-1})_{lk}\right]$ for the linear array.

20: The determinant arises from the fact that when the action is Gaussian with the form, $S = \frac{1}{2}\phi_i G_{ij}^{-1}\phi_j$ where $\phi_j$ are bosonic variables, the path integral becomes $\int D\phi_i \exp(iS) = 1/\det(G_{ij}^{-1}) = \det(G_{ij})$. In particular, renormalizing $G_{ij}$ to $G_{ij}^{\text{eff}} = \sum_k \Xi_{ik} G_{kj}$ implies $\int D\phi_i \exp(iS) \to \int D\phi_i \exp(iS_{\text{eff}}) = \det(\Xi_{ij}) \int D\phi_i \exp(iS)$ Using the identity, $\det(\Xi_{ij}) = \exp(\text{Tr} \ln \Xi_{ij})$, where Tr is the trace, we find the action $S$ is shifted by $-i\text{Tr}\ln \Xi_{ij}$, similar to eq. (5.18).

## 3. Soliton Field Theory Origin of the Lehmann Weight in an Infinite Array

Consider the charge soliton lagrangian of the array,

$$\mathscr{L}_{\text{sol}} = \frac{1}{2\pi}\int dx \left\{\frac{1}{2}\left(\frac{\partial \chi}{\partial t}\right)^2 - \frac{1}{2}\left(\frac{\partial \chi}{\partial x}\right)^2 - \frac{2}{\Lambda^2}\sin^2(\chi/2)\right\},$$

(5.33a)

where the co-ordinates $x \equiv x/a, t \equiv v_0 t/a$ are dimensionless, and $a = 1$ is the length of the islands (lattice constant), $v_0 = 1$ is the velocity of electromagnetic radiation along $x$. The Euler-Lagrange equations of eq. (5.33a) yield the solution,

$$\chi = 4 \arctan \exp\left\{\Lambda^{-1} \gamma (x \pm vt)\right\} + 2\pi n, \qquad (5.33b)$$

with $\gamma = 1/\sqrt{1-v^2}$ the Lorentz factor. Plugging in eq. (5.33b) into eq. (5.33a), we can eliminate $\partial/\partial t$ in favor of $\partial/\partial x$,

$$\left(\frac{\partial \chi}{\partial t}\right)^2 = v^2 \left(\frac{\partial \chi}{\partial x}\right)^2 \qquad (5.34a)$$

$$\mathscr{L}_{\text{sol}} = \frac{-1}{\pi} \int dx \left\{\frac{1}{4\gamma^2}\left(\frac{\partial \chi}{\partial x}\right)^2 + \Lambda^{-2}\sin^2(\chi/2)\right\}. \qquad (5.34b)$$

Observing that since the integrand is quadratic, we can apply the Bogomol'nyi inequality[89] ($A^2 + B^2 \geq 2|AB|$) to evaluate the mass, $M$ given by,

$$\mathscr{L}_{\text{sol}} \geq M = \frac{1}{\pi} \int dx \left|\frac{1}{2\gamma \Lambda}\left(\frac{\partial \chi}{\partial x}\right)\sin(\chi/2)\right|$$

$$= \frac{1}{\gamma \Lambda \pi}\left|\int dx \frac{\partial \cos(\chi/2)}{\partial x}\right| = \frac{1}{\gamma \Lambda \pi}\left|[\cos(\chi/2)]_{-\infty}^{+\infty}\right|$$

$$= \frac{1}{\gamma \Lambda \pi}|\cos(\pi) - \cos(0)| = \frac{2}{\gamma \Lambda \pi} \equiv \frac{E_0}{\gamma \Lambda} \qquad (5.34c)$$

with $E_0 = 1/a\pi$. Treating the solitons as charged dust of mass density, $M/\mathscr{V}$ where $u^\mu = dx^\mu/d\tau = (\gamma, \pm v\gamma, 0, 0)$ is the four-velocity, $\mathscr{V} = \mathscr{A}l$ is the volume and $l$ the length of the array, we can introduce the energy-momentum tensor,

$$T^{\mu\nu} = \frac{2M}{\mathscr{V}} u^\mu u^\nu + \varepsilon_0 \varepsilon_r \left\{F^{\mu\sigma}F^\nu_{\ \sigma} - \frac{1}{4}F^{\sigma\rho}F_{\sigma\rho}\eta^{\mu\nu}\right\}. \qquad (5.35)$$

The total energy is given by,

$$E = \int_\mathscr{V} d^3 x \langle T^{00}\rangle = \frac{1}{\mathscr{V}} \int_\mathscr{V} d^3 x \langle 2M u^0 u^0\rangle$$

$$+ \frac{1}{\mathscr{V}} \sum_{s=\pm 1} \int_\mathscr{V} d^3 x\, \omega \left(\langle a_s(\omega) a_s(-\omega)\rangle_\omega + \frac{\text{sgn}(\omega)}{2}\right), \qquad (5.36a)$$

with $s = \pm 1$ the photon polarization states. Note that,

$$\langle M u^0 u^0 \rangle = \frac{E_0}{\Lambda} \langle \gamma \rangle \simeq \frac{E_0}{\Lambda} + \frac{E_0}{\Lambda} \frac{\langle v^2 \rangle}{2} = \frac{1}{\Lambda}\left(E_0 + \frac{1}{2\beta}\right),$$
(5.36b)

where $\langle \cdots \rangle \equiv \left(\int_{-\infty}^{+\infty} dv \, \exp -\beta E_0 v^2/2\right)^{-1} \int_{-\infty}^{+\infty} dv \, (\cdots \exp -\beta E_0 v^2/2)$ is the Boltzmann average for a gas of (anti-)solitons in $1+1$ dimensions. Comparing eq. (5.36) to eq. (5.18) (neglecting the vacuum evergy $E_0/\Lambda$ and considering only one photon polarization mode), we conclude that the array is an effective single junction with $\Xi_A = \exp(-\Lambda^{-1}) \exp(2\pi m i)$.

## 4. Discussion

In the case of the single Josephson junction, the renormalization of the amplitude of applied oscillating electromagnetic fields is implemented by linear response. This entails the excitation of 'particles' appearing as a mass gap $M - i\varepsilon_m = -\beta^{-1} \ln \Xi(\omega)$ in the thermal radiation spectrum, where $\varepsilon_m$ is the Matsubara frequency[116], $\Xi(\omega) = [1 + y(\omega)Z(\omega)]^{-1}$ is the linear response function, $Z(\omega)$ is the environmental impedance of the junction and $y(\omega) \simeq i\omega C$ is the impedance of the junction. Likewise, when an infinitely long array[32, 51] is modeled as half the infinite array interacting with two junctions, one at the array edge and the other at the center, as illustrated in Fig. 5.3, we find an additional Lehmann weight $\Xi_A = \exp(-\Lambda^{-1})$. This requires that applied oscillating electric fields are damped by the same factor by a finite range of electric fields along the array. This finite range arises due to the presence of charge solitons in the array, and is dual to the Meissner effect where the Cooper-pair order parameter leads to a finite range of the magnetic field.

A Josephson junction circuit that exhibits a large Coulomb blockade voltage is ideal for the observation of the renormalization effect. In particular, for the single junction, power renormalization is negligible ($\Xi(\Omega) \simeq 1$) only for extremely low microwave frequencies satisfying $1/RC \gg \Omega$. However, for samples exhibiting Coulomb blockade that also satisfy the Lorentzian-delta function approximation $\text{Re}\{Z_{\text{eff}}\} = R/(1 + \Omega^2 C^2 R^2) \sim \pi C^{-1} \delta(\Omega)$, the conductance $R^{-1}$ is extremely small ($1/RC \ll \Omega$) and thus we should expect power renormalization for virtually all applied frequencies. In the case of long arrays ($N_0 \gg \Lambda$) with $\Xi \simeq 1$, RF amplitude renormalization should be readily observed due to the additional factor $|\Xi_A(\Omega)| \sim \exp(-\Lambda^{-1})$.[61] Finally, a list of the electromagnetic quantities and their rescaled formulae is displayed in Table 5.1.

The form of the admittance $y(\omega)$ given in eq. (4.42e) neglects the back-action of the Josephson junctions on the environment (with the bath and the junction becoming entangled) which has been reported to dramatically change the predictions of the $P(E)$ theory.[31, 34, 35] This back-action manifests through the non-linear inductive response of the junction where the Josephson coupling energy is renormalized, $E_J^{\text{eff}} = E_J \langle \cos(\Delta 2\phi_J) \rangle = E_J \exp(-\langle \Delta 2\phi_J^2 \rangle)$. In our work, we have made an implicit assumption that, whenever the Josephson coupling energy $E_J$ is considerably small compared to all relevant energy scales such as the charging energy $E_c$, and the current-voltage characteristics of the junction do not exhibit a superconducting branch, this back-reaction can be taken to be small. Nonetheless, considering this back-reaction in our theoretical framework especially in the case of a single Josephson junction is certainly warranted since the back-reaction has been suggested to dramatically alter the insulator-superconductor transition conditions for the Josephson junction.

Within our path integral approach, considering such effects entails performing the path integral for the rescaled $P(E)$ function given by $P_{\kappa=2}^{\text{eff}}(E) = \int dt\, P_{\kappa=2}^{\text{eff}}(t) \exp(-iEt)$ where,

$$P_{\kappa=2}^{\text{eff}}(t) = \int D\phi_J \cos 2\Delta\phi_J(t) \exp\left(iS_{\text{CL}}\right) \quad (5.37)$$

$$\times \exp\left(-i \int dt\, E_J \cos 2\phi_J(t)\right), \quad (5.38)$$

which is challenging to carry out successfully to all orders of perturbation. Typically, the exponent is linearized as $\cos(2\phi_J) \simeq 1 - (2\phi_J)^2/2 + (2\phi_J)^4/4 + \cdots$ which becomes the $\phi^4$ theory.[121] In turn, at order $\phi^2$, the renormalized $E_J$ is expected to enter the usual Caldeira-Leggett expression as an inductance, $L_J^{\text{eff}} = 1/(2e)^2 E_J^{\text{eff}}$, $P_{\kappa=2}(E) = \int dt\, P_{\kappa=2}(t) \exp(-iEt)$ where, $P_{\kappa=2}(t) = \int D\phi \cos \Delta\phi(t) \exp\left(iS_{\text{CL}}^{\text{eff}}\right)$, yields a $P_{\kappa=2}(E)$ function with a linear inductive part in the impedance as given by $y(\omega)$ in eq. (4.42e). Further discussion on the exper-imental results of $E_J$–based renormalization and the back-action is beyond the scope of this work.

We have employed path integral formalism to derive the Cooper-pair current and the BCS quasi-particle current in small Josephson junctions and introduced a model which transforms the infinitely long array[51] into an effective circuit with a rescaled environmental impedance, $Z_{\text{eff}}^A \sim \exp(-\Lambda^{-1})\Xi(\omega)Z(\omega)$, where $Z(\omega)$ is the environmental impedance as seen by a single junction in the array. As is the case for the single junction, we expect that $\Xi(\omega) = \exp[-\beta M(\omega) - \Lambda^{-1}]\exp(i\varepsilon_m)$ also acts as a linear response function for oscillating electromagnetic fields, and can be interpreted as the probability amplitude of exciting a 'particle' of mass $M + \beta^{-1}\Lambda^{-1}$ from the junction ground state by the electromagnetic field,[115] with the quantum statistics of this 'particle' determined by the complex phase $\varepsilon_m$ identified as the Matsubara frequency.[116] In the case of the infinite array, this 'particle' corresponds to a bosonic charge soliton injected into the array. [21] This analysis does not take into account random offset charges which are known to act as static or dynamical background charges in the islands of the array, resulting in shifting of the threshold voltage $V_{\text{th}}$ and noise generation affecting the soliton flow along the array.[52, 84] Since the quasi-particle current naturally reduces to the normal current and the supercurrent vanishes with the superconducting gap ($\Delta = 0$), eq. (5.29) is essentially the time averaged current result previously proposed in ref. [27]. In the classical limit when the frequency $\Omega$ is small compared to the amplitude of the alternating voltage ($\kappa e V_{\text{ac}}^{\text{eff}}/\Omega \gg 1$), multi-photon absorption occurs. Setting, $\kappa e V_{\text{ac}} \sin\theta = n\Omega$, the sum over photon number can be approximated by an integral formula that corresponds to a classical detection of the RF field,[49]

$$I(V) = \frac{1}{\pi} \int_{-\pi/2}^{\pi/2} I_0\left(V - |\Xi(\Omega)|V_{\text{ac}} \sin\theta\right) d\theta. \qquad (5.39)$$

where $I_0(V)$ is given by eq. (4.45). This result offers a way to measure the magnitude of the Lehmann weight $\Xi(\Omega)$, where $|\Xi(\Omega)|$ is proportional to the *sensitivity* of the detector to RF power[61]. Conversely, this implies that our results are indispensable in dynamical Coulomb blockade experiments where linear arrays are used as microwave power detectors.[33, 54]

21: Appendix 3.

# Conclusion

# Summary and Perspectives | 6.

Research with Josephson junctions have come to a critical juncture. On one hand, several key theoretical works on the behaviour of Cooper pair tunneling in single small Josephson junctions as well as linear arrays and their interactions with the electromagnetic field have been proposed over the years in an effort to understand the quantum fluctuative nature of the Josephson junction.[15, 16, 27, 34, 49] On the other, over the years precision measurement techniques have enable crossroad experiments to be performed with these devices.[24, 32, 33, 36, 37, 41] At the heart of this success is the successful application of $P(E)$ theory[16] progressively complex quantum circuits ranging from superconducting single electron transistors where incoherent Cooper pair tunneling is prevalent in the Josephson quasi-particle cycle (JQP),[36, 37, 44, 76] to single Josephson junctions as well as arrays where photon assisted tunneling by microwave irradiation have been successfully carried out.[32, 33, 41] These experiments and others,[24, 28, 29] show that direct application of microwaves to single Josephson junctions in the regime of Cooper-pair Coulomb blockade produces photon-assisted tunneling effects in accordance to standard photon-assisted tunneling expressions in ref. [49] and [48], in accordance to standard $P(E)$ theory.

## Summary

In chapter 2, we described an experiment where microwaves of the sub-gigahertz range were irradiated on a linear array of small Josephson junctions exhibiting clear Coulomb blockade characteristics, dual to the Josephson effect.[1] The Coulomb blockade of tunneling Cooper-pairs was steadily diminished under irradiation, independent of frequency. The observed diminishing of Coulomb blockade with microwave radiation.[2] . This trend is typical for photon-assisted tunneling of Cooper-pairs in the Coulomb blockade regime.

---

1: The linear array of small Josephson junctions satisfying $0.1 < E_J/E_c < 1$ and $R_T > R_Q$ and $\hbar\omega \leq k_B T$. Under these conditions, the tunneling of charges at small voltages is dominated by Cooper pairs, and the characteristics exhibited are in the charge regime, dual to the phase regime. In this regime, Cooper-pair tunneling is precluded by the electromagnetic environment of the array, leading to Coulomb blockade of Cooper pairs.

2: This is dual to the enhanced phase diffusion effect by microwaves observed in ref. [41] for a one dimensional array of Josephson junctions in the regime, $E_J/E_c > 1$

However, on comparing the response obtained in the experiment with well-known equations[15, 48, 49] of photon-assisted tunneling, the results deferred by a loss factor of 0.87, suggesting the possibility of other effects such as electron heating in the islands of the array may play a significant role in the non-equilibrium tunneling processes in the array. However, the quasi-particle contribution calculated is negligibly small to account for this mismatch for the microwave amplitude range considered.[79, 80] Moreover, we found that experimental results, with substantial non-varying magnetic field applied perpendicular to the unirradiated array in order to raise the value of the Coulomb blockade (threshold) voltage $V_{cb}$ to its maximum Cooper-pair Coulomb blockade voltage value,[38, 47] exhibited the exact same mismatch with theory, thus raising the confidence level of our findings.[3]

We proceeded in the discussion section of chapter 2 to consider a voltage division effect arising from the difference in response between the dc and ac voltages applied to the array using an equivalent circuit of the array in the semi-infinite model of the array.[32, 51][4] Consequently, we conclude that a possible voltage division effect in the array occurs resulting in the factor observed, which we refer to as a renormalization effect. The equivalent circuit used relates the the factor to a characteristic length scale of the array, $\Lambda$, over which the applied microwave is damped from the edge into the array (soliton length).[41, 50–52] The soliton length is determined by measuring a device of the same built by using the gate effect,[86] to yield $\exp(-\Lambda^{-1}) \sim 0.89$, comparable to the mismatch observed. Even though such a dc/ac response analysis yields the desired result, a need arises to theoretically place it on an even firmer footing. The essential result here is that the dc and ac responses are related by a factor, which is simply the impedance ratio, $|Z_{\text{eff}}(\omega)/Z(\omega)| \sim 0.87$, where $Z(\omega)$ is the impedance when the circuit is dc biased and $Z_{\text{eff}}(\omega)$ ac biased.[5] ).

In chapter 3, we apply this important observation to introduce fluctuation-dissipation in the context of Einstein's $A(\omega)$ and $B(\omega)$ coefficients[91] [6] and the Caldeira-Leggett model. These coefficients are related to each other by $A(\omega)/B(\omega) = e^2\omega[Z(\omega) - Z(-\omega)]/2\pi$, through the fluctuation dissipation theorem,[9, 10] where $Z(\omega)$ is the environmental impedance of the junction. In this paradigm, we find the rate of stimulated emission $B(\omega) = \Gamma(2eV - \omega)/\omega$ proportional to the

---

3: In addition, the transmission line used in the measurement was calibrated in advance and the calibration uncertainties calculated were less than 4%, well below the 13% uncertainty needed to explain the results.

4: The array is biased by a dc and ac voltage, where the ac voltage represents the irradiated microwaves in the sample.

5: Such ratios multiply the power output of the circuit and can also lead to amplification effects not observed in our experiments. In fact, experimental results and theoretical analyses with single junctions in e.g. refs. [28, 53] show that such a factor can be used for near quantum amplification. However, these references focus on the Josephson junction, whose results are straigh-forward to theoretically analysed compared with arrays.

6: The coefficients represent the photon spontaneous and stimulated emission/absorption rates respectively by the Josephson junction.

tunneling rate $\Gamma$ for a Cooper-pair to tunnel across the junction. Therefore, the rate of spontaneous emission is given by,[7]

$$A(\Omega) = \frac{e^2}{2\pi}[Z(\Omega) + Z(-\Omega)]\Gamma(V - \Omega/2e)$$

The tunneling rate was calculated by Fermi's golden rule within the context of $P(E)$ theory.[16] Thus, the dc/ac division analysis of the experiment in chapter 2 corresponds to a renormalization of the $A(\omega)$ coefficient, since $Z(\omega) \to Z_{\text{eff}}(\omega)$. [8]

In the latter parts of the thesis, we introduced $P(E)$ theory in the context of path integrals in chapter 4 and in chapter 5, we considered the origin of the aforementioned renormalization effect using path integral formalism[102] and Green's function of the Caldeira-Leggett action.[10, 11] In our analyses we find that since the admittance of the circuit is equivalent to the Green's function in the Caldeira-Leggett model, and the Green's function is the Fourier transform of the circuit wavefunction, the aforementioned renormalization is simply a wavefunction renormalization, where the renormalization factor is the Lehmann weight representing the probability of exciting a fictitious quasi-particle from the junction vacuum. This interpretation is useful for the extension of the $P(E)$ theory to the case of the linear array, where the 'particle' in question is a charge soliton.[9] Moreover, we find that a topological flux stored in the circuit reproduces the impedance ratio responsible for the renormalization effect.

## Perspectives

Based on the above discussion, we note that $\Xi(\omega)$ can be interpreted as the amplitude of the junction to radiate by spontaneous emission into the environment and $\Xi(\omega)$ the amplitude of the environment to radiate into the junction by spontaneous emission. The corresponding stimulated emission rates are given by the photon-assisted expressions given by $P(E)$ theory. This observation has straightforward implications for the Josephson junction, when used as microwave detectors and near quantum amplifiers.[33, 53, 54]

In particular, the experimental results demonstrate pristine

---

7: Note that since $\omega$ is a Fourier transform parameter, which is allowed to have negative frequencies. This means that there is a decay rate for the junction with negative frequency, corresponding to absorption. This is in stark contrast to QED, where the photon frequency is always positive.

8: Or equivalently, the renormalization is of the ratio $A(\omega)/B(\omega) = e^2\omega[Z(\omega) + Z(\omega)]/2\pi \to e^2\omega[Z_{\text{eff}}(\omega) + Z_{\text{eff}}(\omega)]/2\pi = A_{\text{eff}}(\omega)/B(\omega)$. This sort of renormalization in quantum electrodynamics has been predicted for the case of the static versus dynamic systems with photon emission and absorption such as Unruh-Hawking radiation[106, 107] within the context of e.g. the Casimir effect.[109]

9: The amplitude for such an excitation process is given by $\Xi(\omega) = Z_{\text{eff}}(\omega)/Z(\omega) = |\Xi(\omega)|\exp(i\beta\epsilon_m)$ where $\epsilon_m$ is the 'Matsubara' frequency defining the statistics of the 'particle'. Further analysis on this is beyond the scope of this thesis.[See [115] for a discussion on the topological nature of the 'particles' in the case of the large junction.]

Josephson junction arrays are poised for microwave detection applications in a wide range of environments such as on-chip detection schemes[54] due to their high sensitivity to low-power, of order $10^6$ V/W.

Moreover, in chapter 1, we reviewed experiments whereby the environment impedance of of the Josephson junction embedded in leads of linear arrays of superconducting quantum interference devices (SQUIDs) can be tuned by an external magnetic field, $B$.[12, 36, 37] Since the environment depends largely on the property $Z_{\text{eff}} \propto \Xi \propto |E_J \cos(eBA)|$, application of finite magnetic fields to Josephson junction array should be able to tune $\Xi$. In chapter 2, the effect is negligible since we used Josephson junctions instead of SQUIDs, with a small Josephson coupling energy, $E_J$ which varies extremely slowly with magnetic fields. This type of magnetic field tuning allows for the the renormalization effect to be exploited to configure 'opaque' $\Xi = 1$, 'translucent' $0 < \Xi < 1$ or 'transparent' $\Xi = 0$ quantum circuits to microwave radiation. Finally, the theoretical work affirms the possibility to use Josephson junctions for microwave amplification schemes by exploiting the $A$ and $B$ coefficients similar to microwave amplification by stimulated emission of radiation (maser).[10]

10: See e.g. ref. [53] for an example of amplication using small Josephson junctions.

# APPENDIX

# A.

# Duality of charge and phase fluctuations

## Josephson Hamiltonian

In Sec. 4 and 1, we introduced two seemingly independent energies $H_{cp}$ and $E$ for the large Josephson junction. Here, we shall demonstrate the equivalence of both energies, discovering old results in the process. Consider the Hamiltonian $H_{cp} = eV_x\sigma_3 + m_0\sigma_1$ in eq. (4.6c). First we compute the bilinear $-\psi^\dagger H_{cp}\psi$ to find,

$$E = -\psi^\dagger H_{cp}\psi = -eV_x\langle n\rangle - m_0\Delta n \cos(2\phi_x) = \frac{Q_x^2}{2C} - E_J\cos(2\phi_x) \quad (A.1)$$

where $\psi^\dagger = \psi_1^*(1,0) + \psi_2^*(0,1)$, $\psi_1 = \sqrt{n_1}\exp(i\phi_1)$, $\psi_2 = \sqrt{n_2}\exp(i\phi_2)$, $n_1 = \langle n^2\rangle^{1/2} + \langle n\rangle$, $n_2 = \langle n^2\rangle^{1/2} - \langle n\rangle$ and we have re-defined the charge stored by the junction as $n$, and its variance as $\Delta n = 2\sqrt{n_1 n_2}$, and simply used $2e\langle n\rangle = Q_x = -CV_x$ and $m_0\Delta n = E_J$ in the last term thus arriving at (4.9a).

From the substitutions above, it is clear that the current $I_S$ is proportional to $\partial\langle n\rangle/\partial t$ as expected. In addition, Josephson energy $E_J \propto \Delta n$ is proportional to charge fluctuations. Thus, treating $E_J$ as a small parameter as we did in the perturbation expansion in Sec. 6 is tantamount to introducing large phase fluctuations, hence small charge fluctuations by the phase-charge Heisenberg relation $[n,\phi_x] = -i$ or $\Delta n\,\delta\phi_x \geq 1/2$, where $\delta\phi_x$ is the variance of the phase. It follows that $\langle n\rangle$ tends to a good quantum number[*] when $E_J$ is taken to be small. Notably, $I_S = 2e\partial\langle n\rangle/\partial t = 2e\langle[n,E]\rangle = 2eE_J\langle\sin(2\phi_x)\rangle$ does not vanish for large $E_J$ since the quantum average is over well defined phase states, thus $\langle\sin 2\phi_x\rangle = \sin 2\phi_x$. Conversely, since $I_S$ is an odd function of $\phi_x$, it vanishes in the charge regime where the average $\langle\cdots\rangle$ is a Gaussian path integral over $\phi_x$. This is simply the Coulomb blockade of Cooper pairs, equivalent to the vanishing of the first term $\langle I_J(0)\rangle = 0$ in the perturbation series [eq. (6)]. The next term in the perturbation series gives rise to the supercurrent in eq. (4.45) and is proportional to $E_J^2 \propto (\Delta n)^2$. It also vanishes when the Cooper pair charge is extremely well defined, $E_J \to 0$, albeit less rapidly than the first term.

---

[*] A good quantum operator commutes with the Hamiltonian $H_{cp}$ of the system and thus can be simultaneously diagonalized.

# Estimation of the Coulomb Blockade Voltage at Zero Temperature

When $\langle n \rangle$ is a good quantum number, Cooper pair charge fluctuations at the capacitor are given by $Q_x^2/2C - (Q_x + 2e\langle n \rangle)^2/2C = -2e\langle n \rangle Q_x/C - 4\langle n \rangle^2 E_c = 2e\langle n \rangle V_x - 4\langle n \rangle^2 E_c \geq 0$, where we have used $Q_x = -CV_x$ as before. It is clear that fluctuations, and hence charge tunneling is precluded for $V_x \leq 4\langle n \rangle E_c/2e$. For small $E_J$, large charge fluctuations are suppressed, and we need only consider single charge fluctuations $\langle n \rangle = 1$, thus $V_x > 4E_c/2e = V_{cb}$ for a finite supercurrent.

Thus, $2eV_{cb}$ plays a dual role to $2eE_J$ in Sec. A. Since the phase is $2\pi$ periodic, we cannot naively take $2eV_{cb} \propto \delta\phi_x$. Instead, motivated by $2e\partial\langle n \rangle/\partial t \propto 2eE_J$ above, we compute

$$\delta\phi_x \propto \lim_{R\to\infty,\,\beta\to\infty} i\frac{1}{4e}\frac{\partial}{\partial t}\ln\langle\cos 2\Delta\phi_x(t)\rangle_{\phi_x} \tag{A.2}$$

to yield,

$$\begin{aligned}
\lim_{R\to\infty,\,\beta\to\infty}\int \{Z_{\text{eff}}(\omega) + n.f.\}\frac{\exp(-i\omega t)}{1-\exp(-\beta\omega)}d\omega \\
= \frac{e}{2\pi}\lim_{R\to+\infty}\int \{Z_{\text{eff}}(\omega) + n.f.\}\exp(-i\omega t)d\omega \\
= \frac{e}{\pi}\lim_{R\to+\infty}\int \frac{1/R}{1/R^2 + \omega^2 C^2}\exp(-i\omega t)d\omega \\
= \frac{e}{C}\int d\omega\,\delta(\omega)\exp(-i\omega t) = 4E_c/2e = V_{cb}, \quad (A.3)
\end{aligned}$$

where $Z_{\text{eff}}(\omega) = [1/R + i\omega C]^{-1}$. The two limits $1/R, \beta = 1/k_B T \to +\infty$ do not commute, and thus should be taken in the order shown above. Equation A.2 says that *phase fluctuations, at the ground state of the system, are proportional to* $V_{cb}$. Consequently, since the (Gaussian) ground state saturates the Heisenberg relation, $\Delta n\delta\phi_x \sim 1/2$, we have $4E_J E_c \sim \omega_p^2/2$ where the plasma frequency $\omega_p$ here plays the role of a conversion factor needed to guarantee the correct units.

# B.

# Gaussian Functional Integrals

For completeness, this section summarizes how to compute correlation functions with Gaussian functional integrals such as the ones used in Sec. 6 in the derivation of the propagator $D_{+\infty}(t)$ in eq. (4.43). Our approach differs from typical procedures with imaginary time [11]. We work with real time instead since the finite temperature propagator is *trivially* related to the zero temperature propagator [eq. (4.43b)].

Consider a quadratic action $S(X,Y)$ with $X$ as the coordinate variable, $Y$ as a fluctuation force, $a$ as a mass term and $g$ a coupling constant. The computation procedure is then as follows:

1. Take the Fourier transform of the action by substituting the Fourier or inverse Fourier transforms below

$$X(t) = \int d\omega\, X(\omega) \exp -i\omega t, \quad X(\omega) = \frac{1}{2\pi} \int dt\, X(t) \exp i\omega t,$$

$$Y(t) = \int d\omega\, Y(\omega) \exp -i\omega t, \quad Y(\omega) = \frac{1}{2\pi} \int dt\, Y(t) \exp -i\omega t$$

in the action,

$$S(X,Y) = \int dt \left[ \frac{a}{2} \left( \frac{\partial X(t)}{\partial t} \right)^2 - \frac{a}{2}\omega_0^2 X^2(t) + g X(t) Y(t) \right]$$

$$= 2\pi \int d\omega \left[ \frac{1}{2} X(\omega) G_X^{-1}(\omega) X(-\omega) + g X(\omega) Y(-\omega) \right], \quad (B.1)$$

where $a^{-1} G_X^{-1}(\omega) = (\omega + i\varepsilon)^2 - \omega_0^2$ and $G_X(t) = \int d\omega\, G_X(\omega) \exp(-i\omega t)$;
2. Perform the functional integral $\int DX \exp iS(X,Y) \propto \exp iS'(Y)$ emulating a typical

Gaussian integral

$$\int dx \exp i[a\frac{x^2}{2} + gyx] \propto \exp[i\frac{(ig)^2 y^2}{2a}] = \exp[-i\frac{g^2 y^2}{2a}]$$

$$\to S'(Y) = 2\pi \int d\omega \left[\frac{(ig)^2}{2} Y(\omega) G_X(\omega) Y(-\omega)\right]$$

$$= \frac{(ig)^2}{2 \times 2\pi} \int dt ds\, Y(s) G_X(s-t) Y(t); \quad \text{(B.2)}$$

3. Compute the correlation functions with the quadratic part of the action as follows,

$$\langle X(t_1) \cdots X(t_n) \rangle = \mathcal{Z}^{-1} \int DX\, [X(t_1) \cdots X(t_n)] \exp iS(X, Y=0)$$

$$= \left[\frac{1}{(ig)^n} \frac{\delta}{\delta Y(t_n)} \cdots \frac{\delta}{\delta Y(t_1)} \mathcal{Z}^{-1} \int DX \exp iS(X, Y \neq 0)\right]_{Y=0, \mathcal{Z}=1}$$

$$= \left[\frac{1}{(ig)^n} \frac{\delta}{\delta Y(t_n)} \cdots \frac{\delta}{\delta Y(t_1)} \exp iS'(Y)\right]_{Y=0}$$

$$= \left[\frac{1}{(ig)^n} \frac{\delta}{\delta Y(t_n)} \cdots \frac{\delta}{\delta Y(t_1)} \sum_{m=0}^{m=\infty} \frac{\{iS'(Y)\}^m}{m!}\right]_{Y=0}. \quad \text{(B.3)}$$

We require the variation $\delta/\delta Y(t)$ and the delta function $\delta(t)$ to satisfy

$$\frac{\delta}{\delta Y(t)} \frac{\delta}{\delta Y(s)} + \frac{\delta}{\delta Y(s)} \frac{\delta}{\delta Y(t)} = 0, \quad \text{(B.4a)}$$

$$\frac{\delta Y(s)}{\delta Y(t)} = \delta(t-s). \quad \text{(B.4b)}$$

Note that the anti-commutation rule in eq. (B.4a) accounts for time ordering. Since $S'(Y)$ is quadratic in $Y$, the integral vanishes for odd number of variables $n = 2N - 1$ where $N$ is a positive integer. For even number of variables $n = 2N$ we have the continuation,

$$\langle X(t_1) \cdots X(t_n) \rangle = \left[\frac{i^N}{(ig)^n N!} \frac{\delta}{\delta Y(t_n)} \cdots \frac{\delta}{\delta Y(t_1)} S'^N(Y)\right]_{Y=0, n=2N}$$

$$= \frac{(ig)^{2N}}{(ig)^{n=2N}} \frac{i^N}{N!} \frac{1}{(2 \times 2\pi)^N} \frac{\delta}{\delta Y(t_n)} \cdots \frac{\delta}{\delta Y(t_1)} \times$$

$$\prod_{m=1}^{m=N} \int ds_{2m-1} ds_{2m}\, Y(s_{2m-1}) G_X(s_{2m-1} - s_{2m}) Y(s_{2m}); \quad \text{(B.5)}$$

4. For illustration, we compute the case $N = 1$,

$$\begin{aligned}
\langle X(t_1)X(t_2)\rangle_{t_1 \neq t_2} &= \frac{i}{2 \times 2\pi} \frac{\delta}{\delta Y(t_2)} \frac{\delta}{\delta Y(t_1)} \int ds_1 ds_2 Y(s_1) G_X(s_1 - s_2) Y(s_2) \\
&= \frac{i}{2 \times 2\pi} \int ds_1 ds_2 \frac{\delta Y(s_1)}{\delta Y(t_2)} G_X(s_1 - s_2) \frac{\delta Y(s_2)}{\delta Y(t_1)} \\
&\quad - \frac{i}{2 \times 2\pi} \int ds_1 ds_2 \frac{\delta Y(s_1)}{\delta Y(t_1)} G_X(s_1 - s_2) \frac{\delta Y(s_2)}{\delta Y(t_2)} \\
&= \frac{i}{2 \times 2\pi} \{G_X(t_2 - t_1) - G_X(t_1 - t_2)\} \\
&= \frac{i}{2 \times 2\pi} \int d\omega \, [G_X(\omega) - G_X(-\omega)] \exp(-i\omega t) \quad \text{(B.6)}
\end{aligned}$$

and

$$\begin{aligned}
\langle X(0)X(0)\rangle \equiv \langle X(t_1)X(t_2)\rangle_{t_1 = t_2} &= \frac{i}{2 \times 2\pi} \{G_X(0^+) - G_X(0^-)\} \\
&= \frac{i}{2 \times 2\pi} \int d\omega \, [G_X(\omega) - G_X(-\omega)] \neq 0, \quad \text{(B.7)}
\end{aligned}$$

where $t = t_2 - t_1$ and we have used eq. (B.4a). Note that, after Fourier transforming the action given in eq. (4.37), we simply substitute $X(t) \to \phi_z(t)/\sqrt{2}$ and $G_X(\omega) \to G_{\text{eff}}(\omega) = -e^2 i \omega^{-1} Z_{\text{eff}}(\omega)$ to yield eq. (4.44).

# C.

# Perturbation expansion formula

To expand eq. (4.27) into a perturbation series, we use the expansion formula (Baker-Hausdorff lemma)

$$\exp(-A) F \exp(A) = \sum_{n=0}^{\infty} \frac{1}{n!} \underbrace{[...[F,A]...]}_{n} \qquad \text{(C.1)}$$

for non-commuting operators $F(0) = I_J(0)$ and $A(t_0) = -i \int H_J(t_0) dt_0$ together with temporal iteration to yield

$$\exp\left[i \int H_J(t_0) dt_0\right] I_J(0) \exp\left[-i \int H_J(t_0) dt_0\right]$$
$$= \sum_{n=0}^{\infty} \frac{(-i)^n}{n!} \underbrace{[...[}_{n} I_J(0) \prod_{k=0}^{n} \left\{, \int_{-\infty}^{t_k} H_J(t_{k+1}) dt_{k+1}\right]\right\} \qquad \text{(C.2)}$$

The proof of eq. (C.1) goes as follows.

$$\exp(-A) F \exp(A)$$
$$= \left[1 - A + \frac{(-A)^2}{2!} + ... + \frac{(-A)^n}{n!}\right] F \left[1 + A + \frac{A^2}{2!} + ... + \frac{A^n}{n!}\right]$$
$$= F \left[1 + A + \frac{A^2}{2!} + ... + \frac{A^n}{n!}\right] - AF \left[1 + A + \frac{A^2}{2!} + ... + \frac{A^n}{n!}\right]$$
$$+ \frac{(-A)^2}{2!} F \left[1 + A + \frac{A^2}{2!} + ... + \frac{A^n}{n!}\right] + ...$$
$$+ \frac{(-A)^n}{n!} F \left[1 + A + \frac{A^2}{2!} + ... + \frac{A^n}{n!}\right]$$
$$= F + FA - AF + \frac{1}{2!} \left[FA^2 - 2AFA + A^2F\right] + ...$$
$$+ \frac{1}{n!} \left[\binom{n}{0} FA^n - \binom{n}{1} A^1 FA^{n-1} + ... + (-1)^k \binom{n}{k} A^k F A^{n-k}\right]$$
$$= \sum_{n=0}^{\infty} \left\{\sum_{k=0}^{n} \frac{(-1)^k}{n!} \binom{n}{k} A^k F A^{n-k}\right\} \qquad \text{(C.3a)}$$

where,
$$\binom{n}{k} = \frac{n!}{n!(n-k)!} \tag{C.3b}$$
are binomial coefficients. Finally, we note that,

$$\sum_{k=0}^{0} \frac{(-1)^k}{0!} \binom{0}{k} A^k F A^{0-k} = F,$$

$$\sum_{k=0}^{1} \frac{(-1)^k}{1!} \binom{1}{k} A^k F A^{1-k} = FA - AF = [F, A],$$

$$\sum_{k=0}^{2} \frac{(-1)^k}{2!} \binom{2}{k} A^k F A^{2-k} = \frac{1}{2!}[(FA - FA)A - A(FA - FA)] = \frac{1}{2!}[[F, A], A],$$

$$\sum_{k=0}^{n} \frac{(-1)^k}{n!} \binom{n}{k} A^k F A^{n-k} = \frac{1}{n!}[...[F\underbrace{, A]...]}_{n},$$

and thus,

$$\exp(-A)F \exp(A) = \sum_{n=0}^{\infty} \left\{ \sum_{k=0}^{n} \frac{(-1)^k}{n!} \binom{n}{k} A^k F A^{n-k} \right\} = \sum_{n=0}^{\infty} \frac{1}{n!}[...[F\underbrace{, A]...}_{n}]. \tag{C.3c}$$

# D.

# Causal Linear Response

This section is meant for the skimming reader, who wants a quick reference to linear response (and its relevance to microwave power renormalization). Thus, we do not strive to introduce the entire subject of linear response and its subtleties. (For a comprehensive introduction to the subject, see e.g ref. [112])

Within Linear Response Theory, the response $\tilde{R}(t)$ of a system is related to the driving force, $\tilde{F}(t)$ by the central causal relation

$$\tilde{R}(t) = \int_{-\infty}^{t} \chi(t-s)\tilde{F}(s)ds, \tag{D.1}$$

where $\chi(\tau)$ is the response function. The system variable, $\tilde{R}(t)$ obeys some equation of motion,

$$f(\partial/\partial t)\tilde{R}(t) = \tilde{F}(t), \tag{D.2}$$

with $f(\partial/\partial t)$ a function of $\partial/\partial t$. Introducing the Green's function of the system, $G_{\tilde{R}}(t)$, satisfying,

$$f(\partial/\partial t)G_{\tilde{R}}(t-s) = \delta(t-s), \tag{D.3}$$

we see that,

$$f(\partial/\partial t)\tilde{R}(t) = \int_{-\infty}^{t} f(\partial/\partial t)G_{\tilde{R}}(t-s)\tilde{F}(s)ds$$

$$= \int_{-\infty}^{t} \delta(t-s)\tilde{F}(s)ds = \int_{0}^{+\infty} \delta(\tau)\tilde{F}(t-\tau)d\tau$$

$$= \int_{-\infty}^{+\infty} \delta(\tau)\theta(\tau+0^+)\tilde{F}(t-\tau)d\tau = \theta(0^+)\tilde{F}(t) = \tilde{F}(t) \tag{D.4}$$

with $\theta(\tau)$ the Heaviside function. Thus, we can equate the Green's function to the response function: $\chi(t-s) = G_{\tilde{R}}(t-s)$.

Substituting the Fourier transforms,

$$\tilde{R}(t) = \int d\omega \tilde{R}(\omega) \exp(i\omega t)$$

$$\tilde{F}(t) = \int d\omega \tilde{F}(\omega) \exp(i\omega t),$$

into eq. (D.1),

$$\int d\omega \tilde{R}(\omega) \exp(i\omega t) = \int_{-\infty}^{t} \int \tilde{F}(\omega) \exp(i\omega t') \chi(t-t') d\omega dt'$$

$$= \int \left\{ \tilde{F}(\omega) \int_{0}^{+\infty} \chi(\tau) \exp(-i\omega\tau) d\tau \right\} \exp(i\omega t) d\omega = \int \left\{ \tilde{F}(\omega) \Xi(\omega) \right\} \exp(i\omega t) d\omega, \quad (D.5)$$

where $\tau = t - t'$ and,

$$\tilde{R}(\omega) = \Xi(\omega)\tilde{F}(\omega), \tag{D.6a}$$

$$\Xi(\omega) = \int_{0}^{+\infty} \chi(\tau) \exp(-i\omega\tau) d\tau = \int_{-\infty}^{+\infty} \theta(\tau)\chi(\tau) \exp(-i\omega\tau) d\tau \tag{D.6b}$$

$$2\pi\theta(\tau)\chi(\tau) = \Xi(t) \tag{D.6c}$$

Finally, that $\Xi(\omega) = \pm \exp(-\beta M) \exp i\eta(\omega)$ acts as the response function to the applied oscillating electromagnetic field is to be understood as the result of the arguments in Sec. 13, and not necessarily the converse. This leaves the possibility that linear response is violated in complicated circuits, where novel physics may lurk.

# E.
# The Array as an Effective Single Junction

It is prudent to highlight the ingredients that went into deriving the $I$–$V$ characteristics of the single small Josephson junction given in eq. (4.45):

1) The Caldeira-Leggett action $S_z(\phi'_x)$ that is varied with respect to $\phi = \phi_J + \phi'_x$ to obtain the equation of motion for the single *large* Josephson junction;
2) the correlation $\langle \sin[\kappa \Delta \phi_J(t)] \rangle_{\phi_J}$ calculated with respect to $\phi_J$;
3) The condition $\sum_i \phi_i = e\Phi$ enforced by the circuit.

In the case of a one dimensional array of $N_0$ small Josephson junctions, it is clear that constraint 3) has to include all the phases $\phi_{J=1} \cdots \phi_{J=N_0}$ of the junctions along the array, and the quantum average in 2) taken over each phase where the action in 1) is the sum of the action of individual junctions in the array. To simplify the calculation, one assumes all the junctions have the same structure constant $\alpha_\kappa(t)$ and calculates the quantum average $\langle \sin[\kappa \Delta \phi_{J=1}(t)] \rangle_{\phi_1 \cdots \phi_{N_0}}$ [step 2)]. Since the same current passes through all the junctions in the array, the calculation is carried out at any one of them [e.g. the J = 1 junction] while assuming that the circuit forms a loop that imposes condition 3) as before. Hence, treating the other junctions as environments each with an effective action of the form $S'_z(\phi_J)$, we have,

$$I_A(V) = ie \sum_{\kappa=1}^{2} \int_{-\infty}^{+\infty} dt\, \alpha_\kappa(t) \langle \sin[\kappa \Delta \phi_{J=1}(t)] \rangle_{\phi_z \phi_1 \cdots \phi_{N_0}}$$

$$= -ei \sum_{\kappa=1}^{2} \int_{-\infty}^{+\infty} dt\, \alpha_\kappa(t) \sin\left[\kappa e \int_0^t A(\tau) d\tau + \kappa \Delta \phi_x(t)\right] \times$$

$$\int \prod_{z'=1}^{N_0} D\phi_{z'} \exp iS'_z(\phi_{z'}) [\cos \kappa \Delta \phi_{z'}(t)], \quad \text{(E.1)}$$

where,

$$-\langle \sin[\kappa\Delta\phi_{J=1}(t)]\rangle_{\phi_z\phi_1\cdots\phi_{N_0}} = \langle \cos[\kappa\Delta\phi_z(t)]\rangle_{\phi_z} \times$$

$$\left\langle \sin\left[\kappa\sum_{J=2}\Delta\phi_J(t) + \kappa e\int_0^t A(\tau)d\tau + \kappa\Delta\phi_x(t)\right]\right\rangle_{\phi_2\cdots\phi_{N_0}}$$

$$= \prod_{z'}^{N_0} \langle\cos[\kappa\Delta\phi_{z'}(t)]\rangle_{\phi_{z'}} \sin\left[\kappa e\int_0^t A(\tau)d\tau + \kappa\Delta\phi_x(t)\right]$$

$$= \exp\left[-\frac{\kappa^2}{2}\sum_{z'}^{N_0}\langle\Delta\phi_{z'}^2(t)\rangle_{\phi_{z'}}\right]\sin\left[\kappa e\int_0^t A(\tau)d\tau + \kappa\Delta\phi_x(t)\right],$$

$$= \exp\kappa^2\left[\sum_{z'}^{N_0}\mathcal{J}_{z'}(t)\right]\sin\left[\kappa e\int_0^t A(\tau)d\tau + \kappa\Delta\phi_x(t)\right]$$

$$= \prod_{z'}^{N_0} P_\kappa^{z'}(t)\sin\left[\kappa e\int_0^t A(\tau)d\tau + \kappa\Delta\phi_x(t)\right].$$

Evidently, the current depends on the product

$$P_\kappa^A(t) = \prod_{z'} P_\kappa^{z'}(t)$$

as expected, since, it is comprised of individual tunneling events at each junction. Since, $1/2\pi\int dt P_\kappa^A(t)\exp iEt$ is the probability that the array will absorb energy $E$ from the environment, we discover that the tunnel current obeys the product rule of probabilities. For identical junctions of capacitance $C$ and impedance $Z(\omega)$,

$$\langle\phi_z(t)\phi_z(0)\rangle \equiv \langle\phi_{z'=2}(t)\phi_{z'=2}(0)\rangle = \cdots = \langle\phi_{z'=N_0}(t)\phi_{z'=N_0}(s)\rangle,$$

we have $\prod_{z'}^{N_0} P_\kappa^{z'}(t) = [P_\kappa(t)]^{N_0}$ where

$$[P_\kappa(t)]^{N_0} = \exp\left(N_0\kappa^2\langle[\phi_z(t) - \phi_z(0)]\phi_z(0)\rangle_{\phi_z}\right) = \exp\left(N_0\kappa^2\mathcal{J}_A(t)\right), \quad (E.2)$$

where $Z_{\text{eff}}^A(\omega)$ in $\mathcal{J}_A(t) = \langle[\phi_z(t) - \phi_z(0)]\phi_z(0)\rangle_{\phi_z}$ will differ from $Z_{\text{eff}}(\omega)$ in eq. (4.46a) due to possible interaction terms. Neglecting these interactions by setting $\mathcal{J}_A(t) \simeq \mathcal{J}(t)$, eq. (E.2) implies that at zero temperature, Cooper pair Coulomb blockade threshold voltage $V_{\text{cb}}^A = N_0 V_{\text{cb}}$ for the array is a factor $N_0$ larger than for the single junction.

However, the rest of the array ($\phi_{J=2}\cdots\phi_{J=N_0}$) acts as the environment for the single junction ($\phi_{J=1}$) thus introducing interaction terms. In particular, the single junction interacts with the rest of the array electromagnetically. Since, in the presence of Cooper pair solitons[32, 50] and the Meisner effect, the electromagnetic field has a finite range within an infinitely long array ($N_0 \gg 1$), the array has a cut-off number of junctions beyond which no electromagnetic interactions occur. Consequently, the effective number of junctions $N_c \leq N_0 - 1$ acting as the environment will be determined by the range of the electromagnetic field. $N_c(\Lambda)$

is independent of the magnetic field, $H$ when the superconducting islands are shorter than the penetration depth of the magnetic field, $H$. It can be evaluated by equating the Coulomb blockade voltage $V_{cb}$ (estimated by replacing $N_0$ with $N_c(\Lambda)$ and setting $\text{Re}\{Z^A_{\text{eff}}(\omega)\} \simeq \text{Re}\{Z_{\text{eff}}(\omega)\}$ in eq. (E.2) to the standard expression for the soliton threshold voltage[51] of the array, $eV^A_{cb} = eV_{cb} \simeq 2E_c[\exp(\Lambda^{-1}) - 1]^{-1}$, leading to

$$N_0 \to N_c(\Lambda) = \frac{1}{\exp(\Lambda^{-1}) - 1}, \tag{E.3a}$$

which approaches the soliton length $N_c(\Lambda) \to \Lambda$ when $\Lambda \gg 1$. However, the infinite array effectively has

$$N_c(\Lambda) + 1 = \frac{1}{1 - \exp(-\Lambda^{-1})} \tag{E.3b}$$

junctions. This means that $N_0$ in eq. (E.2) is instead rescaled to $N_c(\Lambda) + 1$. Since $V^A_{cb}(\Lambda)$ should be invariant under the transformation $N_c(\Lambda) \to N_c(\Lambda) + 1$, we find

$$\lim_{R \to +\infty} \text{Re}\{Z^A_{\text{eff}}(\omega)\} \simeq \lim_{R \to +\infty} \text{Re}\{Z_{\text{eff}}(\omega)\}$$
$$\to \lim_{R \to +\infty} \text{Re}\{Z^A_{\text{eff}}(\omega)\} \simeq \exp(-\Lambda^{-1}) \lim_{R \to +\infty} \text{Re}\{Z_{\text{eff}}(\omega)\}, \tag{E.4a}$$

we discover that switching on electromagnetic interactions leads to a rescaled impedance and a rescaled response function given by $\Xi(\omega) \to \Xi_A(\omega)\Xi(\omega) = \exp(-\Lambda^{-1})\Xi(\omega)$.

Finally, $N_c$ [eq. (E.3b)] is given by [confer: eq. (5.20b)],

$$N_c(\Lambda) = \sum_{m=-\infty}^{+\infty} \frac{1}{\Lambda^{-1} - 2\pi m i} - \frac{1}{2} = \frac{1}{2}\coth\left(\frac{1}{2\Lambda}\right) - \frac{1}{2}. \tag{E.5}$$

This result is not surprising, since we have determined the $I$–$V$ characteristics of the array by treating it as a single junction with the rest of the array acting as its environment. This means that the $N_0 - 1 = k$ junctions themselves act as bosonic excitations whose (average) number $\langle k \rangle = N_c$ determines the electromagnetic cut-off number, which is also the effective number of junctions that can be approximated as the environment of the effective single junction.

# F.
# Preliminary results: irradiated array

This appendix presents the preliminary results for the experiment described in chapter 2. Here, the experiment was carried out to examine the effect of RF electromagnetic fields on an array of 10 small Josephson junctions satisfying $0.1 < E_J/E_c < 1$ and $R_T > R_Q = 2\pi/4e^2 \simeq 6.45$ k$\Omega$, as can be seen in Table F.1, by measuring its $I$–$V$ characteristics, displayed in the inset of Fig. F.1. The measurement/simulation procedures, analyses and interpretation of results are same as the measured/simulated results provided in chapter 2. However, due to the large scatter of the experimental results, the renormalization factor could not be determined to a reasonable degree of accuracy. In Fig. F.1, a renormalization factor of $|\Xi_A| = 0.8$ is included for ease of comparison with experimental data.

**Table F.1.:** Average parameters per junction the array of 10 Aluminium (Al)/Aluminium Oxide (Al$_x$O$_y$)/Aluminium (Al) Josephson junctions whose measured characteristics, $I_0(V)$ have been used in the simulation with eq. (2.3). The parameters consecutively are, the capacitance $C$, tunnel resistance $R_T$, Josephson coupling energy $E_J$, charging energy $E_c$ and $E_J/E_c$ ratio.

| $C$ [fF] | $R_T$ [k$\Omega$] | $E_J$ [$\mu$eV] | $E_c$ [$\mu$eV] | $E_J/E_c$ |
|---|---|---|---|---|
| 0.8 | 10 | 62.5 | 96 | 0.65 |

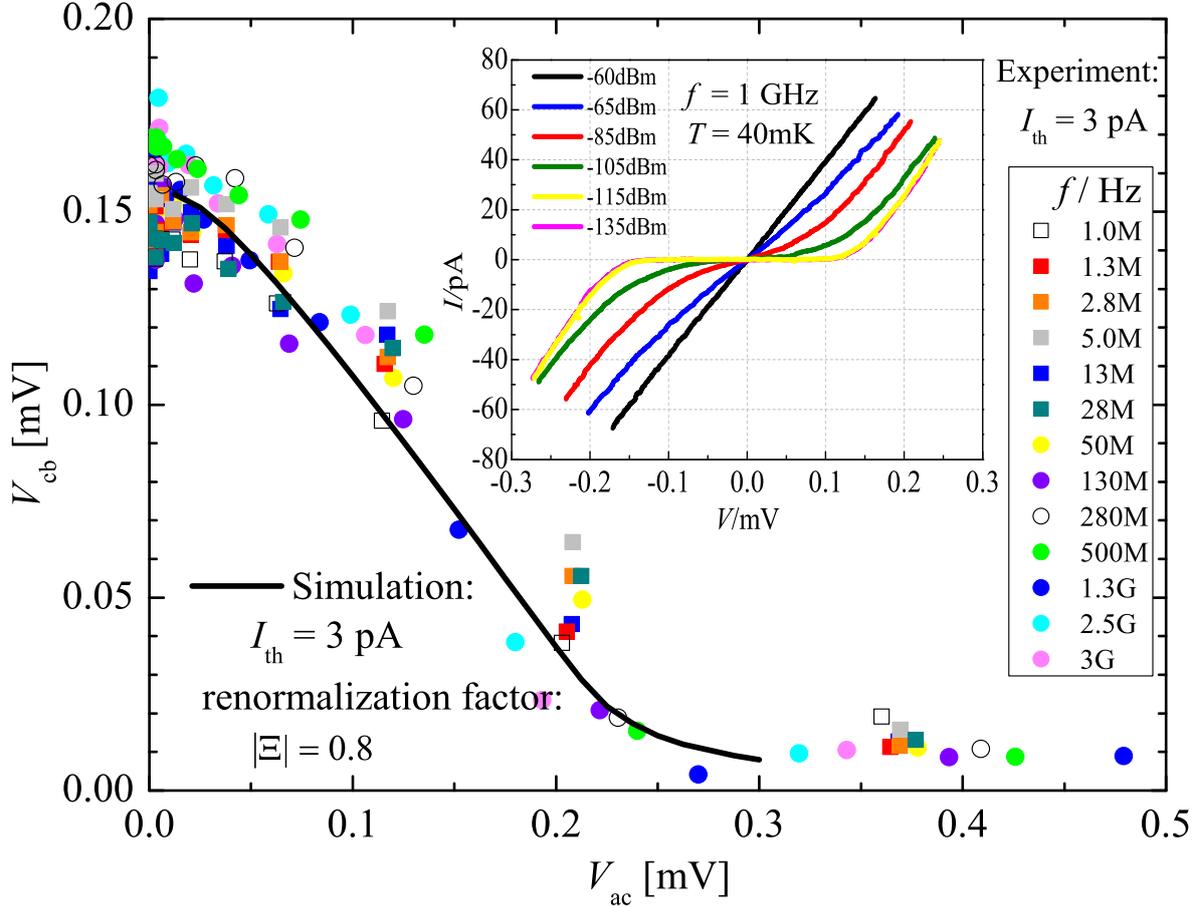

**Figure F.1.:** The dependence of Coulomb blockade voltage, $V_{cb}$ to microwave amplitude, $V_{ac}$ plotted from the measured $I$–$V$ characteristics (inset) of an array with parameters displayed in Table F.1 for magnetic field $H = 0$ Oe and frequency range $1$ MHz $\leq f \leq 3$ GHz. The coulomb blockade voltage is determined at the current value, $I_{th} = 3$ pA. The plot for the $H = 0$ Oe simulated curve using eq. (2.3) given by the black curve is represented alongside the experimental results. The simulated curve exhibits a steeper gradient than the experimental results by a factor of $|\Xi_A| = 0.80$, suggesting the need to consider effects beyond the conventional $P(E)$ theory[15, 16] in order to successfully explain the experimental results.[27, 61] Inset: The array $I$–$V$ characteristics measured at 40 mK for $H = 0$ Oe magnetic field. Different curves correspond to different values of applied microwave power range $-135$dBm $\leq P \leq -60$dBm for microwave frequency $f = 100$ MHz.

# G.
# Photon-assisted tunneling simulation

This C++ program was used to perform successfully the Cooper-pair photon-assisted tunneling integral given in eq. (2.3) using Simpson's rule and spline interpolation. The description for the code at each stage of the code is given. The "$Vac_{lower} = \cdots$" and "$Vac_{upper} = \cdots$" values can be chosen as the lower and upper bounds of the microwave power to be simulated. The default is set to $1 \times 10^{-6}$ V and $3 \times 10^{-6}$ V respectively. The measured $I$–$V$ characteristics of the unirradiated array are inputted via a data file named "Vac = 0.dat" and outputted into "Vac = $\cdots$.dat" files after computation. The default setting to output only the '$Vac_{lower} = \cdots$ .dat" and "$Vac_{upper} = \cdots$ .dat" files is setting intervals = 2. Files for the intermediate $V_{ac}$ values can be outputted by setting this interval to any integer value > 2.

```cpp
#include <fstream> //for read/write data file
#include <iostream> // for read/write data file
#include <vector> // for vector pairs data structure
#include <iomanip> // for setting precision
#include <cmath> // for sin and PI
#include <math.h> // alternative for sin and PI
#include <string> // for working with strings/text
#include <sstream> // for renaming printed files with Vac=
#include <ctime> // for timing

using namespace std;

#define M_PI 3.14159265

// create a structure/datatype to hold a single co-ordinate
struct VI {
    long double V;
    long double I;
    long double Iac;
};

struct SimpsonResult {
    long double I_simp;
    bool valid;
};

/** take a V value, and vector of points(VIs) and use linear spline
 * interpolation to find the corresponding I value
```

```cpp
    **/
long double linear_spline(long double v_find, vector <VI> pairs) {

    // if we got here then v_find is within bounds
    // find relevant indices
    int before = 0;
    int after = 0;

    // search for the relevant spline in the pairs vector
    for (int i = 0; i < pairs.size(); i++){
        if (pairs[i].V == v_find) {
            // point exists in our data set already
            return pairs[i].I;

        } else if (pairs[i].V > v_find) {
            // found our before and after
            if (i == 0) {
                // means v_find is less than lowest value
                // so we use the first spline
                after = i + 1;
                before = 0;
            } else {
                // v_find falls within bounds
                after = i;
                before = i - 1;
            }

            break;
        } else {
            // means v_find is less than current voltage
            if (i == pairs.size() - 1) {
                // means we got to end of the loop without encountering
                // a value larger than v_find
                // we use the last spline then
                after = i;
                before = i - 1;
            }
        }
    }

    // get before and after values for our spline
    long double x_before = pairs[before].V;
    long double y_before = pairs[before].I;
    long double x_after = pairs[after].V;
    long double y_after = pairs[after].I;

    long double gradient = (y_after - y_before) / (x_after - x_before);

    // do the math and return
    return y_before + (gradient * (v_find - x_before));

}
```

```cpp
/**
 * takes a filename, reads it and returns a vector of VIs
 * with the volatages and currents read into it
 * */
vector <VI> readPoints(string filename) {
    // create the vector to return
    vector <VI> VI_0;
    // open the file in read mode.
    ifstream infile;
    infile.open(filename);

    // read every line and save as V and I
    long double x_read, y_read;
    while (infile >> x_read >> y_read) {

        // create VI and save V, I for each pair
        VI temp;
        temp.V = x_read;
        temp.I = y_read;

        // save VI to Vector data structure
        VI_0.push_back(temp);
    }

    // close the opened file.
    infile.close();

    return VI_0;
}

/**
 * take an even number of intervals, Vdc to integrate, a Vac for the shift and
 * a vector of VIs for the interpolation and return the corresponding I value
 * */
SimpsonResult simpsons(int intervals, long double Vdc, long double Vac,
    vector<VI> VI_0, bool infiniteSpline) {
    // result value that includes validity
    SimpsonResult result;

    // NOTE: intervals must be even for simpson's to work
    long double d_theta = M_PI / intervals;

    long double theta = -(M_PI / 2);
    long double Iac = 0;

    // loop through all the points from -PI/2 to PI/2
    for (int i = 0; i <= intervals; i++) {

        // first shift Vdc using Vac and theta, then interpolate
        long double Vdc_shift = Vdc - (Vac * sin(theta));
        // check validity
```

```cpp
            if (!infiniteSpline && (Vdc_shift < VI_0[0].V || Vdc_shift > VI_0[
    VI_0.size() - 1].V)) {
                result.I_simp = 0.0;
                result.valid = false;
                return result;
            }
            long double Idc_shift = linear_spline(Vdc_shift, VI_0);

            if (i == 0 || i == intervals) {
                // first and last
                Iac += Idc_shift;
            } else if (i % 2 == 0) {
                // even-th points counting from 0
                Iac += 2 * Idc_shift;
            } else {
                // odd-th points counting from 0
                Iac += 4 * Idc_shift;
            }

            // increment theta accordingly
            theta += d_theta;
        }

        // finish simpson's calculation and correct value with PI
        Iac *= (d_theta / (3 * M_PI));

        result.I_simp = Iac;
        result.valid = true;

        return result;
    }

    vector <VI> multiple_Simpsons (int intervals, long double Vac, vector <VI>
    VI_0, bool infiniteSpline) {

        vector <VI> after_simpsons;

        // loop through the input values
        // apply simpson's integration to every x value
        for (int i = 0; i < VI_0.size(); i++) {
            long double Vdc = VI_0[i].V;
            long double Idc = VI_0[i].I;
            SimpsonResult result = simpsons(intervals, Vdc, Vac, VI_0,
    infiniteSpline);

            if (result.valid) {
                long double I_new = result.I_simp;

                VI temp;
                temp.V = Vdc;
                temp.I = Idc;
                temp.Iac = I_new;
```

```cpp
                after_simpsons.push_back(temp);
            } else {
                // invalid result, do nothing and skip
            }

    }

    return after_simpsons;
}

/**
 * pass a vector <VI> and print it to file with optional headers in the file
 * */
void printToFile(string filename, vector <VI> VI, bool incColHd) {

    //open file to write to
    ofstream outfile;
    outfile.open(filename);

    if (incColHd) {
        outfile << setw(20) << left << "Voltage, V (V)" <<
        /*setw(20) << left << "Current, I (A)" <<*/ setw(20) << left << "Current, Iac (A)" << endl;
    }

    // save new values to file
    for (int i = 0; i < VI.size(); i++) {
        outfile << setprecision(5) << scientific << setw(20) << left
        << VI[i].V /*<< setw(20) << left << VI[i].I */<< setw(20) << left << VI[i].Iac << "\r\n";

    }

    // close the file
    outfile.close();

    return;
}

void simpsons_range (int intervals, long double Vac_lower, long double Vac_upper,
                     int num_points, vector <VI> VI_0, bool infiniteSpline, bool incColHd) {

    // calculate step between each Vac
    long double Vac_step;
    if (num_points <= 1) {
        Vac_step = 0;
    } else {
        Vac_step = (Vac_upper - Vac_lower) / (num_points - 1);
    }

```

```cpp
        // first Vac value to compute
        long double Vac = Vac_lower;

        // loop from lower Vac to upper Vac, print out file every time
        for (int i = 1; i <= num_points; i++) {
            // get the start time
            clock_t startTime = clock();

            // get the Vac= part in the right format
            stringstream Vac_string;
            Vac_string << setprecision(1) << scientific << Vac;

            // combine two parts of the filename
            string filename_out = "Vac=" + Vac_string.str() + ".dat";

            // print our progress report
            cout << endl << "Creating " << filename_out << "... (" << i << " of " << num_points << ")" << endl;

            // do the simpson's
            vector <VI> VI_Vac = multiple_Simpsons(intervals, Vac, VI_0, infiniteSpline);

            // finally, print to file
            printToFile(filename_out, VI_Vac, incColHd);

            // increment Vac
            Vac += Vac_step;

            // get the end time and calculate
            clock_t endTime = clock();
            int seconds = (endTime - startTime) / CLOCKS_PER_SEC;

            // convert time
            int total, hours, minutes;
            total = seconds * (num_points - i);
            minutes = total / 60;
            hours = minutes / 60;
            cout << filename_out << " created successfully in " << seconds << " seconds. " << endl
                << "Total time left to create all files: " << int(hours) << " hours " << int(minutes % 60)
                    << " minutes " << int(total % 60) << " seconds." << endl;

        }

        // progress report when done
        cout << endl << "Done." << endl;

        return;
    }

int main () {
```

```cpp
    // get the original values into vector
    string filename = "Vac=0.dat";
    vector <VI> VI_0 = readPoints(filename);

    // get the ranges
    long double Vac_lower = 1e-6;
    long double Vac_upper = 3e-4;
    int num_points = 2; // number of points to do the simpson's on in the above
     (inclusive) range
    int intervals = 100; // for simpson's calculations, must be an even number!
    bool infiniteSpline = false;
    bool incColHd = false;

    // start the simpson's process for all the ranges
    simpsons_range(intervals, Vac_lower, Vac_upper, num_points, VI_0,
     infiniteSpline, incColHd);

    return 0;
}
```

# Bibliography

References in citation order.

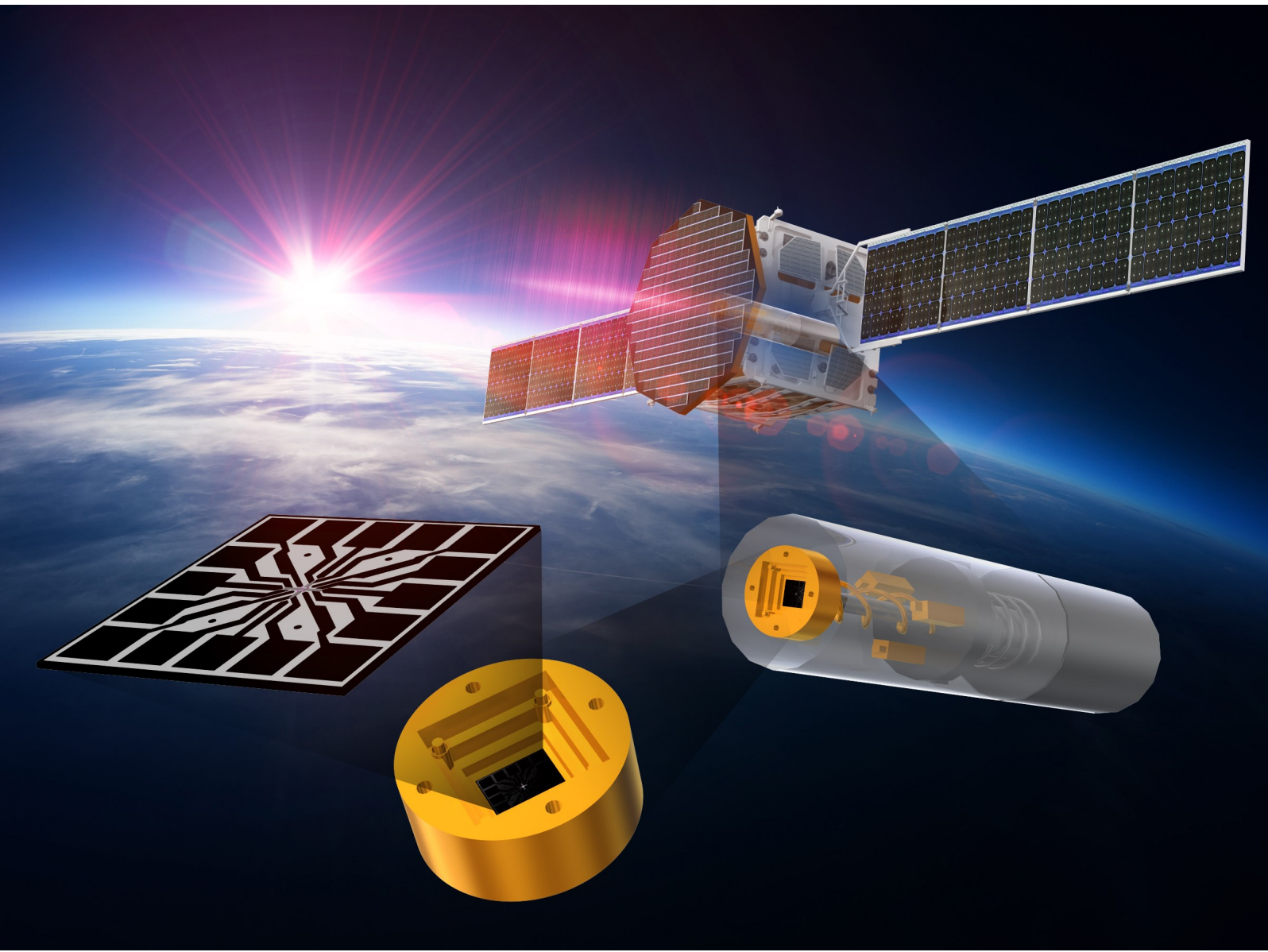